\documentclass{ws-ijmpa}
\usepackage{graphicx}
\usepackage{epstopdf} 
\usepackage{xspace}
\usepackage{tabularx}
\usepackage{hyperref}
\usepackage[super,compress]{cite}
\newcommand{\ignore}[1]{}

\newcommand{\etal}{{\it et al.}}

\newcommand{\pt}{p_T}
\newcommand{\infb}{fb$^{-1}$}

\newcommand{\ppbar}{p{\bar{p}}}

\newcommand{\phistar}{\phi^*}
\newcommand{\phiaco}{\mbox{$\phi_{\rm acop}$}}

\newcommand{\gt}{\rightarrow}

\begin{document}
\markboth{}{Review of Physics Results from the Tevatron: Electroweak Physics}

%
\catchline{}{}{}{}{}
%

\title{REVIEW OF PHYSICS RESULTS FROM THE TEVATRON: ELECTROWEAK PHYSICS}

\author{ASHUTOSH V. KOTWAL}

\address{Department of Physics, Duke University,\\
Durham, NC 27708, USA \\
ashutosh.kotwal@duke.edu}

\author{HEIDI SCHELLMAN}

\address{Department of Physics, Northwestern University, \\
Evanston, IL 60208, USA\\
schellman@fnal.gov}

\author{JADRANKA SEKARIC} 

\address{Department of Physics and Astronomy, University of Kansas, \\
Lawrence, KS 66045, USA\\
sekaric@fnal.gov}

\maketitle

\begin{history}
\received{Day Month Year}
\revised{Day Month Year}
\end{history}

\begin{abstract}
We summarize an extensive Tevatron  ($1984-2011$) electroweak physics program that involves 
a variety of $W$ and $Z$ boson precision measurements. The relevance of these studies using single 
and associated gauge boson production to our understanding of the electroweak sector, quantum 
chromodynamics and searches for new physics is emphasized. We discuss the importance of the $W$ 
boson mass measurement, the $W/Z$ boson distributions and asymmetries,  and diboson studies. We 
highlight the recent Tevatron measurements and prospects for the final Tevatron measurements.

\end{abstract}

\ccode{PACS numbers:~2.15.-y, 13.38.Be, 14.70.Fm, 12.15.Ji, 13.85.Qk, 3.38.Be, 14.60.Cd, 14.60.Ef, 
2.38.Qk, 14.70.Hp, 12.38.Qk, 14.80.Bn, 13.85.Rm,12.60.Cn, 2.15.Ji, 12.60.Cn, 13.38.Dg, 13.40.Em, 
14.70.-e, 13.85.Ni, 13.85.Ql}

\clearpage
\tableofcontents

\clearpage
\section{Introduction}	

The $SU(3)_c \times SU(2)_L \times U(1)_Y$ gauge structure~\cite{GWS} and the fermion multiplets~\cite{GWS} 
and mixing in the standard model (SM) have been impressively motivated and confirmed by the generations of 
fixed-target and collider experiments. In the area of electroweak-symmetry breaking, decades of theoretical 
and experimental effort has recently culminated in the observation of the Higgs boson at the LHC~\cite{lhchiggs}, 
as predicted by the Higgs mechanism~\cite{ewsb}. Precision measurements of Higgs boson properties, including 
its mass (which determines the quartic Higgs self-coupling coefficient in the SM) and its branching ratios 
(which test the fermion Yukawa couplings and the gauge-boson couplings predicted in the SM), are in progress 
at the LHC. 

\par Study of  electroweak  vector bosons at the Tevatron has led to major advances in Standard Model physics.  
In strong interaction physics,  the Tevatron measurements of $W$ and $Z$ production and decay have served as 
major tests of next-to-next-to-leading order (NNLO) QCD calculations, non-perturbative effects and significant constraints on parton distribution 
functions.  In weak interaction physics the $W$ mass and width have been directly measured with unprecedented 
precision while measurements of the effective Weinberg angle $\sin^2\theta^\ell_{\rm eff}$ have reached levels of 
accuracy formerly only achieved at LEP and SLD.

\par The study of electroweak vector boson self-interactions complements both direct and indirect searches for new physics 
that may exist at some energy scale $\Lambda$. 
The gauge boson self-interactions are studied via trilinear and quartric gauge boson couplings. In the presence of 
a New Physics scenario these observables are expected to deviate from their Standard Model predictions.


\section{Experimental Overview}

\subsection{Early History and Tevatron Run 1}

The weak vector bosons have been studied at the Tevatron since the first measurement of the $Z$ boson mass by the CDF collaboration in 1989 \cite{CDFZmass1989}.  That original measurement used 123 $Z^0\gt\mu^+\mu^-$ and 65 $Z^0\gt e^+e^-$  events recorded in an integrated luminosity of 4.7 pb$^{-1}$ to obtain a  $Z$ boson mass at $90.9 \pm 0.3$(stat.)$\pm 0.2$(syst.) GeV.

\label{wMassHistory}
Initial measurements of the $W$ boson mass were performed by   UA1 and UA2  after the $W$ and $Z$ boson discoveries~\cite{WZdiscovery} by these experiments at the $S p \bar{p} S$ at CERN. Increasingly more 
 precise measurements were
 performed at the CDF experiment using the Tevatron Run 0 data, and the CDF and D\O\ experiments using the Tevatron Run 1 data~\cite{CDF,DZERO}. In parallel with the latter, the electron-positron
 collider LEP II above the $Z$-boson pole started producing $W$ boson pairs, first at threshold and later above threshold. The threshold scan of cross section as a function of collider
 center-of-mass energy yielded the first LEP II measurements of $M_W$. More precise measurements resulted from higher statistics at higher energies where final-state reconstruction was
 employed for the semi-leptonic and all-hadronic decay channels.

One highlight of Run I was the  CDF and D\O\ measurements of $M_W$ that yielded~\cite{run1combo}
\begin{equation}
M_W = 80454 \pm 59 \; {\rm MeV}
\end{equation}
 and LEP II concluded with a final combined result~\cite{lepewwg} from ALEPH~\cite{ALEPH}, DELPHI~\cite{DELPHI}, L3~\cite{L3} and OPAL~\cite{OPAL} experiments of
\begin{equation}
M_W = 80376 \pm 33 \; {\rm MeV} \; .
\end{equation}

As this review attempts to summarize the field at the end of more than twenty
years of data taking, the latest measurements are generally shown instead of the first
but we reference the earlier measurements and   highlight major innovations as well as the final outcomes which
build on them.

\subsection{Evolution of the Apparatus}

The CDF detector went through a series of upgrades~\cite{NIMS}, most notably the addition of a silicon tracking detector, and was joined at the Tevatron by the D0 detector~\cite{D0NIMS} in 1992.  Both experiments were extensively upgraded between Run I, which ended in 1996 and Run II which began in 2001~\cite{RunIINIMS}. The D0 detector acquired a magnetic tracking system with silicon and scintillating fiber tracking. By the end of Run II in 2011, both detectors had recorded close to 10 \infb\ of integrated luminosity with over 500,000 reconstructed $Z\gt\ell\ell$ and millions of $W\gt \ell \nu$ decays observed in both the muon and electron decay channels. 

The two experiments now have complementary capabilities.  The CDF tracking system has a significantly larger radius, allowing very high precision momentum measurements for charged particles in the central region while the D0 tracking system includes silicon disks, which extend the angular coverage of the tracker down to 3 degrees from the beamline, and had a reversible magnetic field, allowing precision charge asymmetry measurements out to very large pseudo-rapidities. 

\subsection{Data-driven Efficiencies and Calibrations}

All of these measurements have been greatly aided by the use of data-driven measurements of detector efficiencies and calibrations. The reasonably large $Z$ boson samples at the Tevatron and our extremely precise knowledge of the $Z$ boson mass and width from LEP\cite{LEPEWWG} have made very precise calibrations and rate measurements possible.

The well known mass and identical lepton pairs from $Z$ boson decays allow the use of `tag-and-probe' measurements of detector response. The general method is to identify a sample of $Z$ boson decays in which one leg (an electron, muon or $\tau$) is very cleanly identified and then find a loosely defined second leg which is consistent with coming from a $Z$ decay.  A concrete example is a measure of tracking efficiency for electrons where the 'tag' leg is required to have both a track and an electromagnetic shower and the second 'probe' leg has an electromagnetic shower.  The 4-vectors derived from the two legs are required to be consistent with a $Z$ boson decay.  The efficiency for track finding can then be estimated from the fraction of probe legs which also have a charged track associated with them.  In practice, the  kinematic $Z$ boson requirement biases the efficiency measure by a few percent so the efficiencies derived by this method cannot be used directly.  Instead the tag-probe method is used to measure differences between data and simulation due to detector effects.  The kinematic effects cancel in the data to simulation ratio and can be used to correct simulated distributions on an event by event basis.  Reference~\citen{D0ElectronID} describes this method in more detail.
Similar methods can be used to determine  charge-misidentification  probabilities and to measure the detailed energy response of the electromagnetic calorimeter at module boundaries.  The use of these methods has reduced the systematic uncertainties due to detector efficiency in total cross section measurements below 0.5\% and have allowed the calibration of absolute electromagnetic energy scales at the 0.02\% level.


\section{Total and Differential Cross Section Measurements}

\subsection{Cross Section Measurements}

The inclusive cross section for vector boson production via the Drell-Yan process\cite{DrellYan}
is a convolution of
partonic hard scattering cross sections  with parton distribution functions (PDF's), which carry information about the momentum fraction of the proton carried by each parton type. 
\begin{eqnarray}
\sigma(p_1\overline{p}_2\gt V+X) = \sum_{ij} \int f_i(x_1,Q^2) \bar{f}_j(x_2,Q^2) dx_1 dx_2
\hat{\sigma}(q_i+\bar{q}_j \rightarrow V+X) 
\end{eqnarray}
where the $f_i$ are parton distribution functions, $x_1, x_2$ are the fractional
momenta carried by the partons $q_i$ and $\bar{q}_j$ in the initial  proton $p_1$ and anti-proton $\bar{p}_2$, $Q^2$ is
the momentum transfer squared, and $\hat{\sigma}$ is the parton level cross section.  $q_i$ and $\bar{q}_j$ are generally valence quarks at leading order but higher order diagrams with gluons and sea quarks  in the initial state also play a  role.  The totally inclusive vector boson cross section was predicted to NNLO  in the early 90's \cite{Hamberg1990}, with very small theoretical uncertainties aside from those from the PDFs.

Measurements of the total cross section thus provide constraints on parton distribution functions but also, through cross section ratios, can be used to set indirect limits on the width of the $W$ boson.  The measured observable is not the total production cross section but the cross section times branching ratio into the observed final state.  $e^+e^-$, $\mu^+\mu^-$ or $\tau^+\tau^-$ for $Z$ bosons, $e, \mu$ or $\tau$ + missing  $E_T$ for $W$ bosons.  This also provides a test of lepton universality when different decay channels are compared.

\begin{figure}[htpb]
\centerline{\includegraphics[width=12cm]{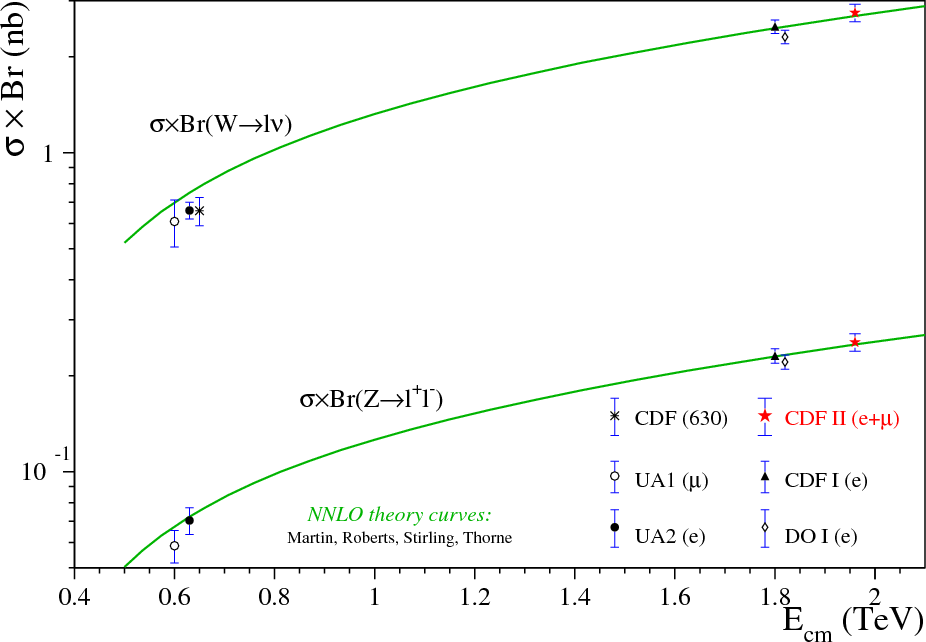}}
\caption{$W\gt l\nu$ and $Z\gt ll$ cross section measurements as a function of the $\ppbar$ center-of-mass energy. The solid lines correspond to the theoretical NNLO SM calculations. \label{fig:Total}}
\end{figure}

Both the CDF \cite{CDFCross1992,CDFCross1994, CDFCross1995,CDFCross1996, CDFCross1998,CDFCross1999} and D0\cite{D0Cross1995, D0WZCross1999, D0WZWidth1999} collaborations measured the total production cross sections  in Run I.  The most precise Run II Tevatron total cross section measurement in the electron and muon channels was performed by CDF \cite{CDFCross2004, CDFCross2005} using 72 pb$^{-1}$ of data taken at center of mass energy $\sqrt{s}$ = 1.96 TeV. Figure \ref{fig:Total} summarizes the Tevatron cross section measurements.  The uncertainties are dominated by the estimated integrated luminosity, not the $Z$ or $W$ signals.  

\subsubsection{Indirect Measurement of the $W$ Boson Width}

The ratio of the measured cross sections time branching ratios $R= 10.84\pm 0.15$~(stat.)~$\pm 0.14$~(syst.).This can be recast in terms of the boson decay widths

\begin{equation} R = \frac{\sigma(\ppbar \gt W+X)\times B(W\gt\ell\nu)}{\sigma(\ppbar\gt Z+X)\times B(Z\gt \ell\ell)} \end{equation}

\begin{equation} R =\frac {\sigma(\ppbar \gt W+X)\times \Gamma(W\gt\ell\nu)}{\sigma(\ppbar\gt Z+X)\times\Gamma(Z\gt \ell\ell)} \times \frac{\Gamma_{tot}(Z)}{\Gamma_{tot}(W)}\end{equation}

If an NNLO QCD calculation is used to obtain the total cross section ratio, the Standard model is used to obtain the leptonic decay widths, and the LEP measurement of the $Z$ boson width is used, $R$ can be used to estimate $\Gamma_{tot}(W)= 2092 \pm 42$ MeV.

\subsubsection{PDF Constraints}

\begin{figure}[htpb]
\centerline{\includegraphics[width=12cm]{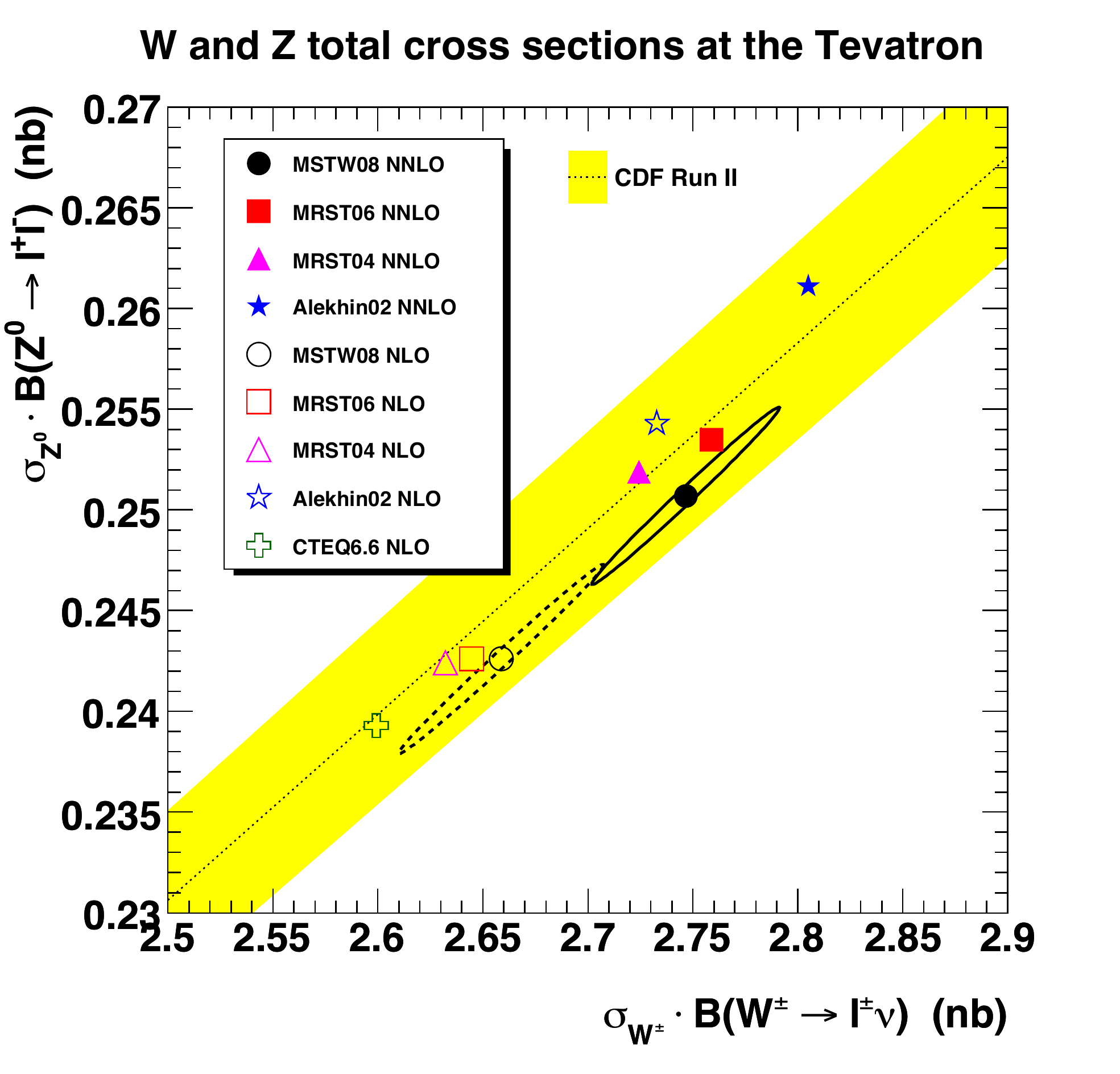}}
\caption{$W\gt l\nu$ and $Z\gt ll$ cross section ratio measurement  compared to NNLO SM calculations with different PDF sets.  The yellow band  is the experimental measurement while the points denote predictions from different PDF sets. The ellipses illustrate the estimated input errors in the PDF fits for MSTW08 NLO (dashed) and MSTW08 NNLO (solid). From Ref.~\citen{MSTW2008}\label{fig:MSTWCROSS}.}
\end{figure}

Alternatively, if SM decays for the $W$ and $Z$ bosons are assumed and NNLO QCD cross sections are used, the total cross section times branching ratio measurements can be used to constrain parton distribution sets. Figure \ref{fig:MSTWCROSS} shows a comparison of theoretical predictions from different PDF sets~\cite{MSTW2008,MRST2004,CTEQ61,CTEQ66,Alekhin02} to the $R$ value determined by CDF \cite{CDFCross2005}.

\subsection{Differential Distributions for Vector Boson Production}

The differential Drell-Yan cross sections provide a significant test of perturbative and non-perturbative QCD and of PDF sets.  Both CDF and D0 have published differential distributions for $W$ and $Z$ boson production as a function of the boson transverse momentum $\pt$\cite{CDFCross1999,CDFCross1989,CDFCross1990,CDFZPT1991, D0WPT1998,D0ZPT1999,D0ZPT2000, D0WPT2001, D0WZPT2001,D0ZPT2008,D0ZPT2010,CDFPT2012} and rapidity $y$\cite{D0ZY2007,CDFZY2010}.

\subsubsection{Rapidity Distributions}

\begin{figure}[htpb]
\centerline{\includegraphics[width=10cm]{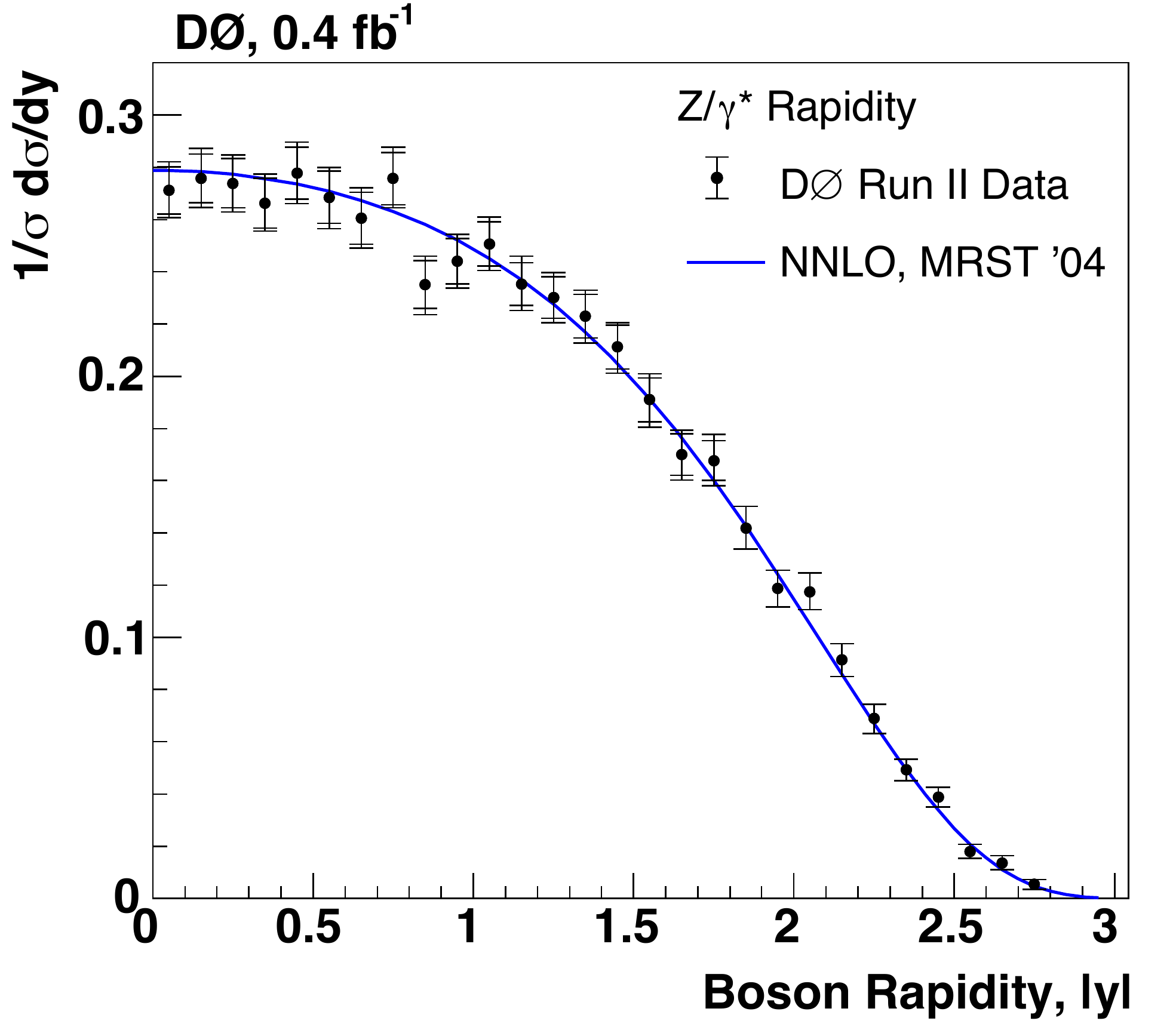}}
\caption{D0 data on the $Z$ boson rapidity from 0.4 \infb of data at $\sqrt{s} = 1.96 $ TeV in the $e^+e-$ decay channel compared to theoretical predictions at NNLO from Ref.~\citen{ADMP} using PDF's from Ref.~\citen{MRST2004}. }\label{fig:D0Y}
\end{figure}

\begin{figure}[htpb]
\centerline{\includegraphics[width=12cm]{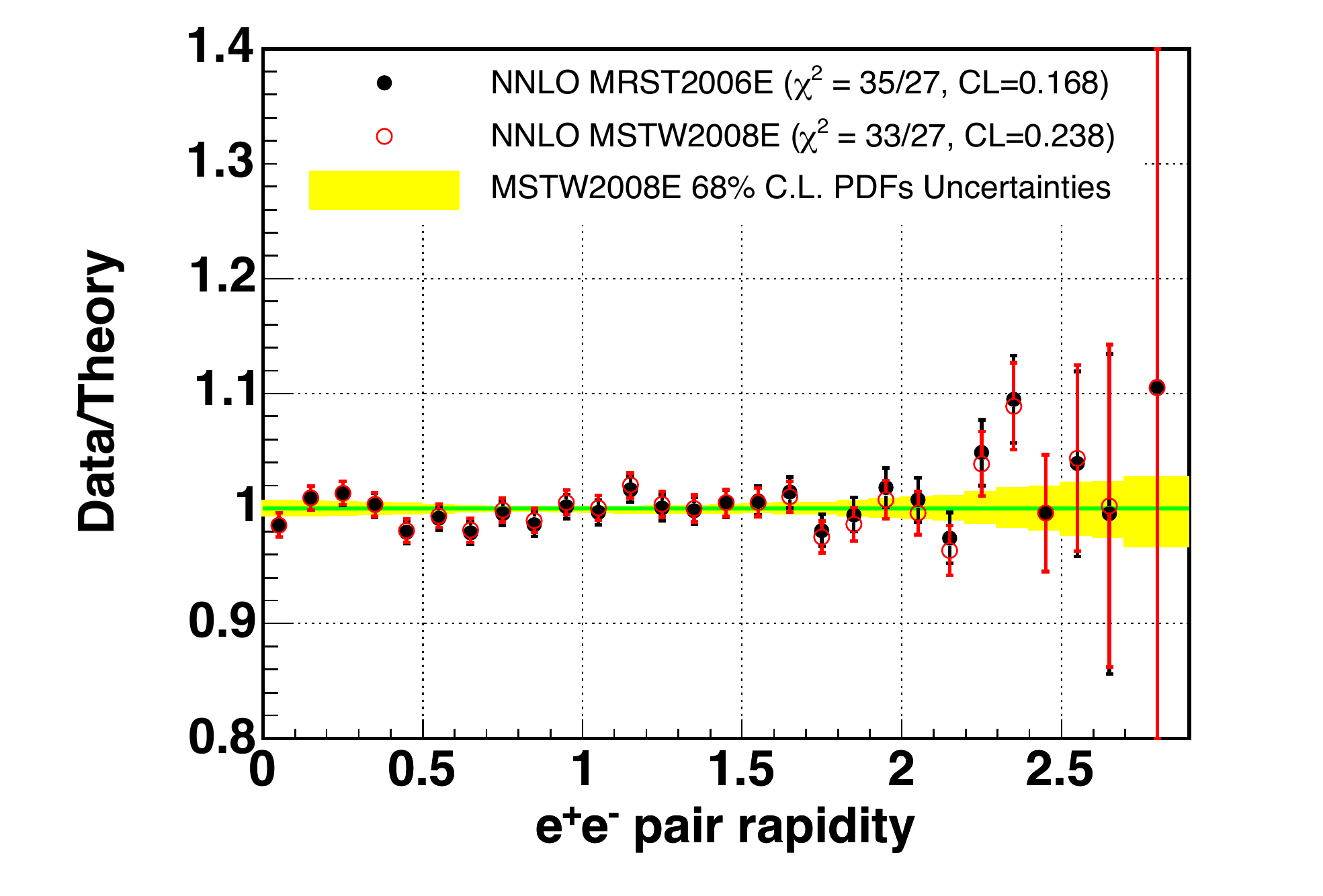}}
\caption{Comparison of the ratio of higher statistics  CDF data for the $Z$ boson rapidity from  2.1 \infb of data   to theoretical predictions at NNLO using PDF's from Ref.~\citen{MSTW2008}.}\label{fig:CDFY}
\end{figure}

The advent of reasonably fast computational techniques for calculating differential distributions for vector boson production at NNLO \cite{ADMP, Catani2007} in the mid-2000's led to increased interest in precision measurements of the rapidity distribution for $Z$ boson production.   With the advent of high statistics measurements from D0 \cite{D0ZY2007} and CDF \cite{CDFZY2010} the rapidity distribution became a testing ground for NNLO QCD calculations and for new PDF sets. Two such measurements are shown in Figure~\ref{fig:D0Y} and Figure~\ref{fig:CDFY}.

\subsubsection{$\pt$ Dependence}

The $\pt$ dependence of vector boson production at colliders was one of the first NNLO QCD predictions for hadron colliders\cite{ArnoldReno}.  Over the past two decades both measurements and theoretical predictions have grown increasingly precise with the state-of-the-art now including NNLO calculations at high virtuality but requiring a significant non-perturbative component at small transverse momentum.

The most precise studies have been done using fully leptonic decays of the $Z$ boson. 
As the statistical precision of the data has improved,  experimental limitations due to finite resolution in the $Z$ boson transverse momentum $\pt^Z$, have led to the introduction of new variables\cite{aT_papers}  with better experimental sensitivity. In particular, the D0 collaboration in Ref.~\citen{D0ZPT2010} used the $\phi^*$ variable 

\begin{equation}\phistar = \tan\left(\phiaco/2\right)\sin(\theta^*_{\eta})\end{equation}
 where \phiaco\ is the acoplanarity angle ($\phiaco\ = \pi - \Delta\phi_{\ell \ell}$ and $\Delta\phi_{\ell \ell}$ is the azimuthal separation between the two decay leptons in the transverse plane).
 The variable $\theta^*_{\eta}$ estimates the angle between the scattered leptons in the $Z$-boson center of mass frame and the proton beam direction. 
$\cos(\theta^{*}_{\eta})=\tanh\left[\left(\eta^--\eta^+\right)/2\right]$, 
where $\eta^-$ and $\eta^+$ are the pseudorapidities of the decay leptons. Since the $\phistar$ variable uses only angular information, it is measured more precisely than the boson $p_T$ which depends on the lepton $p_T$. 

\begin{figure}[htpb]
\centerline{\includegraphics[width=13cm]{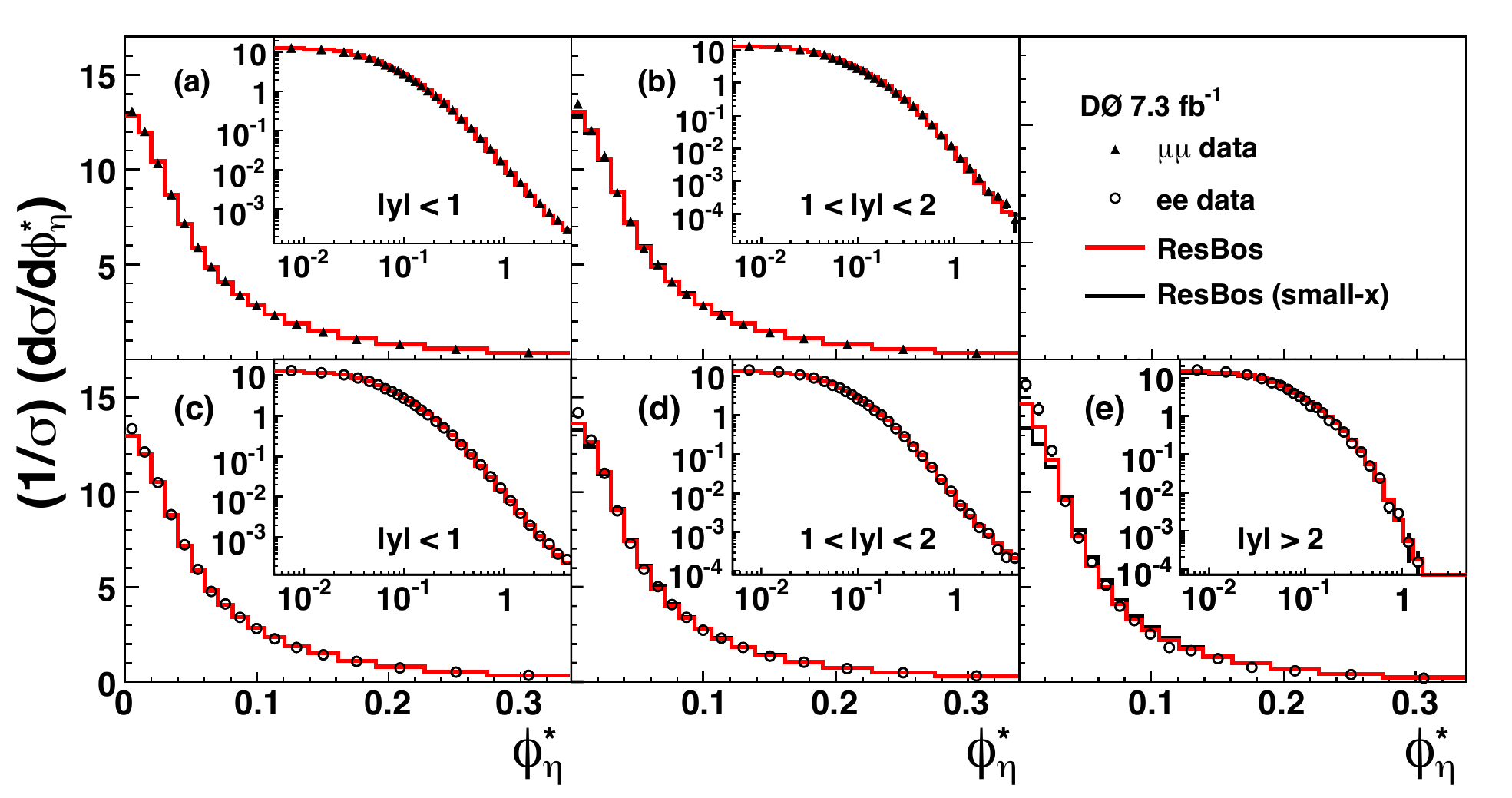}}
\caption{Comparison of D0 data for the $Z$ boson $\phistar$ variable for muons (top) and electrons (bottom) in different rapidity bins compared to the RESBOS NLO model with nonperturbative corrections. \label{fig:phistar}}
\end{figure}

The $\phistar$ variable is more sensitive to the true boson boost than a direct $\pt$ measurement at low transverse momentum and allows more stringent tests of models in that kinematic regime. Figure \ref{fig:phistar} shows the measured distributions from 7.3 \infb\  of data  collected by the D0 detector compared to a standard \textsc{RESBOS} \cite{ResBos} next-to-leading order (NLO) calculation which includes non-perturbative effects and a variant of that calculation with an enhancement at low parton $x$ \cite{smalls}.

\begin{figure}[htpb]
\centerline{\includegraphics[width=13cm]{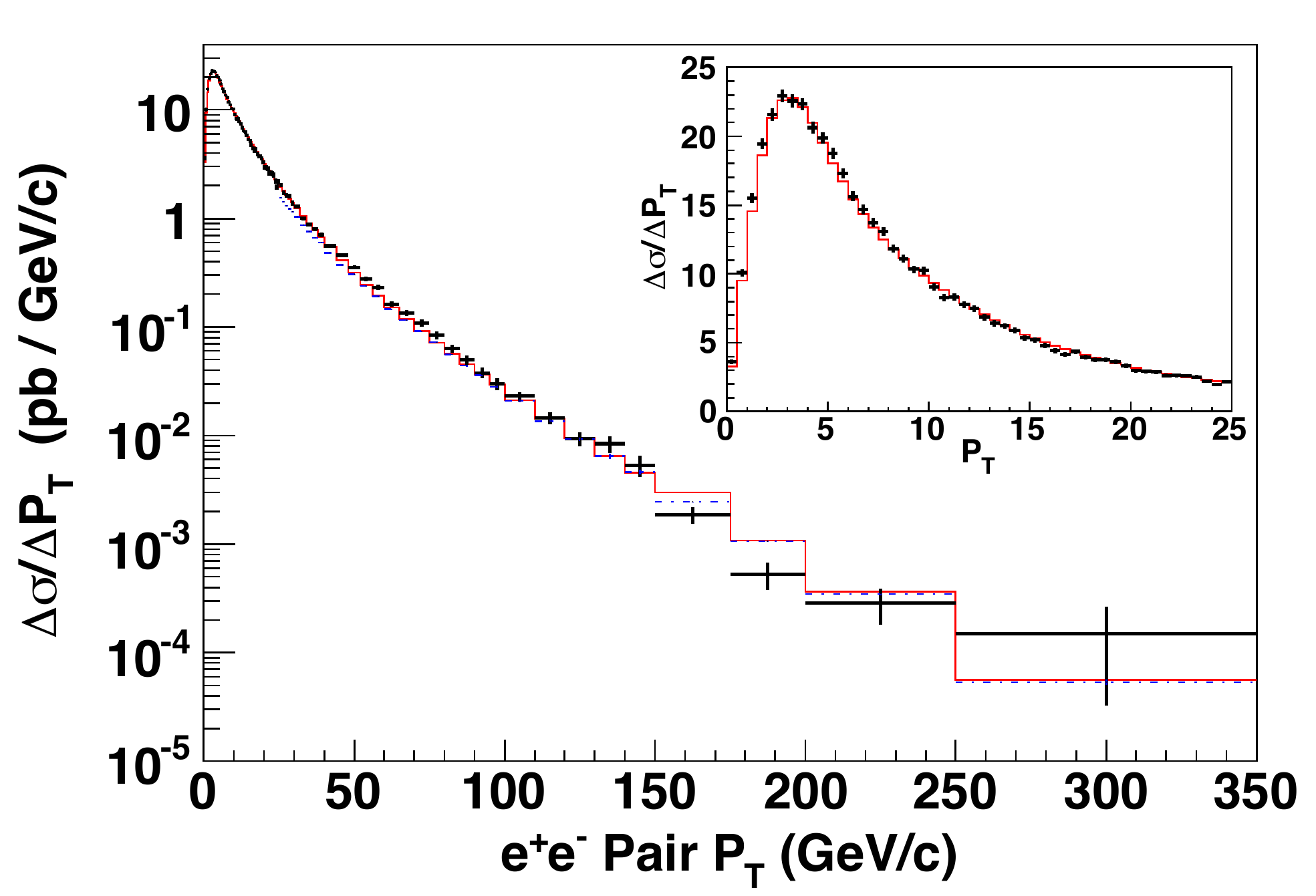}}
\caption{Comparison of CDF data for the $Z$ boson $\pt$ variable for electrons compared to the \textsc{FEWZ2} NNLO QCD calculation (dashed blue histogram above 25 GeV) and the \textsc{RESBOS} NLO+nonperturbative calculation (solid red histogram). \label{fig:CDFpt}}
\end{figure}

At high momentum transfer, the traditional $\pt$ variable is more powerful and provides stringent tests of perturbative QCD at NNLO. Figure \ref{fig:CDFpt} shows recent high statistics results from 2.1 \infb\ of data collected with the CDF detector \cite{CDFZPT2012} compared to the NNLO \textsc{FEWZ2}\cite{FEWZ2} calculation. 

\subsection{Associated jet production}
Studies of $V+$jets (differential in N$_{\rm jets}$, jet flavors, jet $p_T$ and $\eta$ etc.) have been a major theme of the Tevatron program.  These studies have being discussed in the QCD chapter of this review~\cite{ijmpaQCD}.

\subsection{Asymmetry Measurements}

\subsubsection{W Asymmetry}

The charge asymmetry of  $W$ bosons produced at the Tevatron is related\cite{berger} to the $u$ and $d$
quark parton distribution functions as:

\begin{equation} A(y_W) = {{d\sigma\over dy}(W^+)- {d\sigma\over dy}(W^-)\over {d\sigma\over dy}(W^+)+ {d\sigma\over dy}(W^-)}\simeq {{u(x_1)/d(x_1) - u(x_2)/d(x_2)}
\over {u(x_1)/d(x_1) + u(x_2)/d(x_2)}}\label{eq:wasym}
\end{equation}

where $x_1$ and $x_2$ are the momentum fractions carried by the quarks in the
proton and anti-proton respectively and $y_W$ is the boson rapidity. This leading order parton-level expression ignores potential contributions
from flavor asymmetries  in the sea quarks.  

Early Tevatron measurements \cite{CDFWasym1994, CDFWasym1998,
CDFWasym2005, D0MUASYM, D0Wasym2008, D0WAsymMuon2013}
did not directly measure the $W$ charge asymmetry but instead the charge asymmetry of the decay leptons as that can be directly observed where the $W$ signature includes a missing neutrino.  Because the $V-A$ nature of the decay, the lepton tends to go backwards in the boson frame, thus washing out the effect.

\begin{figure}[htpb]
\centerline{\includegraphics[width=12cm]{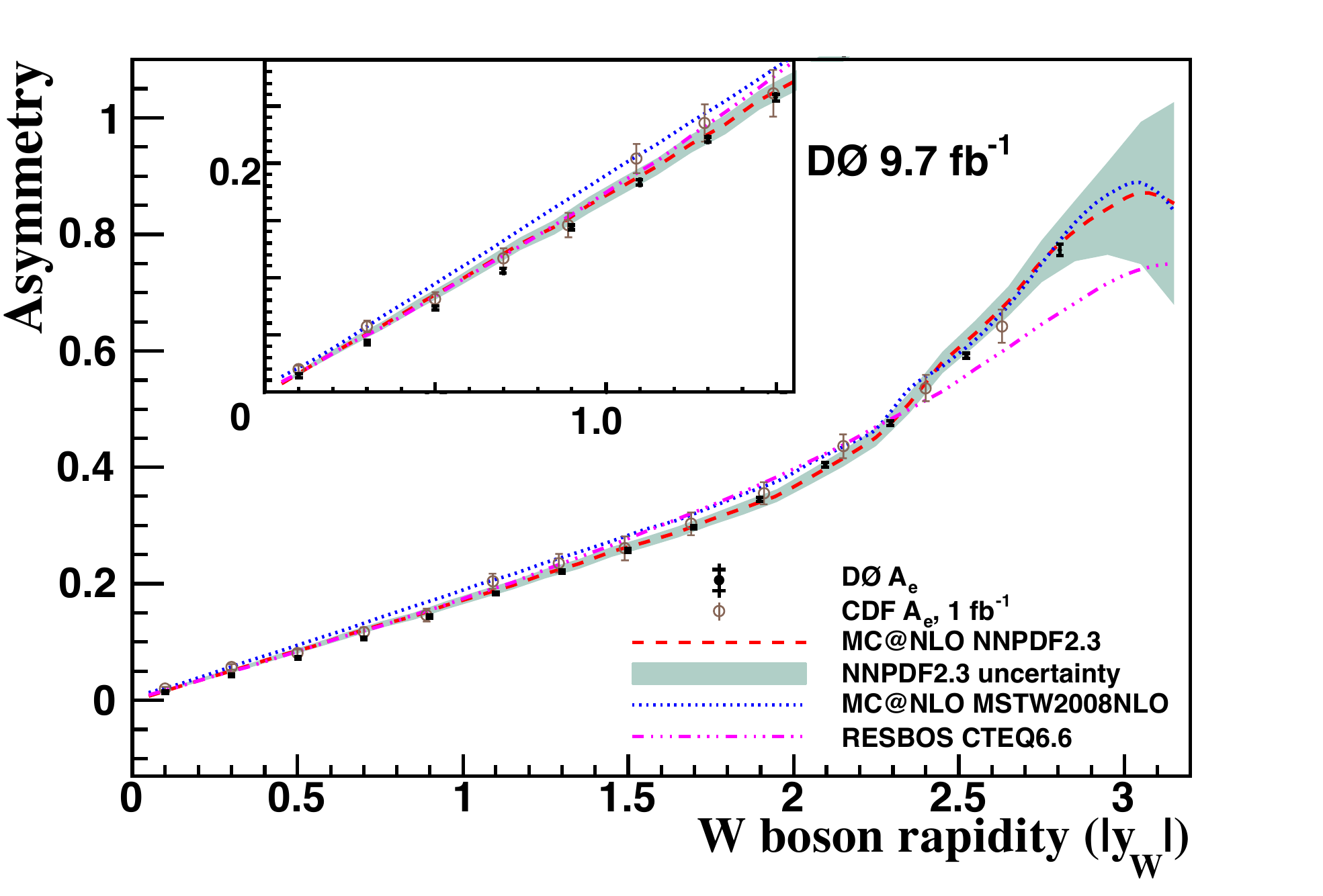}}
\caption{Data from the CDF (open circles) and  D0 (closed circles) measurements of the $W$ boson charge asymmetry as a function of $W$ rapidity $y_W$.   The red-dashed curve shows the NNPDF2.3\cite{NNPDF2.3} PDF set with its error set, the blue-dotted curve shows the MSTWNLO2008 set and the pink-dot-dashed curve shows the CTEQ6.6M PDF set.  The inset shows more detail in the region close to $y_W=0$. \label{fig:wasym}}
\end{figure}

 In 2007, Bodek \etal. \cite{Wasym}  proposed a new method for using a $W$-mass constraint to determine the neutrino momentum, with two solutions for the longitudinal momentum.  Their method depends, to some extent, on theoretical models of $W$ boson production and decay to determine the relative weights for the two neutrino solutions, but allows reconstruction of the $W$ boson rapidity.  Both the CDF\cite{CDFWasym2009} and D0\cite{D0Wasym2014} collaborations have used this technique to make direct measurements of the $W$ boson asymmetry which allows a much more direct estimate of the parton probabilities ratios in Equation \ref{eq:wasym}.   Figure \ref{fig:wasym} shows the CDF and D0 results compared to recent theoretical calculations.  The CDF result used 1~fb$^{-1}$ of integrated luminosity while the later D0 measurement used 9.7~fb$^{-1}$. The $W$ boson charge asymmetry measurement constrains the PDFs needed for precise modeling of $W$ boson production in order
 to measure its mass. The experimental uncertainties are smaller than the current PDF uncertainties, and therefore help to constrain the PDFs.

\subsubsection{$A_{FB}$}

\newcommand{\sineff}{\sin^2\theta^\ell_{\rm eff}}
Both the CDF\cite{CDFAFB1991, CDFAFB1996, CDFAFB2013, CDFAFB2014}
 and the D0\cite{D0AFB2011} collaborations have performed measurements of the effective weak mixing angle $\sineff$ using the forward-backward charge asymmetry in measured in Drell-Yan production around the $Z$-pole.  The standard measurement method used in most of the CDF measurements and  the D0 measurement is to count events with the electron going forward~(F) or backwards~(B) in the Collins-Soper frame\cite{CSframe}.
 
 The asymmetry $A_{FB}$ is then defined as:
 
 \begin{equation} A_{FB} = \frac{\sigma_F-\sigma_B}{\sigma_F+\sigma_B}\label{eq:added1}\end{equation}
 
\begin{figure}[htpb]
\centerline{\includegraphics[width=10cm]{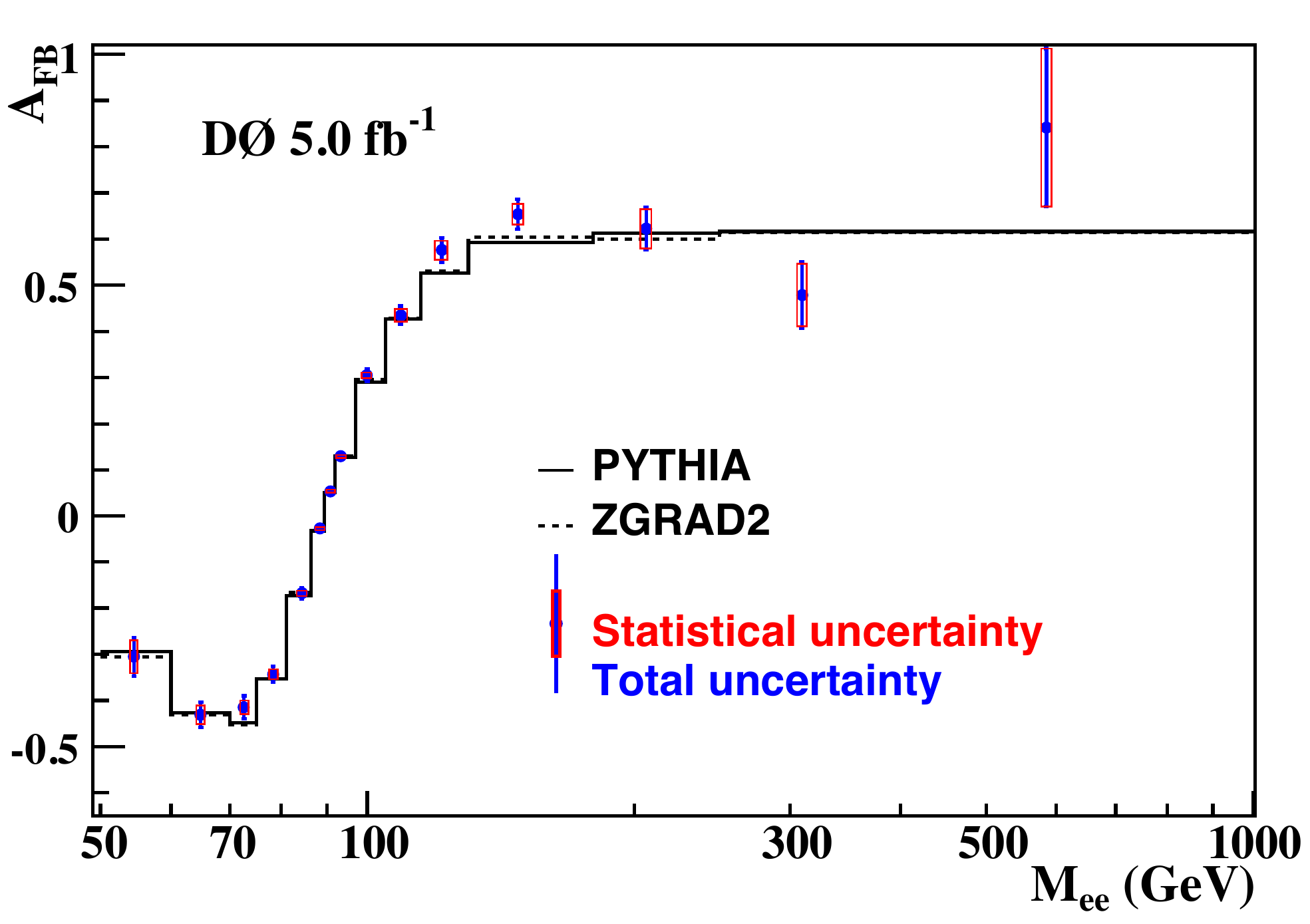}}
\caption{Corrected forward-backward Drell-Yan asymmetry in the electron channel from the D0 collaboration.  The points show the measured asymmetry;  the green (lower) histogram  is the best fit template from a leading order simulation \textsc{PYTHIA}\cite{PYTHIA} while the red curve shows the best fit NLO RESBOS prediction.}\label{fig:D0AFB}
\end{figure}

 Figure \ref{fig:D0AFB} shows $A_{FB}$ as measured by the D0 collaboration in the electron decay channel \cite{D0AFB2011} after correction for detector acceptance and charge dependent efficiencies.

 In these measurements, the raw asymmetry is corrected for detector acceptance, in particular, charge dependent efficiency differences determined (for example) via the tag and probe method.  Monte Carlo simulations are then used to generate templates with differing values of $\sineff$ to find the best fit.
 
 The CDF collaboration have also extracted $\sineff$ from the parity violating angular coefficient $A_4$ in the $Z$ boson decay angular distributions\cite{Mirkes1992}.  Ignoring azimuthal terms,

 \begin{equation}
\frac{dN}{d\cos\theta}
   \propto 
        \: (1 + \cos^2 \vartheta) +  \nonumber 
     A_0 \:\frac{1}{2} \:
             (1 -3\cos^2 \vartheta) + \nonumber 
  A_4 \: \cos \vartheta  \\
\label{eq:added2}\end{equation}

A cross section weighted moment
$\bar A_4$  is then calculated and used to extract $\sineff$.  

\begin{equation}
  \bar{A}_4  = 
	\frac{1}{\sigma}
	\int_{-\infty}^{\infty} dy
        \int_0^{\infty} dP_{\rm T}^2
        \int dM
        \: A_4
	\frac{d^3\sigma}{dy dP_{\rm T}^2 dM} ,
\label{eq:added3}\end{equation}

The measured value of $\bar A_{4}$ from 2 \infb of data~\cite{CDFAFB2013} is $0.1100 \pm 0.0079 \pm 0.0004$ which translates into a value of $\sineff = 0.2328\pm0.0010$.

\begin{figure}[htpb]
\centerline{\includegraphics[width=10cm]{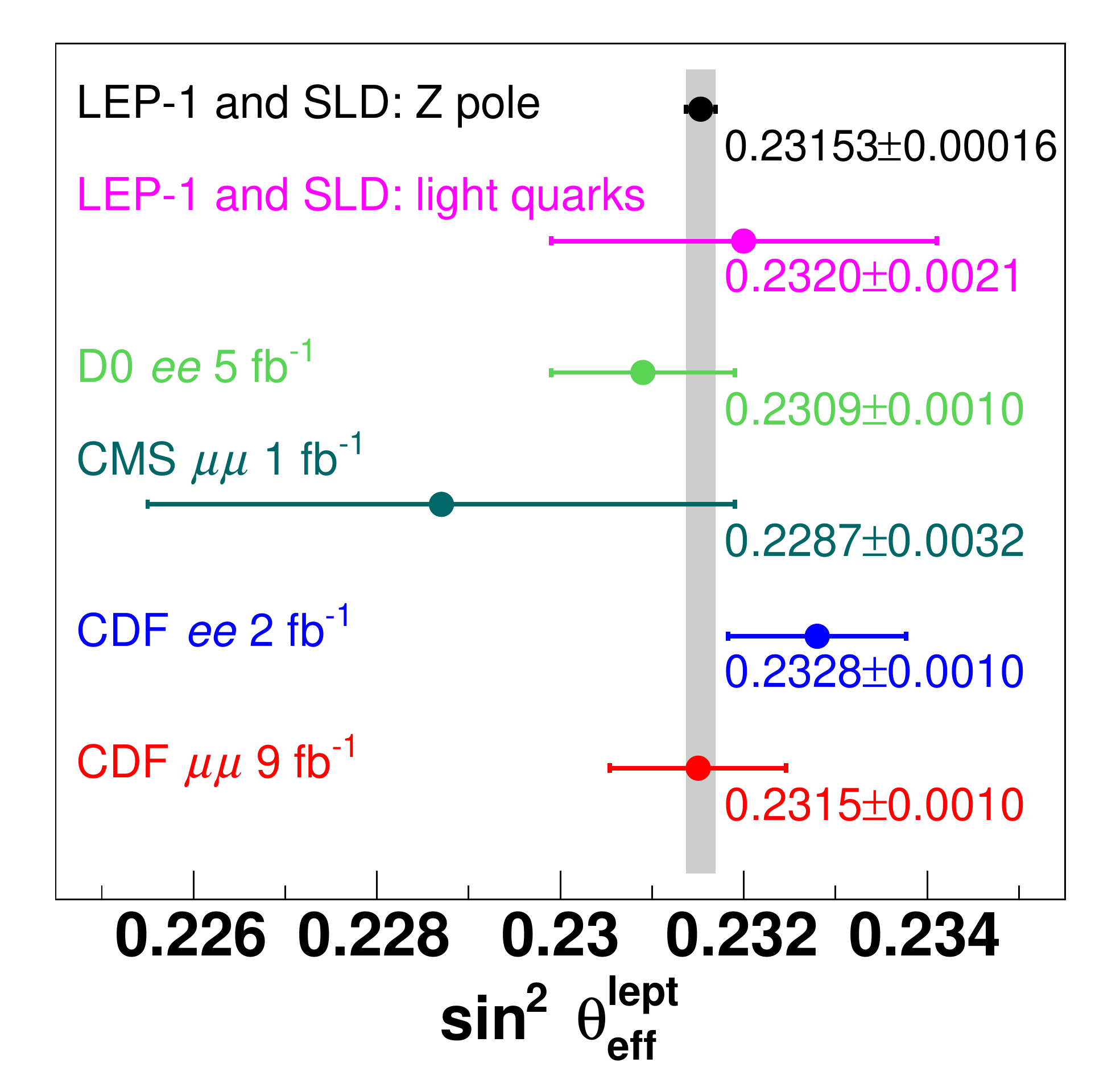}}
\caption{Summary of measurements of $\sineff$ as of early 2014.}\label{fig:AFBSummary}
\end{figure}

Figure \ref{fig:AFBSummary}  from Ref.~\citen{CDFAFB2014} summarizes the status of $\sineff$ measurements in early 2014.
While the precision of these measurements is not yet at the level achieved in the leptonic channels at LEP and SLD, ongoing analyses by both experiments are likely to achieve final uncertainties of order 0.0005 \cite{D0Wasym2014}.  At that point, correlated parton distribution uncertainties of $\approx 0.0004$ begin to dominate over the statistical errors. It is interesting to note that
 the Tevatron results are the most precise for light quarks. In the on-shell renormalization scheme, where $\sineff \equiv 1 - M_W^2 / M_Z^2$, an indirect measurement of $M_W = 80365 \pm 47$~MeV
  can be extracted~\cite{CDFAFB2014} in the context of the SM. 
 

\section{Mass and Width of the $W$ Boson}
 In the arena of precision electroweak measurements, the mass of the $W$ boson $M_W$ and the effective weak mixing angle $\sin^2 \theta^\ell_{\rm eff}$ continue to be very interesting. 
 In particular, after the direct measurement of the Higgs boson mass~\cite{lhchiggs}, all parameters defining the electroweak sector in the SM are now known to fairly
 high precision. As a result, $M_W$ and $\sin^2 \theta^\ell_{\rm eff}$ can now be predicted at loop-level in terms of other known quantities in the SM. Loop-level predictions for these
 observables can also be made in extensions of the SM~\cite{npconstraints}. Therefore, $M_W$ and $\sin^2 \theta^\ell_{\rm eff}$ can provide stringent tests of the SM by over-constraining it, just as multiple
 measurements in the flavor sector have over-constrained the unitarity of the CKM quark-mixing matrix and its CP-violating phase. 

\subsection{Theoretical Considerations of $M_W$}

 At loop-level $M_W$ can be calculated in terms of other known quantities and can be written as~\cite{sirlin}
\begin{equation}
M_W^2 (1 - \frac{M_W^2}{M_Z^2}) = \frac{\pi \alpha}{\sqrt{2} G_F} \frac{1}{1 - \Delta r}\, ,
\label{eqn:mwLoop}
\end{equation}
where setting $\Delta r = 0$ recovers the tree-level relation in the SM, $\alpha$ is the electromagnetic coupling and $G_F$ is the Fermi constant extracted from the muon decay lifetime. 
 The tree-level masses of $W$ and $Z$ boson are directly related to their coupling to the Higgs field's 
 vacuum expectation value $v$. The term $\Delta r$ contains the radiative corrections, which in the SM are dominated by (i) the running of the electromagnetic coupling due to 
 light-quark loops, (ii) the contribution due to the loop involving top ($t$) and bottom ($b$) quarks in the $W$ boson propagator, and (iii) the loops in the $W$ boson
 propagator involving Higgs bosons. The $t \bar{b}$ loop contributes to a splitting between the $W$ and $Z$ boson masses because of the large difference in the masses of these quarks
 due to their different Yukawa couplings to the Higgs field. This difference breaks the "custodial'' $SU(2)$ symmetry which maintains the tree-level relationship
 between the $W$ and $Z$ boson masses. The Higgs boson loops also cause a splitting between the $W$ and $Z$ boson masses because of the difference in the $WWh$ and $ZZh$ couplings, the
 latter arising from the mixing between the $T_3$ generator of $SU(2)_L$ and the $U(1)_Y$ generator caused by $\sin^2 \theta^\ell_{\rm eff} \neq 0$. It is interesting to note that, if a smaller
 value of  $\sin^2 \theta^\ell_{\rm eff}$ had occurred in nature, the $M_W$ measurement would have been less sensitive to the Higgs boson mass $m_H$.  

 In the approximation that new physics contributes to the precision electroweak observables through loop corrections to the gauge-boson self-energies, i.e.. through propagator corrections, 
 the new physics contributions can be generalized by using the $S, T, U$ {\it oblique} parameters~\cite{STUreference}. 
 In terms of the gauge-boson self-energies $\Pi(Q^2)_{VV^\prime}$ as functions of
 the renormalization scale $Q^2$, these parameters can be described as follows: $S$ is related to the slope ($\Pi^\prime_{VV^\prime}$)
 of $\Pi_{VV^\prime}$ with respect to $Q^2$, $T$ is related to the 
 difference of $\Pi(0)_{WW}$ and  $\Pi(0)_{ZZ}$, and $U$ is related to the difference of slopes $\Pi^\prime_{WW}$ and $\Pi^\prime_{ZZ}$. It is clear that $T$ and $U$ parameterize
 propagator effects of new physics that violate the custodial $SU(2)$ symmetry. New physics contributions to $U$ tend to be of higher order than contributions to $S$ and $T$; as one
 can imagine, it is easier to contribute to the intercept and/or the slope of $\Pi$ than to contribute a {\it difference} in the slopes for the $W$ and $Z$ boson propagators. Hence, in
 the interest of simplicity, it is common to work in the $U=0$ approximation. 

\begin{figure}[b]
\includegraphics[width=6.2cm]{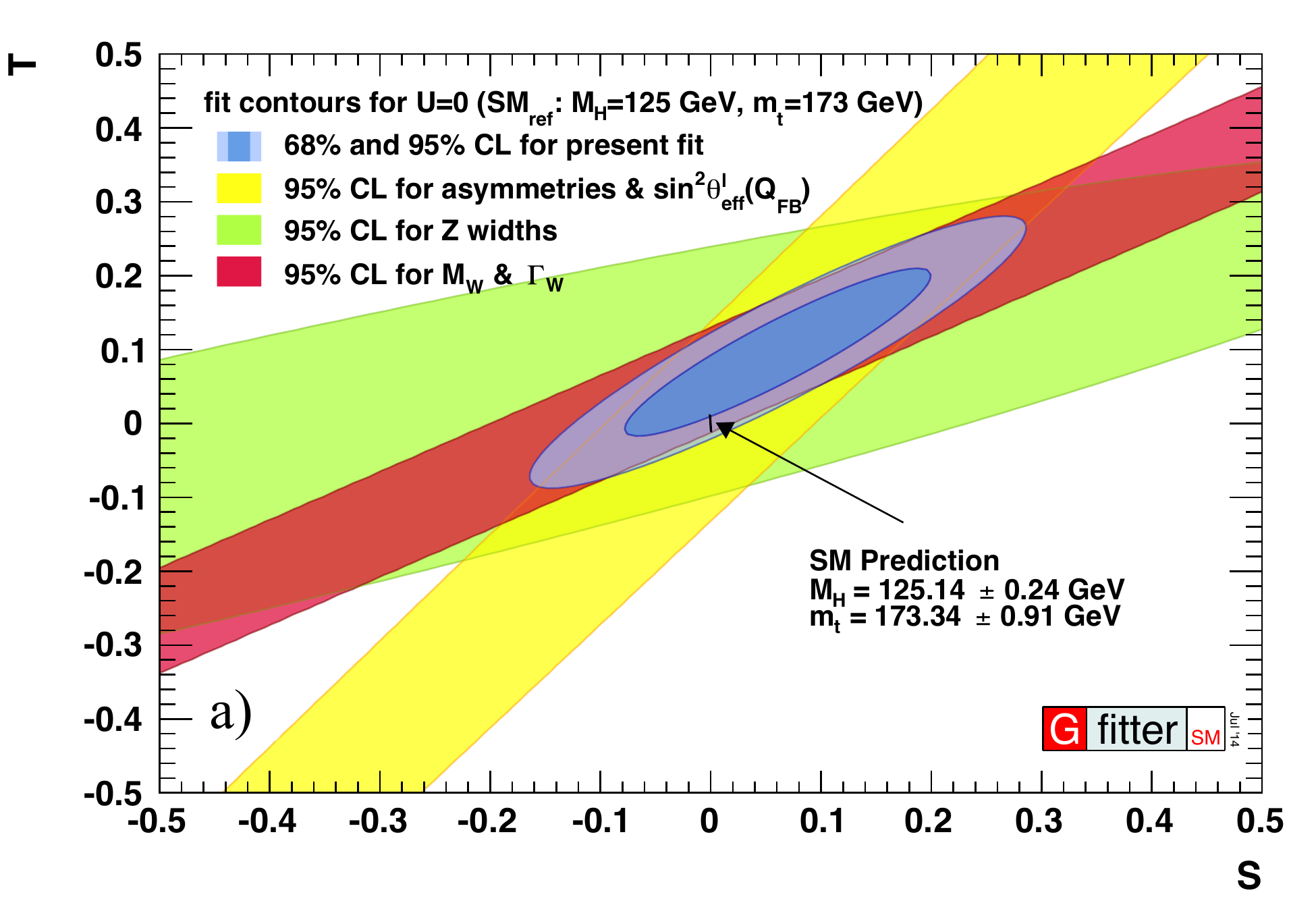}
\includegraphics[width=6.2cm]{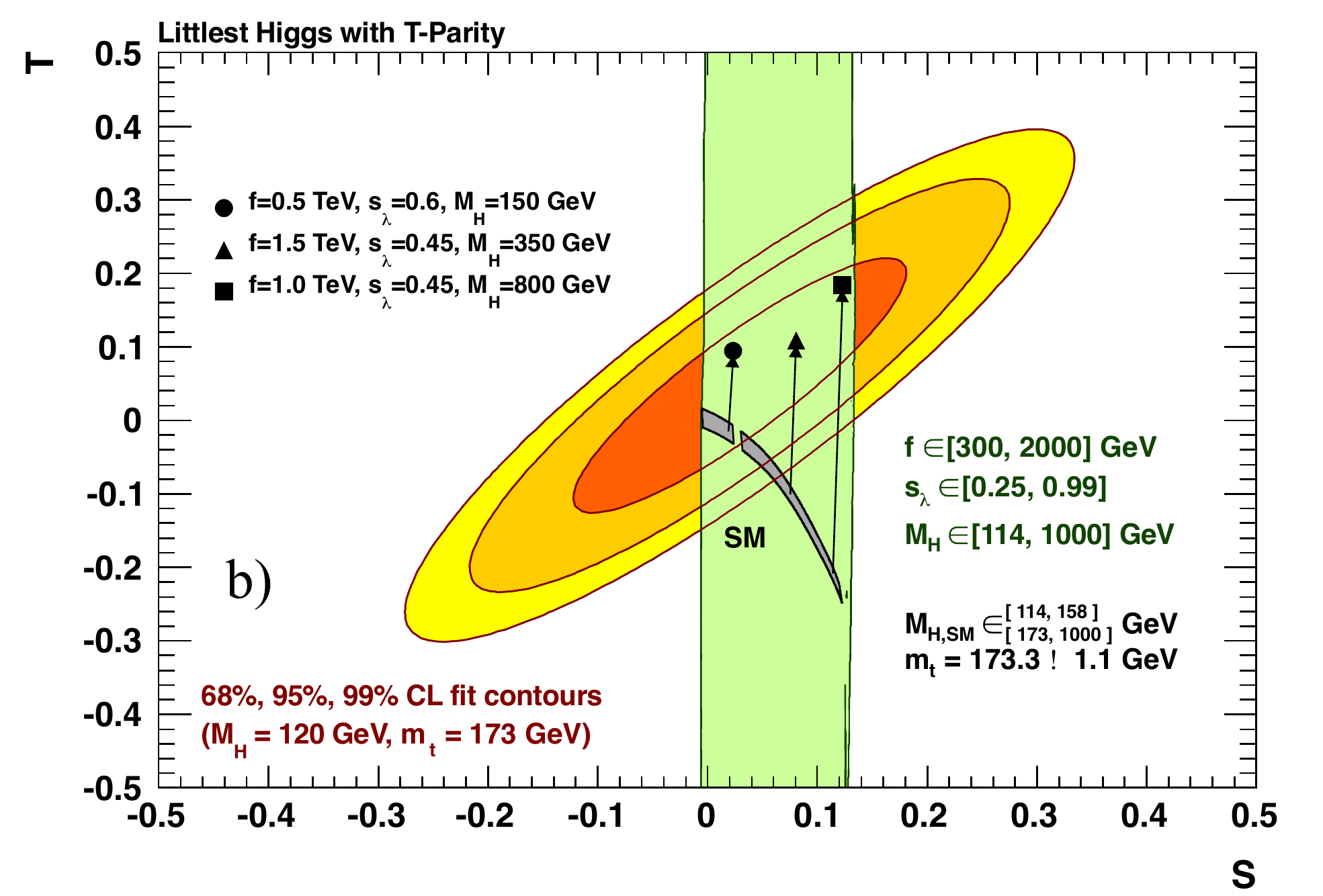}
\caption{(a) Illustration of the constrained region of $ST$ parameter space from measurements, reproduced with permission from Ref.~\citen{Baak:2014ora}, and (b) illustration of the constrained region of $ST$ parameter space from measurements, compared to a range of predictions from Littlest Higgs models (reproduced with permission from Ref.~\citen{npGfitter}).}
\label{ST-data-LH}
\end{figure}

\begin{figure}[htpb]
\includegraphics[width=6.2cm]{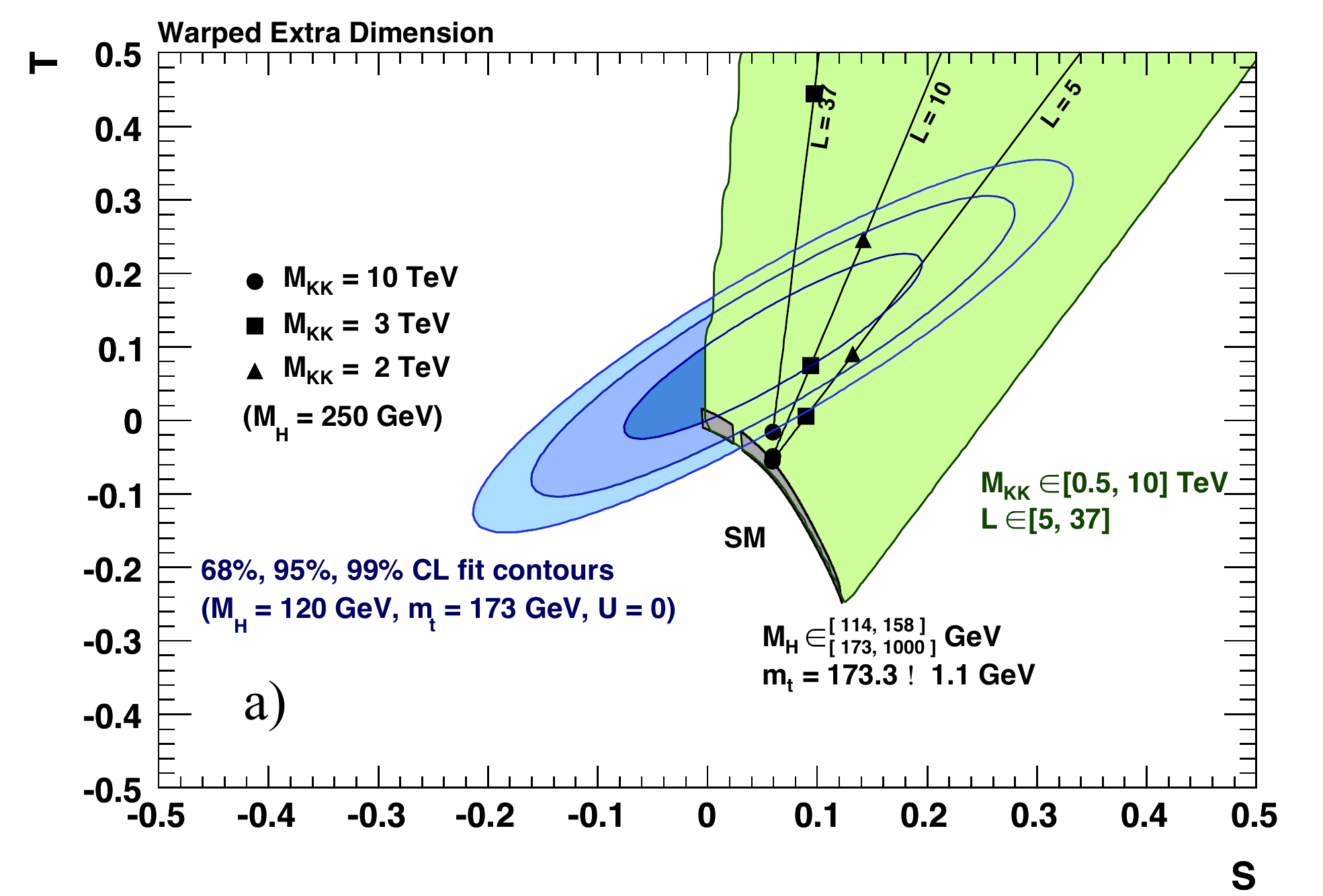}
\includegraphics[width=6.2cm]{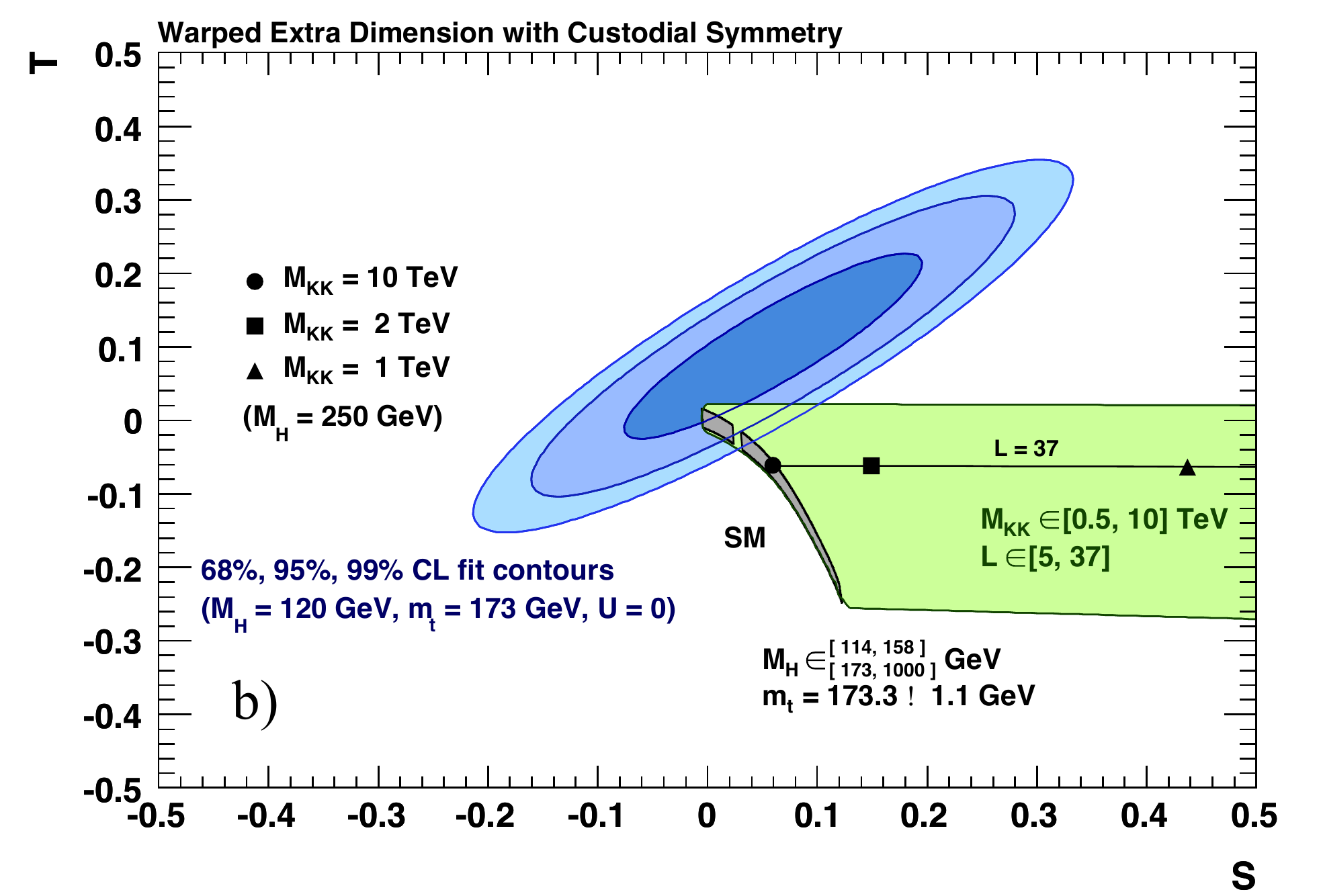}
\caption{Illustration of the constrained region of $ST$ parameter space from measurements, compared to a range of predictions from warped extra-dimensional models without (a) and with (b)
 an additional custodial symmetry introduced.  Figures reproduced with permission from Ref.~\citen{npGfitter}.  \label{ST-WED}}
\end{figure}

 In terms of these oblique parameters (which are defined to be zero in the SM), the radiative corrections to $M_W$ and $\sin^2 \theta^\ell_{\rm eff}$ can be written as
\begin{eqnarray}
\Delta r \approx \Delta r^{SM} + \frac{\alpha}{2s_W^2}S + \frac{\alpha c_W^2}{s_W^2}T + \frac{s_W^2 - c_W^2}{4 s_W^4}U \nonumber \\
\Delta \sin^2 \theta^\ell_{\rm eff} = \Delta \sin^2 \theta^{\ell,SM}_{\rm eff} + \frac{\alpha}{4(c_W^2 - s_W^2)}S +\frac{\alpha s_W^2 c_W^2}{c_W^2 - s_W^2}T 
\label{eqn:STU}
\end{eqnarray}
Note that the coefficients of $S$ and $T$ are different in relations for $\Delta r$ and $\Delta \sin^2 \theta^\ell_{\rm eff}$,
  allowing the measurements of the latter to put a two-dimensional constraint on
 new physics. Constraints in the $ST$ plane from the data are shown in Figure~\ref{ST-data-LH},
  and the range of $ST$ variation from two models of new physics are shown in Figures~\ref{ST-data-LH} and~\ref{ST-WED} respectively. It is
 clear that improving the precision of electroweak measurements can guide the search for new physics and complement direct searches.  

\subsection{Run 2 $M_W$ Measurements}

Run 2 from Fermilab's Tevatron $p \bar{p}$ collider has produced four measurements of $M_W$ so far. The two measurements from CDF~\cite{CDF2,cdf2fbprl}, using 200~pb$^{-1}$ and 2.2~fb$^{-1}$ respectively of
 integrated luminosity, are $M_W = 80413 \pm 48$~MeV and $M_W = 80387 \pm 19$~MeV. The second measurement included the data used for the first measurement and subsumed the latter. The
 two measurements~\cite{DZERO2,dzero5fbprd} from D\O\, using 1~fb$^{-1}$ and 4.3~fb$^{-1}$ respectively of
 integrated luminosity (corresponding to consecutive, independent datasets), are $M_W = 80401 \pm 43$~MeV and $M_W = 80367 \pm 26$~MeV.  The most recent combination of all Tevatron 
 measurements to date~\cite{run2combo} is
\begin{equation}
M_W = 80387 \pm 16 \; {\rm MeV} \; .
\end{equation}
which significantly surpasses the precision achieved by LEP II. The ultra-precise measurement of $M_W$ is now in the realm of hadron colliders. 

\subsection{$M_W$ Measurement Techniques at Hadron Colliders}
 While the simulation of $W$ boson production and decay, the detector response and resolution, and the detector calibrations have become increasingly more accurate and subtle, 
 the essence of the technique has remained the same over the last two decades~\cite{kotwalStark}. Inclusively produced $W$ bosons decay largely to quark-antiquark pairs, however the measurement of 
 the resulting jet energies cannot be performed with sufficient accuracy to be competitive. Furthermore, the QCD dijet background swamps the $W$ boson signal in this channel, both 
 at the online trigger and the offline reconstruction level. On the other hand, the electron and muon decay channels are cleanly identifiable with small backgrounds, and the charged
 lepton momenta can be measured with sufficient accuracy following detailed calibrations. 

 The disadvantage of the leptonic channels is that the presence of the undetectable neutrino in the two-body decay of the $W$ boson prevents the reconstruction of the invariant
 mass distribution. Apart from the need for precise calibration of the lepton momentum, many of the other tasks and associated systematic uncertainties stem from the presence 
 of the neutrino. The transverse momentum ($p_T$) distribution of the leptons have the characteristic feature called Jacobian edge, present in any two-body decay mode, where
 the distribution rises up to $p_T \sim M_W / 2$ and falls rapidly past this value. The events close to the Jacobian edge correspond to those where the $W$ boson decay axis is perpendicular
 to the beam axis. The location of the Jacobian edge provides sensitivity to the $W$ boson mass. 

 The transverse boost of the $W$ boson and the angular distribution of the boson decay in its rest frame also affect the lepton $p_T$ distribution, which therefore need to be measured
 or constrained in the theoretical production and decay model. Two approaches have been followed. In one approach, the boson $p_T$ distribution is measured using $Z$ boson
 decays to dileptons, where the lepton momenta can be measured well. This measurement is used to constrain the theoretical model that predicts the $p_T(W)$ spectrum. 
 In the second approach, the hadronic activity measured in the event is used to obtain information about $p_T(W)$ on an
 event-by-event basis. In most of the events, the hadronic activity recoiling against the $W$ boson has small net $p_T$ and is fairly difffuse, hence reconstruction of collimated jets 
 is not performed. Instead, an inclusive vector sum of transverse energies over all calorimeter towers (excluding towers containing energy deposits from the charged lepton) yields a measurement
 of the recoil $p_T$ vector (denoted by $\vec{u_T}$), and $\vec{p_T}(W) \equiv - \vec{u_T}$.  In this approach, the non-linear response and resolution affecting the $\vec{u_T}$ measurement,
 including the energy flow from the underlying event (spectator parton interactions) and additional $p \bar{p}$ collisions (both synchronous and asynchronous with the hard scatter), have
 to be carefully simulated. A measurement of $\vec{p_T}(\nu) \equiv - \vec{p_T}(\ell) - \vec{u_T}$ can be deduced by imposing transverse momentum balance. The Jacobian edge is also present
 in the transverse mass $m_T$, analogous to the invariant mass but  computed using only the $\vec{p_T}$ of the charged lepton and the neutrino; 
  $m_T = \sqrt{ 2 p_T^\ell p_T^\nu (1 - cos \Delta \phi)} $, where $\Delta \phi$ is the azimuthal opening angle between the two leptons. 
 In practice, the distributions of $m_T$, 
 $p_T (\ell)$ and $p_T (\nu)$ are all used to extract (albeit correlated) measurements of $M_W$, with differing systematic uncertainties.  

\subsubsection{Lepton Momentum and Energy Calibration}
 The precision achievable on $M_W$ is directly related to the precision on the charged lepton's energy/momentum calibration. The D\O\  experimental strategy~\cite{dzero5fbprd} is to use the highly-segmented
 uranium-liquid argon (U-LAr) sampling calorimeter to measure the electron energy, and the scintillator fiber tracker to measure the electron direction from the associated track. The tracker is
 not used to measure the lepton momentum because the resolution was not deemed to be adequate. A related consequence is that the muon channel was not used to measure $M_W$. 

 The U-LAr electromagnetic calorimeter provides readout in four longitudinal segments, with the first two samples corresponding to $\approx 2$ radiation lengths ($X_0$) each, and the third (fourth)
 sample corresponding to 7 (10) $X_0$, all at normal incidence. One of issues studied in detail in the D\O\ analysis is the estimation of the passive material in front of the EM
 calorimeter. The electron energy loss in the upstream material is not proportional to the energy, thus it causes a non-linearity in the EM calorimeter response. The absolute energy scale
 is set by calibrating the measured $Z \to ee$ boson mass to the world-average value~\cite{pdg}. Any non-linearity in the response has to be corrected for or included in the simulation so that the
 calibration can be extrapolated from the $Z$ boson mass to the $W$ boson mass. 

\begin{figure}[htpb]
\includegraphics[width=6.2cm]{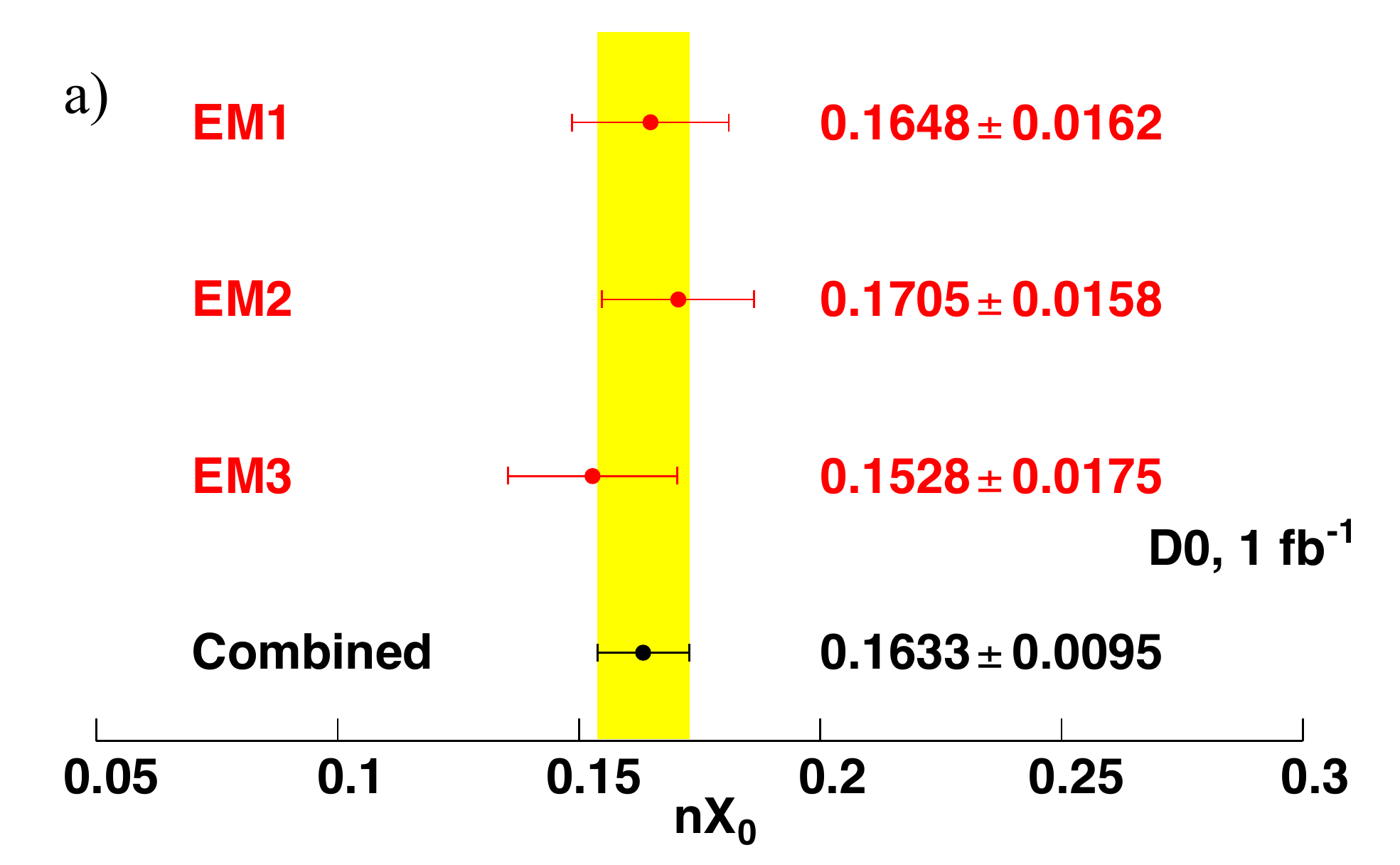}
\includegraphics[width=6.2cm]{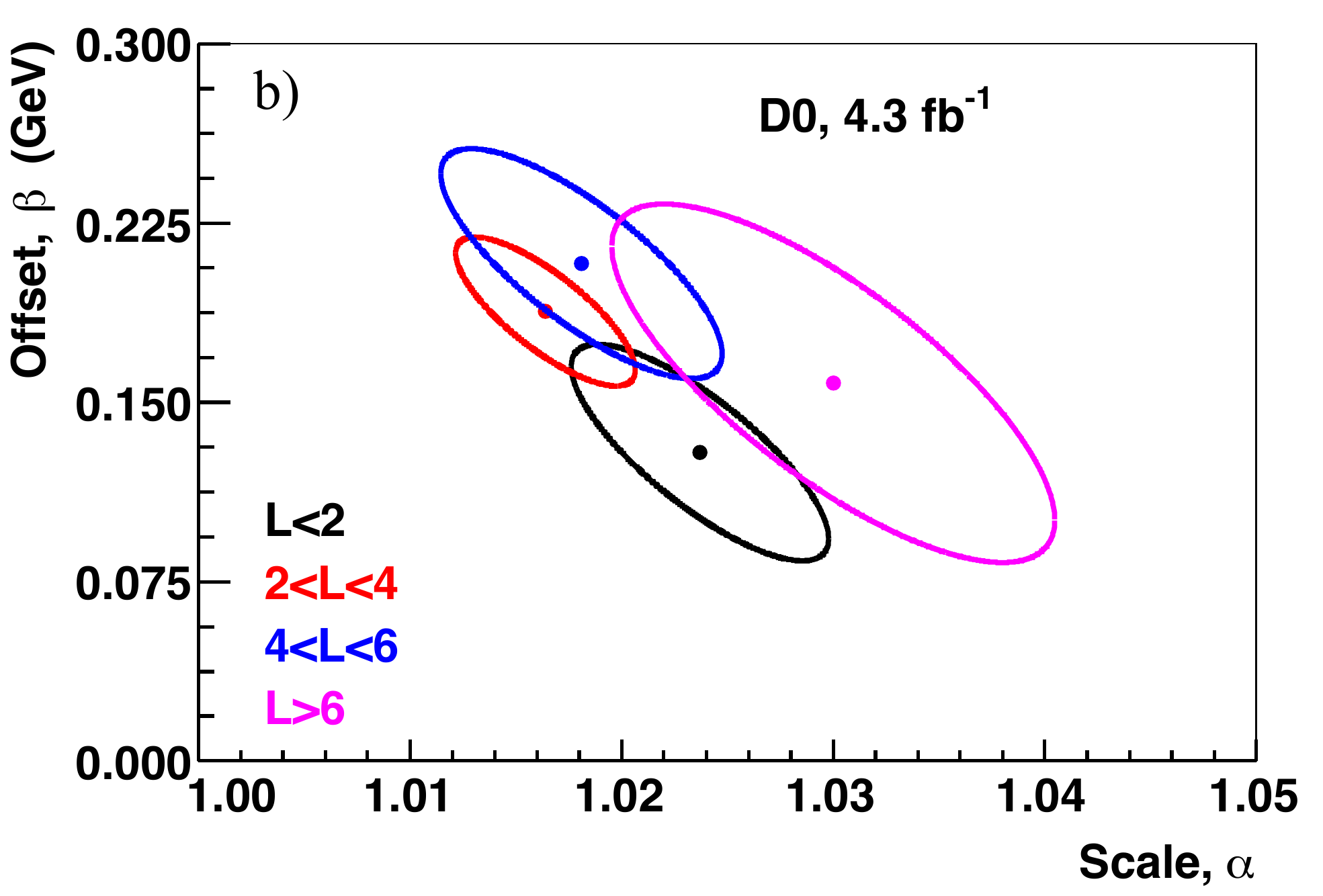}
\caption{D\O\ results for the fit for the amount of uninstrumented material in front of the EM calorimeter, in units of radiation lengths of copper (a), and the fit results
 for the electron energy scale $\alpha$ and offset $\beta$ extracted from the D\O\ data (b). The different ellipses correspond to different bins of instantaneous 
 luminosity as indicated, and the results are consistent. Instantaneous luminosity is shown in units of $36 \times 10^{30}$~cm$^{-2}$~s$^{-1}$. 
 Figures reproduced with permission from Ref.~\citen{dzero5fbprd}. }
\label{d0EMplots}
\end{figure}

 The electron energy fractions measured in the first three longitudinal samples provides information on the shower development which is sensitive to the amount of material traversed
 upstream. An additional passive layer is incorporated in the {\sc geant}-based~\cite{GEANT}
  detector simulation to mimic unaccounted-for material such that the energy fractions predicted by the simulation
 agree with their measurements in $Z \to ee$ data (shown in Figure~\ref{d0EMplots}). 
 These studies are performed in bins of electron pseudo-rapidity, and cross-checked with the $W \to e \nu$ data. Considerable effort 
 is invested in understanding the calibration as a function of pseudo-rapidity and instantaneous luminosity; the latter affects the underlying event energy deposited in the electron cone
 and the loss of high voltage across the LAr gap. The high-voltage lost across the resistive coat on the signal boards depends on the average current, which in turn depends on the
 instantaneous luminosity. The dependence of the electron energy
 resolution on pseudo-rapidity and other factors is also carefully studied and simulated. Additional constraints on non-linear effects are obtained by studying the variation of the measured
 $Z$ boson mass with electron energy. The energy response for electrons is characterized by a scale factor $\alpha$ and an offset $\beta$, with results shown in Figure~\ref{d0EMplots}.  

 The CDF experimental strategy~\cite{CDF2,cdf2fbprd} for the lepton momentum calibration is based on the first-principles calibration of the central drift chamber~\cite{COT}
 and the solenoid magnetic field~\cite{solenoid}.
  Since this allows the momentum
 measurement of both electron and muon tracks, both channels can be used, providing increased statistical precision and systematic cross-checks. The CDF electromagnetic (EM) 
 sampling calorimeter~\cite{CEM,cemresponse}
  uses lead absorber and plastic scintillator with relatively coarse transverse granularity compared to D\O. There is no longitudinal segmentation in the CDF EM
 calorimeter. Due to emission of bremsstrahlung radiation upstream of the drift chamber, the per-electron energy resolution of the calorimeter cluster (where the bremsstrahlung
 photons are coalesced with the electron shower) is better compared to the track-based measurement. Therefore the strategy adopted at CDF is to use the distribution
 of $E_{\rm cal} / p_{\rm track}$ to transfer the tracker calibration to the EM calorimeter by fitting the peak near unity. Electrons from $W \to e \nu$ and $Z \to ee$ decays
 are used for this purpose. 

 A bonus from this strategy is that $Z \to ee$ and $Z \to \mu \mu$ mass measurements can be performed independently of the $M_W$ measurements, and the $Z$ boson mass
 measurements provide independent confirmation of the calibration strategy by proving consistency with the world-average value. For exploiting the full power of the data, the
 $Z$ boson mass measurements are subsequently used as additional calibration points. Incidentally, these are the most precise measurements of the $Z$ boson mass at hadron 
 colliders, though far from being competitive with the LEP measurements. 

 The calibration of the tracker starts with a precise wire-by-wire alignment of the drift chamber (which has $\approx 30,000$ wires) using cosmic ray tracks recorded {\it in-situ} with
 collider data. A special reconstruction algorithm~\cite{cosmic} is used to fit both sides of the cosmic ray trajectory to a single helix. The hit residuals with respect to this fit
 provide information~\cite{cosmicAlignment} on various internal deformations of the drift chamber (relative rotations of radial layers, relative twists of the cylinder end plates). Comparison of the
 track parameters of the diametrically opposite segments of the same cosmic ray track also provides information on the
 gravitational and electrostatic deflections of the wires between the end plates. These effects are studied in detail to minimize the biases in curvature and polar angle 
 measurements and provide a response as linear as possible. After using these alignment constants for track reconstruction, additional tweaks to track parameters are applied
 to equate the $\langle E_{\rm cal} / p_{\rm track} \rangle$ for positrons and electrons. 

 Energy-loss effects such as the Landau-distributed ionization energy loss, bremsstrahlung (including detailed estimation of the Landau-Pomeranchuk-Migdal~\cite{migdal} suppression of soft photon 
 bremsstrahlung), Compton scattering and $e^+ e^-$ conversion of photons, as well as multiple scattering~\cite{bichsel,ms,mstail} are simulated as particles are propagated through a high-granularity
 spatial grid of passive material towards the calorimeters. The grid is built from a detailed accounting of silicon sensors, support and readout structures, the beampipe and the 
 drift chamber's internal construction. The absolute momentum scale, the total amount of passive material and magnetic field non-uniformity are measured using fits to the
 $J\psi \to \mu  \mu$ and $\Upsilon \to \mu \mu$ mass peaks, including the variation as a function of muon momentum and polar angle. 
\begin{figure}[h]
\includegraphics[width=6.2cm]{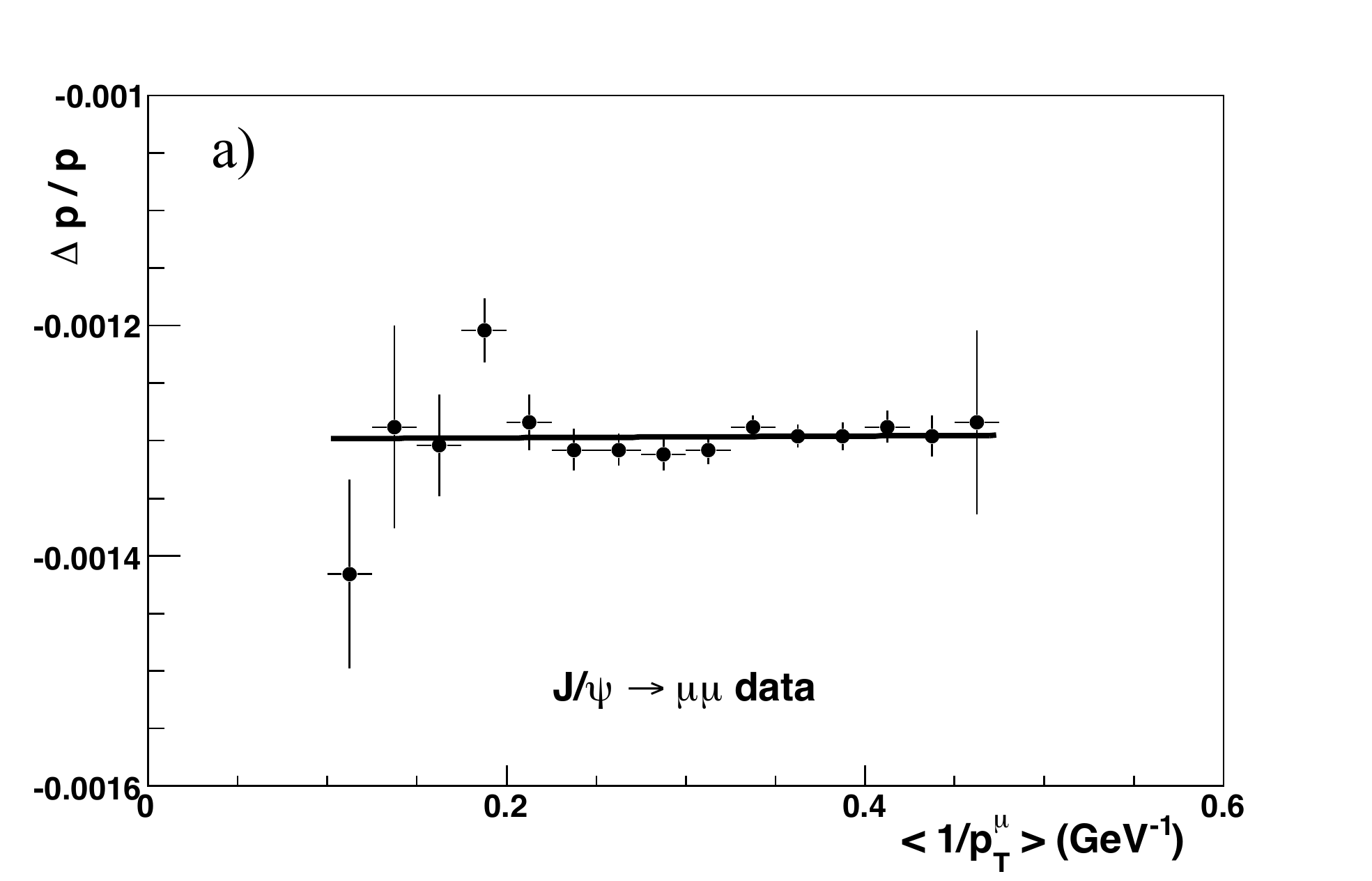}
\includegraphics[width=6.2cm]{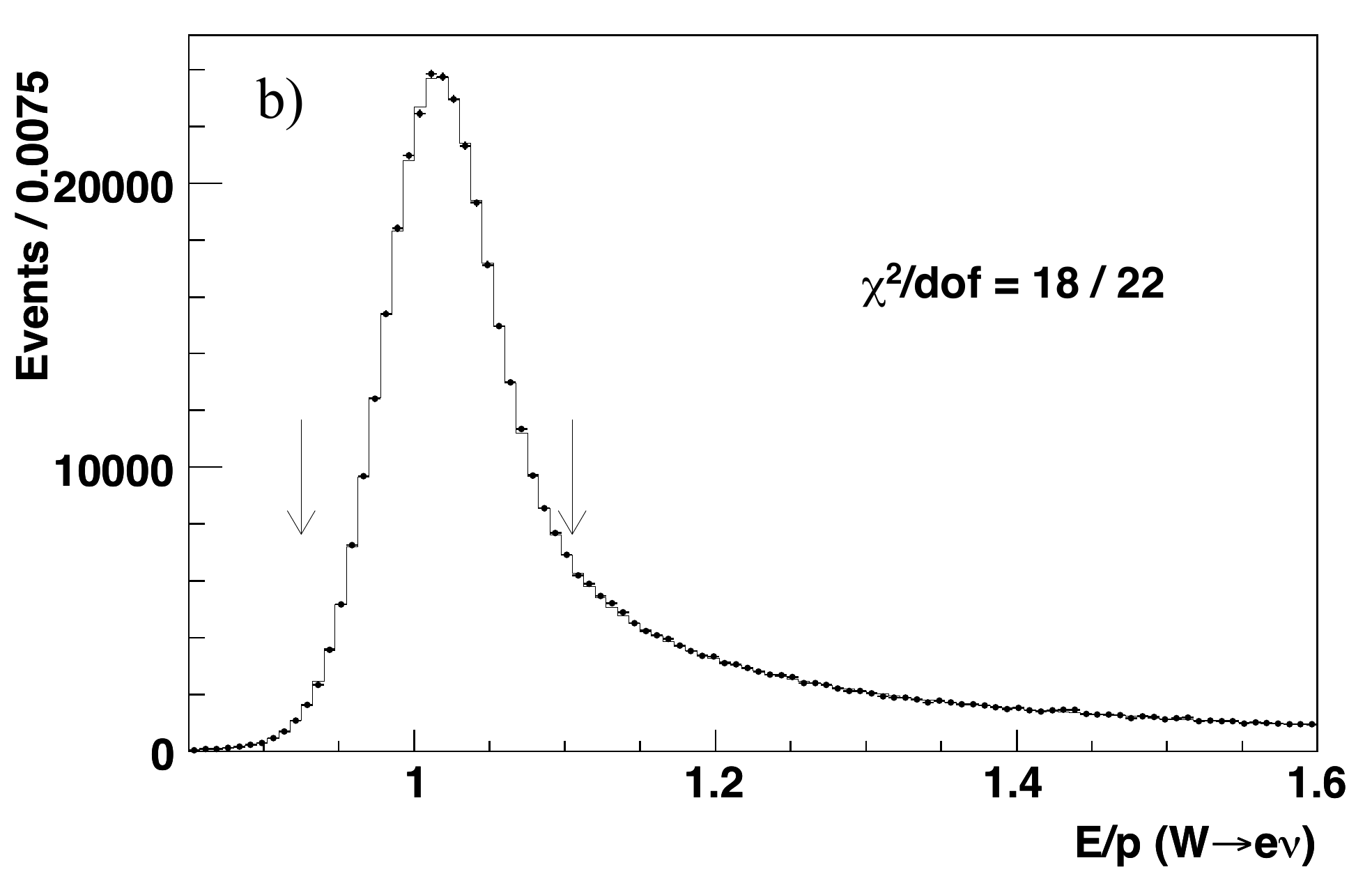}
\caption{The fractional momentum scale correction as a function of the average $p_T^{-1}$ of the muons, extracted from $J/\psi \to \mu \mu$ data on CDF (a), and the distribution
 of $E_{\rm cal} / p_{\rm track}$ overlaid with the best-fit simulation used to calibrate the EM calorimeter on CDF (b). Figures 
  reproduced with permission from Ref.~\citen{cdf2fbprd}.  \label{cdfPplot}}
\end{figure}

 In addition to the tracker, the EM calorimeter response and resolution is also studied at first-principles level using a detailed {\sc geant}4 simulation~\cite{calgeantnim} of electrons and photons 
 propagating though the sampling calorimeter geometry. Low-energy nonlinearity due to absorption of soft shower particles, and high-energy non-linearity and non-gaussian resolution
 due to longitudinal shower leakage, and calculated. The calorimeter thickness is tuned in pseudo-rapidity bins using the rate of events with low values
 of $E_{\rm cal} / p_{\rm track}$, while the radiative material upstream of the calorimeter is tuned using rate of events with high values of $E_{\rm cal} / p_{\rm track}$. 
 Residual non-linearity is measured by performing the $E_{\rm cal} / p_{\rm track}$-based calibration in bins of $E_{\rm cal}$. Additional cross-checks of electron response
 are obtained by fitting for the $Z \to ee$ mass using sub-samples of radiative and non-radiative electrons, separately using calorimeter energies and track momenta. 

\subsubsection{Hadronic Recoil Simulation}

\begin{figure}[h]
\includegraphics[width=7.4cm]{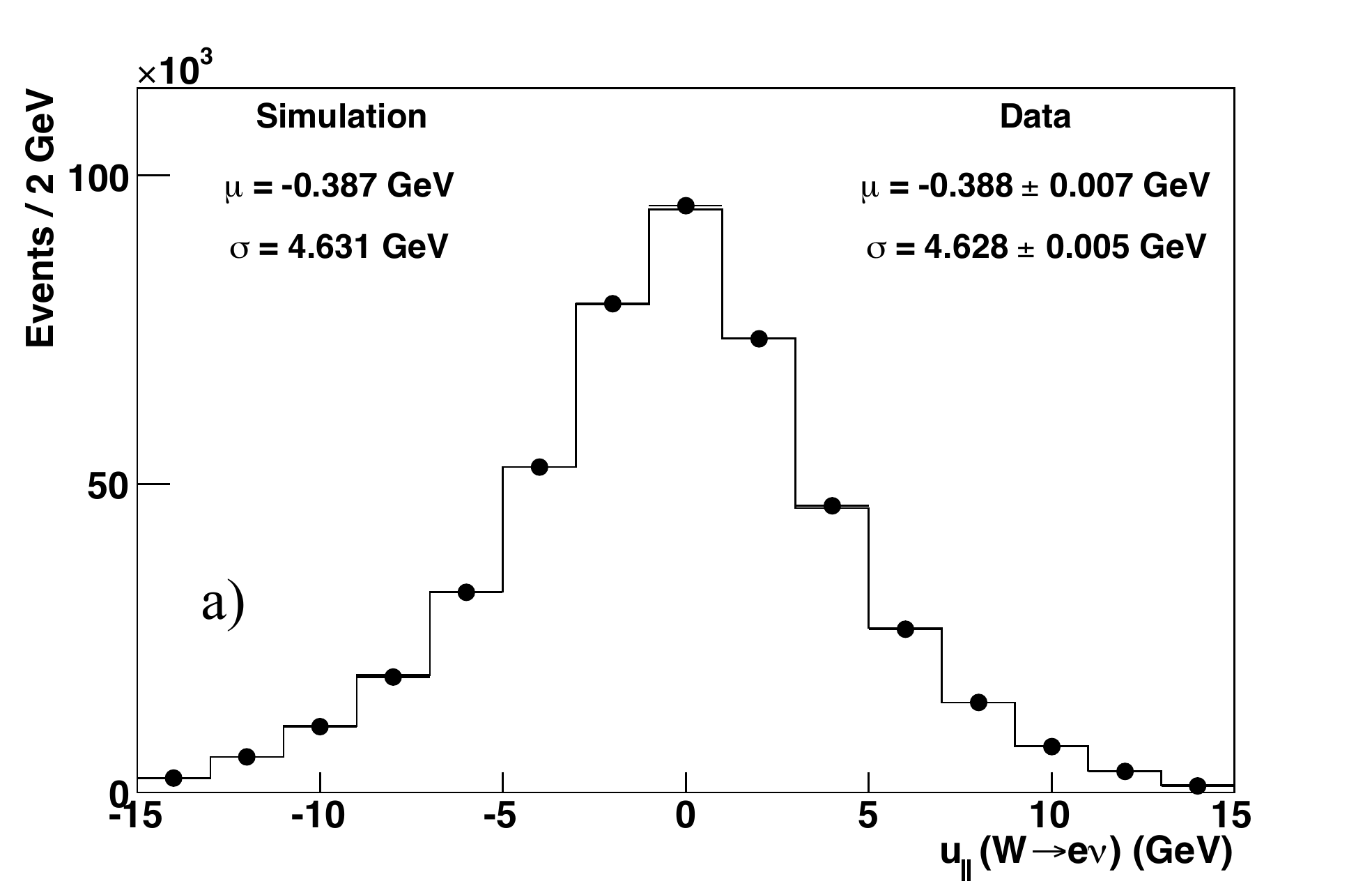}
\includegraphics[width=5.0cm]{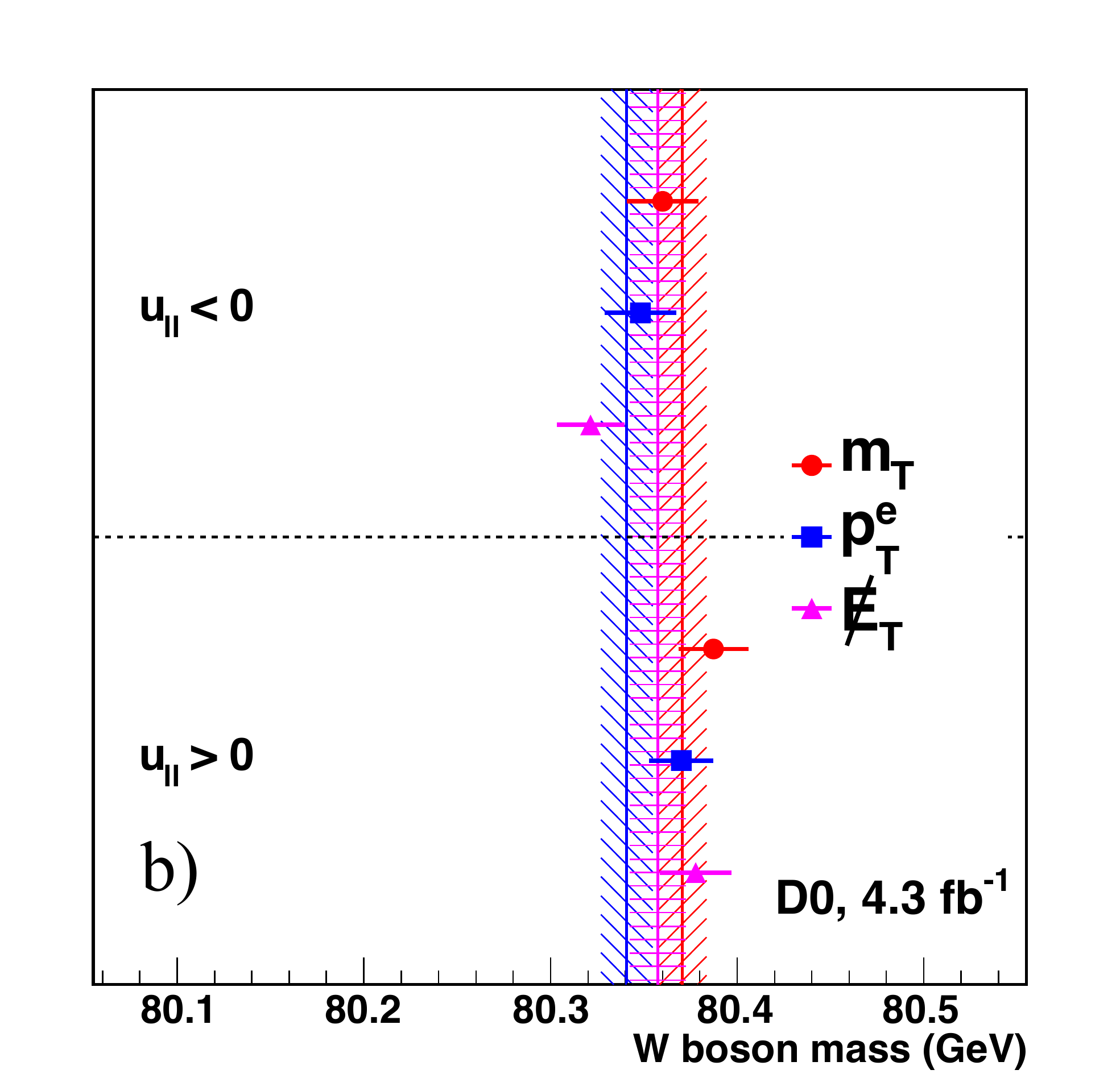}
\caption{The distribution of $u_{||}$ from CDF data and simulation (a), and $W$ boson mass fits from two sub-samples of D\O\ data binned in $u_{||}$ (b). Figures
 reproduced with permission from Refs.~\citen{cdf2fbprd} and~\citen{dzero5fbprd} respectively.   \label{recoilplot}}
\end{figure}

Due to the soft and diffuse nature of the hadronic recoil $\vec{u_T}$, the net calorimeter response is significantly less than one; soft particles with $p_T < 400$~MeV may curl up
 in the magnetic field, soft photons may be absorbed in the upstream material, etc. Further resolution degradation due to the underlying event and additional $p \bar{p}$ 
 collisions imply that applying corrections to the measured $\vec{u_T}$ is not a fruitful strategy. Rather, all of the response and resolution effects are included in the
 custom simulation. The main source of information is the $p_T$-balance between $\vec{p_T}(\ell \ell)$ and $\vec{u_T}$ in $Z$ boson events. Events triggered randomly 
 on beam crossings and on  inelastic $p \bar{p}$ collisions (minimum bias events)
 provide information for the modeling of the underlying event and additional $p \bar{p}$ collisions. An important consideration is the measurement and modeling of the hadronic energy
 deposited in the calorimeter towers which receive large energy deposits from the charged lepton(s). These towers are omitted from the calculation of $\vec{u_T}$, therefore
 the latter misses an amount of hadronic energy $\Delta u_{||}$ whose direction is aligned with the lepton. Since the component of $\vec{u_T}$ along the lepton direction directly
 enters $m_T$, it is important to carefully measure $\Delta u_{||}$ and its dependence on event kinematics. Figure~\ref{recoilplot} shows the $u_{||}$ distribution from CDF and
 the $W$ boson mass measurements in sub-samples separated by $u_{||}$ from D\O.

\subsubsection{Backgrounds}
The three categories of backgrounds are (i) $Z$ boson events in which one of the leptons is outside the acceptance or otherwise not identified as such, and all or most of its energy/momentum
 is not measured, leading to the inference of missing $E_T$, (ii) $W \to \tau \nu \to \ell \nu \bar{\nu} \nu$ which is a small but irreducible background, and (iii) misidentified leptons
 typically from QCD jet events. Except for the $ Z \to \mu \mu$ background for CDF, the backgrounds are small (of $\cal O$(1\%) or less) but at the level of precision pursued by these analyses,
 the background fractions and kinematic shapes have to be determined to high fractional accuracy. 

 In the electron channel, the $Z$ boson events generate background when one of the electrons impacts a poorly instrumented region of the detector, 
 such as between the central and endcap calorimeters or (in the CDF case) between the azimuthal modules in central EM calorimeter. This background is determined from the data by D\O\ and 
 from a combination of simulation and data by CDF. In the muon channel, the CDF central drift chamber extends up to $ | \eta | \approx 1$ and muons at higher pseudo-rapidity are not
 tracked, mimicking a $W \to \mu \nu$ candidate event. This background is essentially geometrical in origin and estimated using simulation. As the lepton $p_T$
 spectrum from $Z$ boson decays is peaked above the $W \to \ell \nu$ Jacobian edge, this background has a larger impact than a monotonically falling background distribution. 
 The $W \to \tau \nu$ background can also
 be estimated reliably from simulation, paying attention to the $\tau$ polarization which determines the $p_T(\ell)$ spectrum. 

 Backgrounds arising from mis-identified leptons in purely hadronic events are typically caused by a combination of reconstruction effects which are rare and difficult to simulate. 
 This necessitates the use of purely data-driven techniques. The typical source of electron mis-identification background is multijet projection, in which at least one jet fragments
 to a relatively isolated, high-$p_T$ $\pi^0 \to \gamma \gamma$, followed by an asymmetric $\gamma \to ee$ conversion in the detector material. If the other jet(s) is simultaneously
 mis-measured, sufficient missing $E_T$ is produced to satisfy the $W \to e \nu$ selection. By loosening or inverting the electron identification criteria, or by requiring small
 missing $E_T$, a background-enriched
 sample is obtained which can be used to extract the background kinematic shapes. A fit to the distribution of the electron identification criteria, or the distribution of missing $E_T$, 
 using pure signal and background templates yields the background fraction. Mis-identification background is probably the most difficult to estimate precisely since guidance from
 simulation is the least reliable. 

 In the muon channel for CDF, the jet-to-muon misidentification background is substantially smaller than the electron channel, arising mostly from punch-through. However, $\pi / K \to \mu$
 decays-in-flight (DIF) lead to another source of background. Due to the kink at the decay vertex, a low $p_T$ meson decaying to a lower $p_T$ muon can be mis-reconstructed as a 
 high-$p_T$ muon if the kink occurs within the tracking volume of the drift chamber. DIF cause minimum bias events to be promoted to $W \to \mu \nu$ candidates. This background
 is estimated by detailed studies of track properties for prompt muons and DIF kinked tracks, including the track-fit $\chi^2$, impact parameter, and the ``seagull''-like
 pattern of hits. The latter is discerned by studying the pairwise correlation between the sign of the residual for consecutive hits. A unique feature of this background is that
 at high $p_T$, the shape of this background distribution is a  relatively hard spectrum since the fitted track-curvature distribution is approximately uniform. 
\subsubsection{Production and Decay Model}
Four properties of the $W$ boson production model are relevant, (i) the longitudinal momentum, controlled by the parton distribution functions (PDFs), which maps the rest-frame 
 lepton $p_T$ to the lab-frame lepton pseudo-rapidity, (ii) the transverse momentum, which smears the Jacobian edge in the lepton $p_T$ spectra, (iii) decay angular distribution
 which affects the lepton $p_T$ spectrum and the correlation between the lepton and boson $p_T$, and (iv) the QED radiative corrections, which affects the sharing of energy
 between the charged lepton and the radiative photons. 

 The transverse kinematics used to fit for $M_W$ are only sensitive to longitudinal boost because they are sculpted by the limited (central) acceptance for the charged lepton. A lepton
 of a given $p_T$ may fall inside or outside the central acceptance depending upon the longitudinal boost. As a result, the fitted $M_W$ depends on the choice of PDF in the production 
 model. The uncertainties in the PDFs have been parameterized by the global fitting groups and have been propagated by reweighting simulated events to the corresponding uncertainty
 in $M_W$.  PDF uncertainties are determined with the PDF error sets provided by the CTEQ~\cite{CTEQ} and MSTW~\cite{MSTW2008} groups. D\O\ has used the CTEQ6.1 PDF set while CDF has used the
 CTEQ6.6 set and cross-checked the latter against the MSTW2008 set, showing that the central values obtained from the two PDF sets are consistent within the error envelope.

 The boson $p_T$ spectrum is calculated by the {\sc RESBOS} program which includes the NLO calculation and the dominant NNLO amplitudes, QCD parton showering and a beam-energy and $Q^2$-dependent
 non-perturbative form factor. Historically, the non-perturbative form factor was tuned using the dilepton $p_T$ spectrum or the shape of the azimuthal opening angle in $Z \to \ell \ell$ events. 
 The $Q^2$-dependence and the perturbative effects were constrained from global fits to data and their contributions were sub-leading in impact on this analysis. As the $Z$ boson 
 statistics have increased, the impact of these external constraints is likely to grow in relative importance. In the most recent analysis, CDF has included $\alpha_S$ as a second parameter,
 in addition to the non-perturbative $g_2$ parameter, 
 in the fit to their $p_T(Z)$ spectrum. D\O\ has propagated the uncertainty in $g_2$ derived from the global fit~\cite{blny}. 

 The decay angular distribution has been calculated at NLO and partial calculation of NNLO effects are estimated to have negligible impact in the published analyses. A complete NNLO calculation
 would be desirable in the future. 

 Considerable effort has been invested in understanding the QED and electroweak radiative corrections, in order to incorporate the most accurate rates and distributions of radiative photons. Both
 experiments use {\sc photos}~\cite{photos} interfaced to {\sc resbos}~\cite{ResBos} to simulate final-state radiation (FSR) of photons. CDF has cross-checked {\sc photos} using {\sc horace}~\cite{horace}, where the latter has two
 modes: an FSR-only mode and  an exact {\cal O}($\alpha$) mode which is interfaced to a photon shower. Furthermore, in the latter mode all photons in the multi-photon shower have a correction
 factor applied, which is extracted from the comparison between the first photon in the shower and the photon in the exact {\cal O}($\alpha$) calculation. CDF calibrates {\sc photos}-FSR
 against {\sc horace}-FSR which is then calibrated against the more complete {\sc horace} calculation. 
\subsubsection{Results}
The measurements from CDF and D\O\ were summarized in Sec.~\ref{wMassHistory}. A summary of their uncertainties is shown in Table~\ref{tbl:combinedsys}.  The Tevatron (world) average
 of $M_W = 80387 (80385) \pm 16 (15)$~MeV~\cite{run2combo} can be compared with its SM prediction: $M_W = 80358 \pm 8$~MeV~\cite{Baak:2014ora}. The agreement puts stringent limits new physics, 
 though the measurement is about $1.6 \sigma$ above the SM prediction. The latter~\cite{Baak:2014ora} uses as inputs the precision $Z$-pole measurements from LEP and SLD, the top quark mass from Tevatron and LHC
 experiments, the Higgs boson mass from LHC, and a recent determination of the hadronic vacuum polarization contribution to $\alpha_{EM} (M_{Z}^{2})$. 
\begin{table}[htpb]
\tbl{Uncertainties in units of MeV on the combined result ($m_T$ fit) on $M_W$ from CDF (D\O) using 2.2 (4.3) fb$^{-1}$ of integrated luminosity. ``na'' denotes the uncertainty is
 not individually tabulated.}
{\begin{tabular}{@{}lcc@{}} \toprule
Source & CDF Uncertainty & D\O\ Uncertainty \\ \colrule
Lepton energy scale and resolution & 7 & 17 \\
Recoil energy scale and resolution & 6 & 5 \\
Lepton tower removal                    & 2  & na \\
Backgrounds                                     & 3  & 2 \\
PDFs                                    & 10& 11 \\
$p_T(W)$  model                         & 5  & 2 \\
Photon radiation                        & 4 & 7 \\
\colrule
Statistical                             & 12  & 13 \\
\colrule
Total                                   & 19 & 26 \\ \botrule
\end{tabular} 
\label{tbl:combinedsys}}
\end{table}

Current measurements from CDF and D\O\ have been obtained from the analysis of about a quarter and a half respectively of their full Run 2 dataset. The analyses of the remainder of the data to obtain the final, most precise
 Tevatron measurements of $M_W$ are in progress. The dominant, correlated uncertainty between both experiments will be due to the PDFs. Measurements of $W$ and $Z$ boson differential distributions
 such as the $W$ boson charge asymmetry and the $Z$ boson rapidity from the Tevatron and the LHC can significantly reduce the PDF uncertainty, such that a combined Tevatron measurement of $M_W$ with a total uncertainty of 
 10~MeV may be possible. 
\subsection{Measurements of the $W$ boson Width}
 The direct measurement of the $W$ boson width $\Gamma_W$ is of interest because it is precisely calculable in the SM. A comparison between the measurement and the theoretical
 prediction can constrain the CKM matrix element $V_{cs}$ and the properties of new heavy particles that can induce loop-level radiative corrections. The prediction for
 the $W$ boson width is 
\begin{equation}
 \Gamma_W = \frac{3G_F M_W^3}{\sqrt{8}\pi}(1 + \delta_{\rm QCD} + \delta_{\rm EW})
\end{equation}
where $G_F$ is the Fermi constant extracted from the muon lifetime, $\delta_{\rm QCD} = 2 \alpha_S / 3\pi$ is the QCD radiative correction at $\cal O$($\alpha_S$), and $\delta_{\rm EW}$ is the electroweak
 radiative correction. The uncertainty in the prediction of $\Gamma_W$ is dominated by the uncertainty in the experimental value of $M_W$ used as input, followed by the uncertainty due to
 higher-order radiative corrections. New physics may enter through $\delta_{\rm EW}$.

The most recent direct measurements of $\Gamma_W$ have been published by CDF~\cite{cdfWwidth} and D\O~\cite{d0Wwidth}
  using 350 pb$^{-1}$ and 1 fb$^{-1}$ respectively of integrated luminosity at $\sqrt{s} = 1.96$~TeV. The analysis techniques are very 
 similar to those used for the measurements of $M_W$, but with an emphasis on understanding the kinematic region of large transverse mass, $90 < m_T < 200$~GeV. As this $m_T$-range is higher than the resonant production
 of $W$ bosons, events in this range are dominated by their off-shell production whose rate depends on $\Gamma_W$. As a result, resolution smearing of events from the resonant region into the high-mass tail of the $m_T$
 distribution is a more important issue for the $\Gamma_W$ analysis, as are the relatively higher background fractions. The PDF uncertainty is also larger due to the wider fit range. The latest measurement from
 CDF is~\cite{cdfWwidth} $\Gamma_W = 2032  \pm 45_{\rm stat} \pm 57_{\rm syst} = 2032 \pm 73$~MeV
  and from D\O\ is~\cite{d0Wwidth} $\Gamma_W = 2028  \pm 39_{\rm stat} \pm 61_{\rm syst} = 2028 \pm 72$~MeV. These measurements are consistent with the SM prediction of $\Gamma_W = 2093 \pm 2$~MeV. 


\section{Diboson Production at the Tevatron}
The simultaneous production of two weak vector bosons ($W\gamma$, $Z\gamma$, $WW$, $WZ$ or $ZZ$) 
has been at the center of a large range of measurements at the Tevatron experiments over the last decade. 
Diboson production at the Tevatron predominantly occurs via $t$-channel exchange. The $s$-channel contributes 
the diboson production via direct interaction of gauge bosons through trilinear gauge boson vertices. 
Both the CDF and D0 experiments developed extensive diboson research programs as more 
and more data were available to analyze. Precise knowledge of diboson processes and their 
proper modeling is highly valuable for various studies. Many diboson processes represent non-negligible 
backgrounds in Higgs boson and top quark production, and production of supersymmetric particles. Therefore 
a complete and detailed understanding of electroweak processes is a mandatory precondition for early discoveries 
of very small new physics signals. Furthermore, several electroweak analyses represent a proving ground for 
analysis techniques and statistical treatments used in the Tevatron Higgs searches during the Run II data taking period. 

The diboson processes have been studied at the Tevatron since the beginning of Run I. Most of the Run I 
studies were statistics-limited and focused on setting limits on anomalous trilinear gauge boson couplings 
(TGCs)~\cite{cdf-runI-tgc1,cdf-runI-tgc2,cdf-runI-tgc3,d0-runI-tgc1,d0-runI-tgc2,d0-runI-tgc3,d0-runI-tgc4,d0-runI-tgc5,d0-runI-tgc6,d0-runI-tgc7,d0-runI-tgc8,d0-runI-tgc9,d0-runI-tgc10,d0-runI-tgc11,d0-runI-tgc12} and diboson production cross 
sections~\cite{d0-runI-tgc11,cdf-runI-xs1,d0-runI-xs1}. The CDF collaboration also reported first 
evidence for $WW$ production and measured the $WW$ production cross section of 
$\sigma_{WW}=10.2^{+6.3}_{-5.1}$~(stat)~$\pm 1.6$~(syst) in $\ell\nu \ell\nu$ final states~\cite{cdf-runI-xs1} .
 
In the early years of Run II diboson production was studied mainly in purely leptonic final states such as 
$W\gamma\rightarrow \ell\nu\gamma$, $Z\gamma\rightarrow \ell\gamma$, $WW\rightarrow \ell\nu \ell\nu$, 
$WZ\rightarrow \ell\nu \ell\ell$ and $ZZ\rightarrow \ell\ell\ell\ell$ ($\ell$ is an electron or muon, $\nu$ is a neutrino). Study of other 
final states were unfavored due to limiting factors such as detector resolution, irreducible background, or 
lack of analysis techniques that would overcome some of these challenges and improve sensitivity of a measurement. 
Some studies such as those of $WW$ and $WZ$ production employed sophisticated analysis techniques that helped to 
extract the significant results for $\ell\nu{jj}$ final states. 

\section{Cross Section Measurements}
Measuring diboson production cross sections addresses the basic physics interest of observing fundamental 
electroweak processes and tests the validity of theoretical predictions. The proper modeling of diboson 
production processes was also important in the context of searches for New Physics and searches for the 
SM Higgs boson at the Tevatron. Table~\ref{xsTheo} summarizes the theoretical NLO cross sections for 
diboson production at the Tevatron used in the analyses. 
\begin{table}[ph]
\tbl{The diboson production cross sections at the Tevatron. Values for $WW$, $WZ$ and $ZZ$ production are calculated 
with MCFM~\cite{mcfm} at NLO for $\sqrt{s}=1.96$~TeV using MSTW2008NLO PDFs.}
{\begin{tabular}{@{}cccccc@{}} \toprule
\multicolumn{6}{c}{Cross Section at NLO [pb]} \\
 $\sigma$ & $W\gamma$ & $Z\gamma$ & $WW$ & $WZ$ & $ZZ$ \\ \colrule
Theory \hphantom{00} & \hphantom{0}$178.4\pm 13$ & \hphantom{0}$134\pm 12$ & $11.7\pm 0.8$ & $3.7\pm 0.3$ & $1.4\pm 0.1$ \\ \botrule
\end{tabular} \label{xsTheo}}
\end{table}

\begin{figure}[htpb]
  \begin{centering}
  \includegraphics[width=10.0cm]{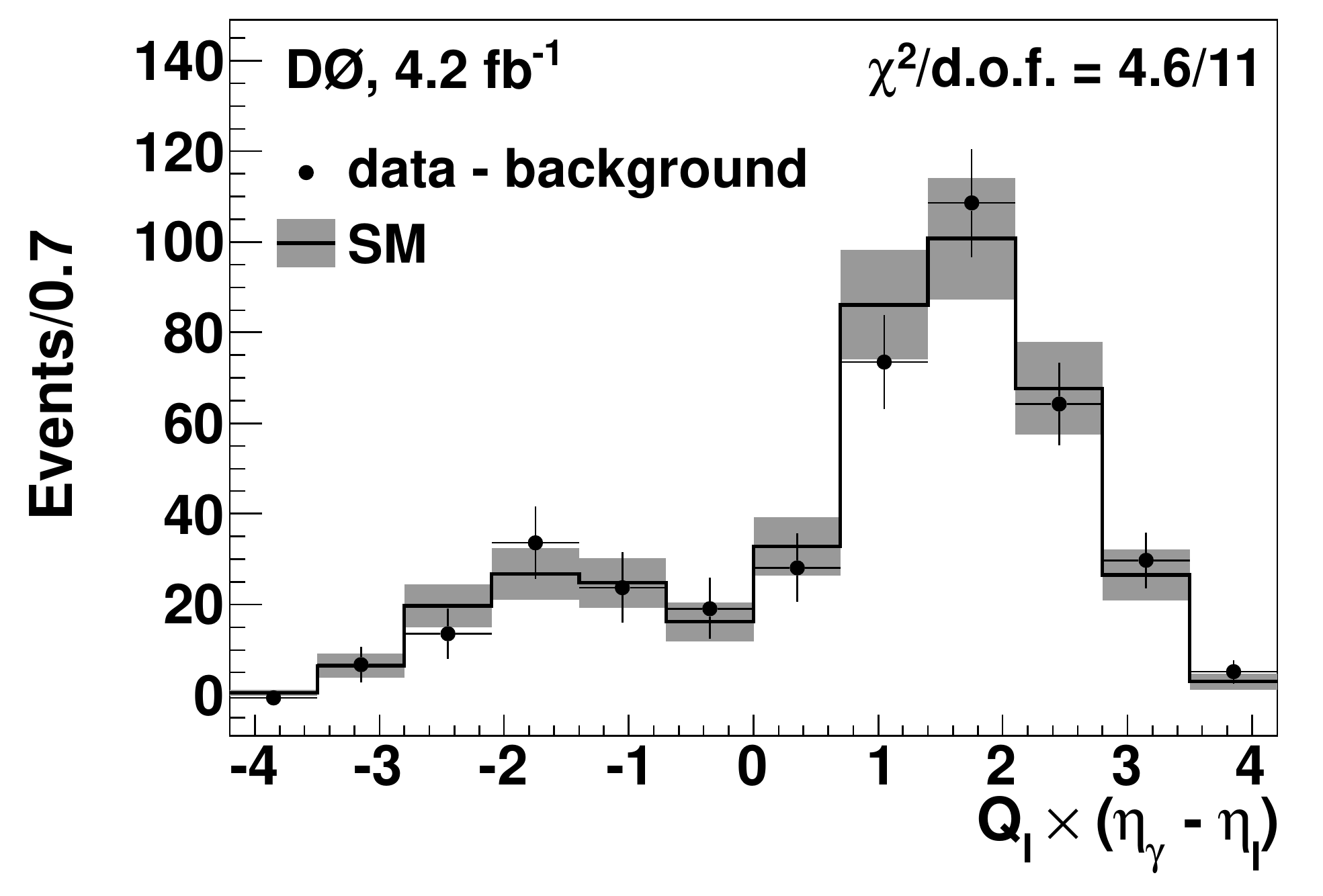}
  \caption{The charge-signed photon-lepton rapidity difference for background-subtracted D0 data compared to the SM 
    $W\gamma$ prediction. The shaded area represents uncertainty on $W\gamma$ prediction (from Ref.~\citen{d0-wgamma3}).}
  \label{fig:d0-raz} 
  \end{centering} 
\end{figure}

The diboson process with the highest production cross section, $W\gamma$, has been studied at the Tevatron since the first data were 
ready for analysis. Samples of 0.20~fb$^{-1}$ and 0.16~fb$^{-1}$ of integrated luminosity collected by CDF and D0 
respectively, confirmed the agreement between the experiment and theoretical predictions~\cite{cdf-wgamma-xs,d0-wgamma1}. 
The cross section was measured with a precision of $\sim$15\% within a phase space defined by $E_{T}^{\gamma}>7$ or 8~GeV 
and $dR_{\ell,\gamma}>0.7$. In following years the radiation amplitude zero was of great interest when studying this process. 
This effect, evident in the charge-signed lepton-photon rapidity difference as a dip around -0.3 shown in 
Figure~\ref{fig:d0-raz}, is a consequence of negative interference among the tree-level diagrams for which the amplitude 
for SM $W\gamma$ production is expected to be zero around cos$\theta =-0.3$ ($\theta$ is an opening 
angle between incoming quark and outgoing $W$ boson). The most precise measurement of $W\gamma$ production cross section 
at the Tevatron yields $\sigma_{W\gamma}\times$~BR$(W\rightarrow{\ell\nu})=7.6\pm 0.4$~(stat)~$\pm 0.6$~(syst)~pb as 
obtained from D0 data of 4.2~fb$^{-1}$ of integrated luminosity~\cite{d0-wgamma3}.

\begin{figure}[htpb]
  \begin{centering}
  \includegraphics[width=6.5cm]{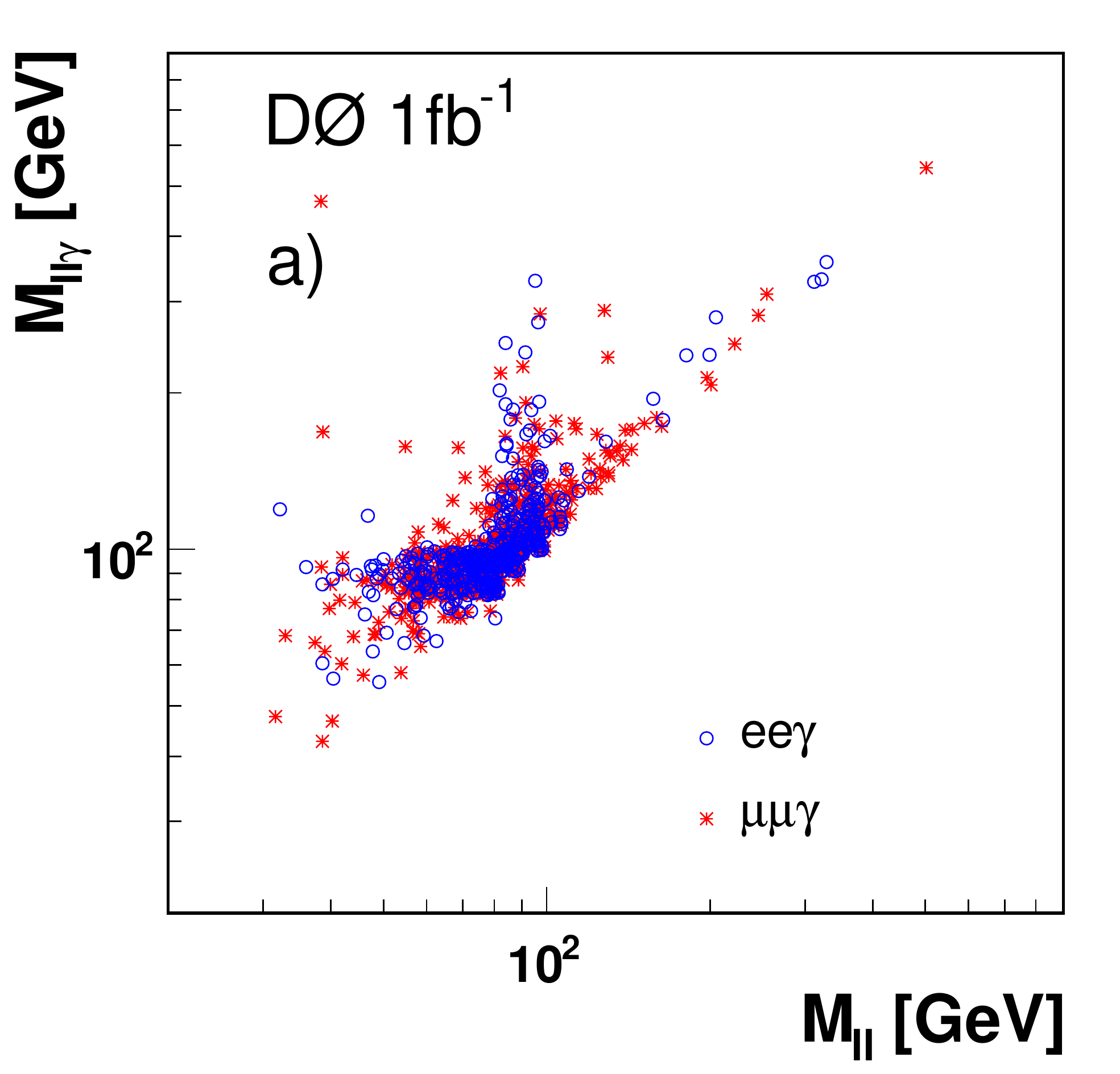}
  \includegraphics[width=6.0cm]{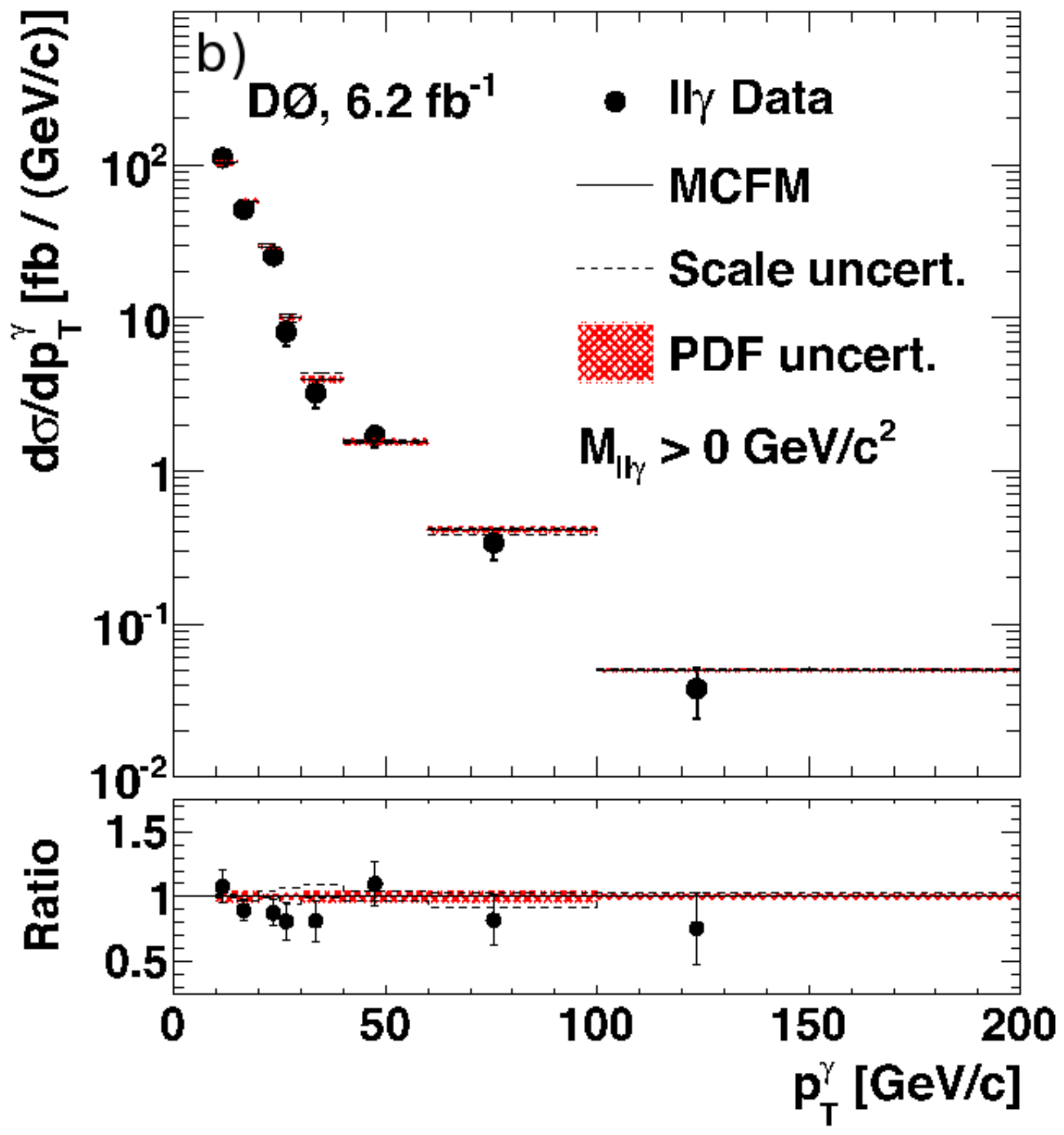}
  \caption{(a) $\ell\gamma$ vs. $\ell$ mass of $Z\gamma$ candidate events~\cite{d0-zgamma2} from the D0 experiment.
    (b) Unfolded $d\sigma/dp_{T}^{\gamma}$ distribution from the D0 experiment with no $M_{\ell\gamma}$ requirement 
    compared to the NLO MCFM prediction (from Ref.~\citen{d0-zgamma4}).}
  \label{fig:d0-zgamma-xs} 
  \end{centering} 
\end{figure}
Another channel with a photon in the final state, $Z\gamma$, was extensively studied by both 
collaborations~\cite{cdf-wgamma-xs,cdf-zgamma1,d0-zgamma1,d0-zgamma2,d0-zgamma3,d0-zgamma4}. The $\ell\gamma$ 
candidates selection is largely optimized based on the structure shown in Figure~\ref{fig:d0-zgamma-xs} (left) where the 
invariant mass of the dilepton pair and photon versus the dilepton invariant mass is shown. The vertically populated region 
is a target group because it represents decays where $M_{\ell}\approx M_{Z}$ and a photon is emitted by one of the interacting 
partons resulting in $M_{\ell\gamma}> M_{Z}$. Over the years improvements that were introduced into analyses of $\ell\gamma$ 
final states such as track isolation, photon identification efficiency, and improved modeling of converted photons helped in 
reaching a precision in the cross section measurement of 5-6$\%$. The most precise measurement of $Z\gamma$ 
production cross section in $\ell\gamma$ final states is reported by the D0 collaboration~\cite{d0-zgamma4}. The same study 
investigates the differential distribution $d\sigma/dp_{T}^{\gamma}$ in $\ell\gamma$ final states, shown in Figure~\ref{fig:d0-zgamma-xs} 
(right) and confirms the agreement between the theoretical prediction for NLO calculations with MCFM. 

The first observation of $Z\gamma\rightarrow\nu\nu\gamma$ final states at the Tevatron in 2009 with a statistical significance 
of 5.1~s.d. yields the most precise cross section times branching ratio measurement of 
$\sigma_{Z\gamma}\times$~BR$(Z\rightarrow{\nu\nu})=32\pm 9$~(stat+syst)~$\pm 2$~(lumi)~fb using 3.6~fb$^{-1}$ of D0 data~\cite{d0-zgamma3}. 
\begin{figure}[htpb]
  \begin{centering}
  \includegraphics[width=8.0cm]{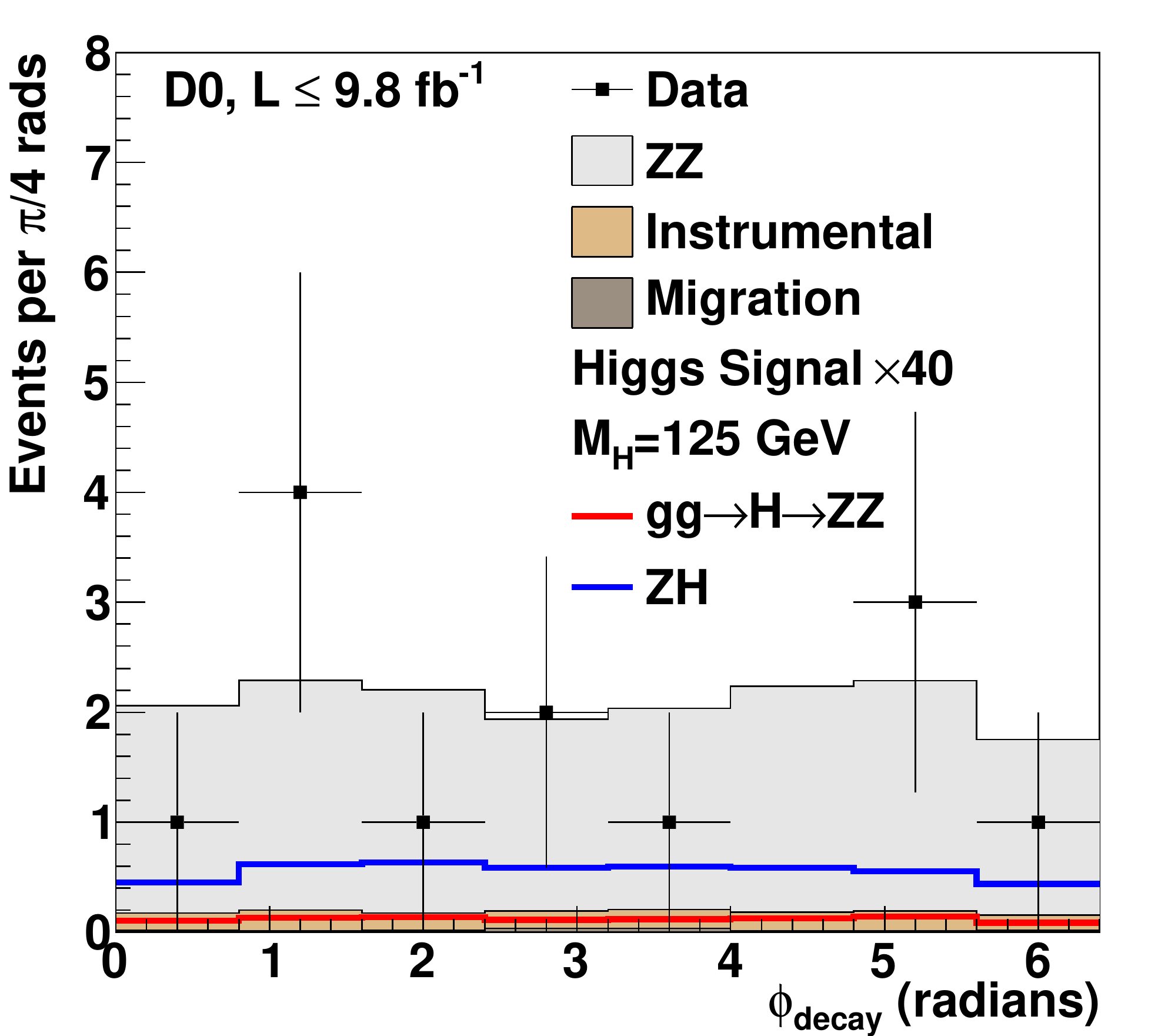}
  \caption{Distribution of the azimuthal angle $\phi_{decay}$ for the decay planes of the $Z/\gamma^{*}$ bosons 
    selected at D0 compared to the expected $ZZ\rightarrow \ell\ell\ell\ell$ signal and background (from Ref.~\citen{d0-zz3-xs}).}
  \label{fig:tevd0-zz-xs} 
  \end{centering} 
\end{figure}

\begin{figure}[htpb]
  \begin{centering}
  \includegraphics[width=8.0cm]{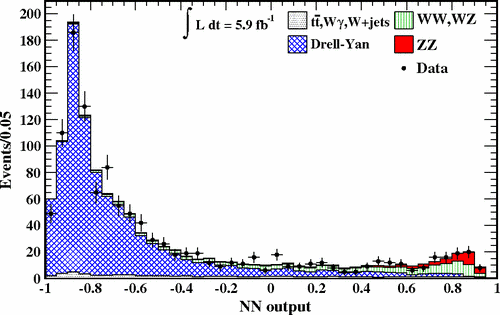}
  \caption{Neural network output distribution for the processes contributing to the $ZZ\rightarrow \ell\ell\nu\nu$ sample at CDF, 
    scaled to the best values of the fit to the data (from Ref.~\citen{cdf-zz2-xs}).} 
  \label{fig:tevcdf-zz-xs} 
  \end{centering} 
\end{figure}

As the process with the smallest production cross section but a negligible amount of background, $ZZ$ production was 
first observed at the Tevatron in 2008 through $\ell\ell\ell\ell$ final states in 1.7~fb$^{-1}$ of integrated luminosity collected 
by the D0 detector~\cite{d0-zz1-xs}. Three events found in data yielded a significance of 5.3~s.d.. Several months 
earlier CDF reported the first evidence for $ZZ\rightarrow \ell\ell\ell\ell$ production with a significance of 4.4~s.d.~\cite{cdf-zz1-xs} 
based on the same number of events selected from 1.9~fb$^{-1}$ of integrated luminosity. Updated D0 analysis of 
$\ell\ell\ell\ell$ final states improved precision of a cross section measurement and provided a significant statistics to 
study various kinematic distributions~\cite{d0-zz-rayan}. This electroweak process is a main background for Higgs 
boson production $H\rightarrow ZZ^{*}$. In particular the $\phi_{decay}$ distribution shown in Figure~\ref{fig:tevd0-zz-xs} 
is sensitive to different beyond the SM models~\cite{zzphi,d0-zz3-xs}. $ZZ$ production in $\ell\ell\nu\nu$ 
final states was studied by both CDF and D0 using 5.9~fb$^{-1}$ and 8.6~fb$^{-1}$ of integrated luminosity, respectively. 
The precision of these measurements (about $\sim 30\%$ with the statistical uncertainty dominating) has been maximized 
by employing neural network (NN) discriminants to separate the $ZZ$ contribution from the dominant Drell-Yan background 
at CDF~\cite{cdf-zz2-xs} (Figure~\ref{fig:tevcdf-zz-xs}) and by canceling out some systematic effects when calculating 
the cross section at D0~\cite{d0-zz2-xs}. Previous measurement in $\ell\ell\nu\nu$ final states at D0 yielded a significance 
of 2.6~s.d. using 2.7~fb$^{-1}$ of data~\cite{d0-zz-emanuel}. The most precise combined cross sections from CDF and D0 are  
$\sigma (p\bar{p}\rightarrow ZZ)=$~1.64$^{+0.44}_{-0.38}$~(stat+syst) pb and 
$\sigma (p\bar{p}\rightarrow ZZ)=$~1.44$^{+0.35}_{-0.34}$~(stat+syst) pb, respectively, after corrections for contribution 
from $\gamma^{*}$ and $Z/\gamma^{*}$ interference. Similar measurements were also performed with a full datasets collected by 
the CDF and D0 detectors~\cite{d0-zz3-xs,cdf-zz3-xs}. 
%

At the Tevatron $WW$ and $WZ$ production in fully leptonic final states is measured with a precision of $15-20\%$.
To achieve the best possible precision experiments employ complex selection criteria and sophisticated analysis techniques. 
At both Tevatron experiments the $WW$ and $WZ$ cross sections are extracted based on the number of selected 
events~\cite{cdf-ww1-xs,cdf-wz-xs,d0-ww1-xs,d0-ww-xs1,d0-wz1-xs,d0-wz2-xs,d0-wz2-lnull}. The first observation of $WZ$ 
production comes from CDF and yields the cross section of $\sigma_{WZ}= 5.0^{+1.8}_{-1.4}$~(stat)~$\pm 0.4$~(syst)~pb.

In addition to that CDF uses NN output and a matrix element likelihood ratio to fit $WW$, $WZ$, and background to 
data~\cite{cdf-ww-xs1,cdf-wz-xs1}. The most precise measurements of these processes at the Tevatron are 
$\sigma_{WW}= 12.1\pm 0.9$~(stat)$^{+1.6}_{-1.4}$~(syst)~pb made by CDF and 
$\sigma_{WZ}= 4.50\pm 0.61$~(stat)$^{+0.16}_{-0.25}$~(syst)~pb by D0.

\begin{figure}[htpb]
  \begin{centering}
    \includegraphics[width=6.0cm]{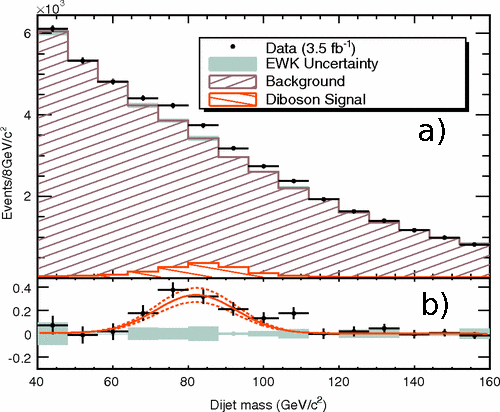}
    \includegraphics[width=6.0cm]{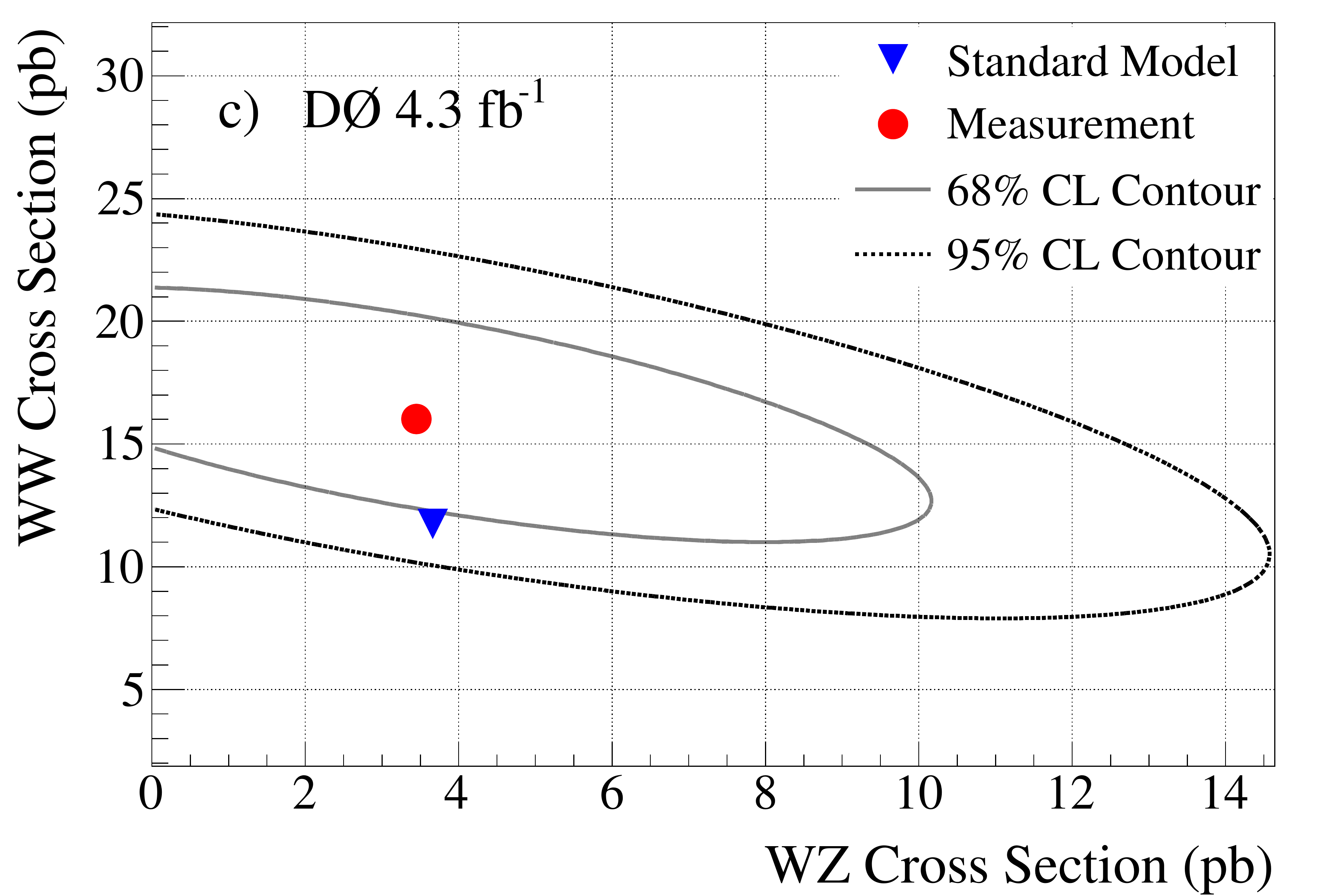}
    \caption{{\it Left}: Dijet invariant mass distribution of reconstructed $W/Z\rightarrow{jj}$ candidates by CDF compared 
    to the fitted signal and background components (a), and for the corresponding background subtracted distribution (b).
    (c) The D0 results from the simultaneous fit of $\sigma_{WW}$ and $\sigma_{WZ}$ using the Random Forest output 
    distributions. The plot shows the best fit value with 68\% and 95\% C.L. regions and the NLO SM prediction (Table~\ref{xsTheo}).}
  \label{fig:wwwz-xs} 
  \end{centering} 
\end{figure}

As the sensitivity to associated Higgs boson production was growing, efforts expanded to identify events in which one 
of the vector bosons decays hadronically. The production of a $W$ boson that decays leptonically, associated with 
a second vector boson $V$ ($V=W$ or $Z$) that decays into pair of jets involves the same final states with a dijet 
resonance near the Higgs mass. The first observation of the dijet final states, associated with a large missing 
transverse energy (MET), was made by CDF with a significance of 5.3~s.d.~\cite{cdf-wwwz-xs1}. The dijet mass distribution 
for the dijet+MET analysis is shown in Figure~\ref{fig:wwwz-xs} (left). Because these final states include invisible 
decays as well, further efforts were made to identify $\ell\nu{jj}$ final states specifically. The first evidence for $WW+WZ$ 
production in $\ell\nu{jj}$ final states from D0 with 4.4~s.d. significance~\cite{d0-wwwz-xs1} was followed by its observation 
at the Tevatron experiments~\cite{cdf-wwwz-xs2,cdf-wwwz-xs3,d0-wwwz-xs2}. Both collaborations used the dijet mass spectra along with 
sophisticated analysis techniques such as Random Forest and matrix element technique to extract the diboson signal and to 
measure precisely the cross section, reporting results 
of $\sigma_{WW+WZ}= 16.0\pm 3.3$~(stat+syst)~pb (CDF) and $\sigma_{WW+WZ}= 19.6^{+3.2}_{-3.0}$~(stat+syst)~pb (D0). 
In addition, both collaborations exploited $b$-tagging algorithm to further separate $WW$ from $WZ$ 
contributions~\cite{d0-wwwz-xs2,cdf-wwwz-xs4}. The two-dimensional representation of measured $WW$ and $WZ$ production 
cross section at D0 is shown in Figure~\ref{fig:wwwz-xs} (right). 

\begin{figure}[htpb]
  \begin{centering}
  \includegraphics[width=10.0cm]{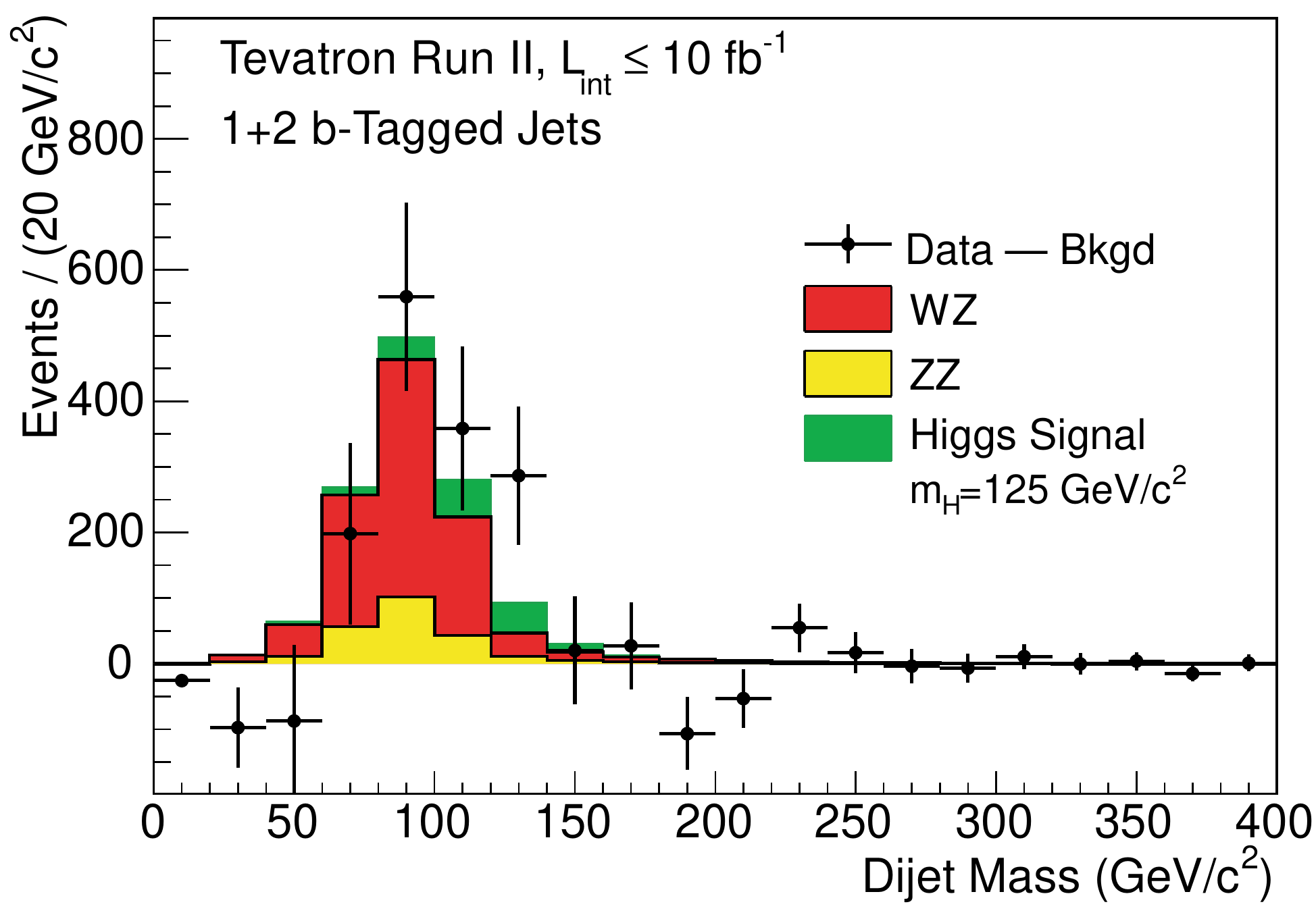}
  \caption{Background-subtracted distribution of the reconstructed dijet mass summed over CDF and D0's channels. 
    The $VZ$ signal and the background 
    contributions are fit to the data, and the fitted background is subtracted. The fitted $VZ$ and expected SM Higgs 
    ($m_{H}=125$~GeV/c) contributions are shown with filled histograms (from Ref.~\citen{tev-combo2}).}
  \label{fig:wh-combo1} 
  \end{centering} 
\end{figure}

In an effort to probe the sensitivity to a Higgs-like small signal in a large background both the CDF and D0 experiments 
searched for $VZ$ production in semileptonic final states with a $Z$ boson decaying into pairs of $b$-quarks. Although 
both experiments measured the $VZ$ cross section individually~\cite{d0-vz-xs1,d0-vz-xs2,cdf-vz-xs2} the best cross 
section measurement of $VZ$ production combining $\ell\ell b\bar{b}$, $\ell\nu b\bar{b}$, and $\nu\nu b\bar{b}$ final states 
yields $3.0\pm 0.6$~(stat)~$\pm 0.7$~(syst)~pb when combining both experiments~\cite{tev-combo1,tev-combo2}. The measured cross 
section agrees well with the SM prediction and clearly demonstrates the ability to extract a small electroweak 
signal in a large background using analysis tools and techniques common in Higgs boson searches at the Tevatron. 
The combined background-subtracted dijet mass distribution for the $VZ$ analysis is shown in Figure~\ref{fig:wh-combo1}.

\par All measured cross sections are generally consistent with SM predictions calculated at next-to-leading order in QCD.

\section{Gauge Boson Self-Interactions}
In the SM the neutral vector bosons, $\gamma$ and $Z$, do not interact among themselves, 
while the charged vector bosons, $W^{\pm}$, couple with the neutral ones and among themselves through 
trilinear and quartic gauge interactions.  The most general $\gamma{WW}$ and $ZWW$ interactions can be 
described using a Lorentz invariant effective Lagrangian~\cite{lagrangian,HWZ}. Assuming charge ($C$) 
and parity ($P$) conservation and electromagnetic gauge invariance ($g^\gamma_1=1$) 
the Lagrangian terms take the form:
  \begin{equation} \begin{array}{ccl} \frac{\mathcal{L}_{VWW}}{g_{VWW}} & = & i g_{1}^{V}
  (W_{\mu\nu}^{\dag}W^{\mu}V^{\nu} - W_{\mu}^{\dag}V_{\nu}W^{\mu\nu}) \\ & +
  & i{\kappa}_{V}W_{\mu}^{\dag}W_{\nu}V^{\mu\nu} +
  i\frac{\lambda_{V}}{M_{W}^{2}}
  W_{\lambda\mu}^{\dag}W_{\nu}^{\mu}V^{\nu\lambda}.
  \label{eq:eq-lag} \end{array}{} \end{equation}
In the SM, the five remaining TGCs are $\lambda_\gamma=\lambda_Z=0$ and $g_1^Z=\kappa_{\gamma}=\kappa_Z=1$. 
Any deviation of these couplings from their predicted values 
would be an indication for new physics~\cite{strong} and could provide information on the mechanism for 
electroweak symmetry breaking. These deviations are denoted as the anomalous TGCs, $\Delta\kappa_V$ and 
$\Delta g_1^Z$, defined as $\kappa_V-1$ and $g_1^Z-1$, respectively. At hadron colliders anomalous TGCs 
would cause divergences of the production cross sections as the center-of-mass energy, $\sqrt{\hat{s}}$, 
of the partonic constituents approaches a high energy $\Lambda$. Thus unitarity is protected by  a form 
factor:
\begin{equation} \Delta{a(\hat{s})}=
  \frac{\Delta{a_{0}}}{(1+\hat{s}/\Lambda^{2})^{n}} \label{eq:alpha}
  \end{equation}
\noindent where $n=2$ for $\gamma{WW}$ and $ZWW$ couplings, and $a_{0}$ is a low-energy approximation 
of the coupling $a(\hat{s})$. Limits on anomalous TGCs are set in terms of $a_{0}$. The 
scale of new physics, $\Lambda$, is usually set to 1.5 or 2~TeV. Limits on anomalous TGCs depend on 
choice of $\Lambda$; as $\Lambda$ increases the sensitivity to the anomalous TGC $a_0$ increases. 
Typically one sets the largest $\Lambda$ value consistent with the preservation of unitarity. 

Due to different interpretations of the effective Lagrangian~[Eq.~\ref{eq:eq-lag}] there are several 
scenarios which can be used in TGC representation. The most meaningful to use is the $SU(2)_L\times{U(1)_Y}$ 
scenario~\cite{su2u1} which we refer to as the ``LEP parameterization''. This scenario 
assumes the following relation between the anomalous TGCs:
\begin{equation}
    \Delta\kappa_Z = \Delta g_1^Z - \Delta\kappa_\gamma \cdot \tan^2\theta_W, \text{ and } 
    \lambda_Z = \lambda_\gamma = \lambda.
    \label{eq:lepparam}
\end{equation}
In the equal couplings scenario~\cite{HWZ}, the $\gamma WW$ and the $ZWW$ couplings are set equal 
to each other and are sensitive to interference effects between the photon and $Z$-exchange diagrams 
in $WW$ production. Electromagnetic gauge invariance requires that $\Delta g_1^Z=\Delta g_1^\gamma=0$ 
and 
\begin{equation}
    \Delta\kappa_Z = \Delta\kappa_\gamma = \Delta\kappa \text{ and } 
    \lambda_Z = \lambda_\gamma = \lambda.
    \label{eq:eqparam}
\end{equation}
Finally, the $SU(2)_L\times{U(1)_Y}$ scenario can take the form of the Hagiwara-Ishihara-Szalapski-Zeppenfeld 
(HISZ) scenario~\cite{hisz} with the following relation between TGCs:
  \begin{equation}
  \begin{array}{ccl}
  \Delta\kappa_{Z} = \frac{1}{2}\Delta\kappa_{\gamma}(1-tan^{2}\theta_{w}), \Delta g^{Z}_{1} = 
  \frac{\Delta\kappa_{\gamma}}{2cos^{2}\theta_{w}} & \text{and} & \lambda_{Z} = 
  \lambda_{\gamma} = \lambda.
  \label{eq:lepeq}
  \end{array}{}
  \end{equation}

Although neutral gauge bosons do not interact among themselves at tree-level, new physics effects can give 
rise to $\gamma{ZZ}$ and $ZZZ$ vertices at low energies. The $Z\gamma{Z}$ vertex with a photon and one $Z$ 
boson on-shell is described by the following Lagrangian~\cite{zgammaBaur}:
\begin{equation} \mathcal{L}_{Z\gamma V}  =  -ie[
(h_{1}^{V}F^{\mu\nu} + h_{3}^{V}\tilde{F}^{\mu\nu}) Z_{\mu}\frac{( +m_{V}^{2})}{m_{Z}^{2}}V_{\nu} +
(h_{2}^{V}F^{\mu\nu} + h_{4}^{V}\tilde{F}^{\mu\nu}) Z_{\alpha}\frac{( +m_{V}^{2})}{m_{Z}^{2}}\partial_{\alpha}\partial_{\mu}V_{\nu}],
\label{eq:zgamma} \end{equation}
\noindent where $h^{V}_{3}$ and $h^{V}_{4}$ are $CP$ conserving couplings and are extensively studied at the Tevatron 
experiments. Values of $n$ from Eq.~\ref{eq:alpha} are set to $n=3$ for $h^{V}_{1,3}$ and $n=4$ for $h^{V}_{2,4}$ when studying 
vertices in $Z\gamma$ production. On the other hand, vertices with two on-shell $Z$ bosons and a virtual photon ($\gamma{ZZ}$) or 
a $Z$ boson ($ZZZ$) are characteristic of $ZZ$ production and were not studied at the Tevatron in great detail 
due to low sensitivity to the corresponding TGCs~\cite{zzBaur}. Values of $n$ are set to $n=3$ for all anomalous 
TGCs from $ZZ$ production.
%

\par Because TGCs introduce terms in the Lagrangian that are proportional to the momentum of the weak boson,
 the anomalous behavior is expected to show up at large production angles or high $p_{T}^{V}$ ($V=W,Z$ 
or $\gamma$) of the weak boson. If the weak boson cannot be reconstructed without ambiguities due to the 
presence of a neutrino, as in leptonic decays of $W$ boson, the lepton transverse momentum $p_{T}^{l}$ 
($\ell=e,\mu$) is used instead. Each Tevatron analysis uses some generator that provides an option for simulation 
of anomalous TGC effects. The most common are MCFM, HWZ~\cite{hwz}, and generators of Ref.~\citen{baurs} 
although their application might differ from analysis to analysis. Usually all of these models that predict  
the shape of corresponding $p_{T}$ distributions in a presence of anomalous TGCs, take into account 
$p_{T}$-dependent efficiency and NLO effects. The likelihood between data and Monte Carlo $p_{T}$ 
distributions has been used to set limits on anomalous TGCs. The one-dimensional limits are set when only 
one TGC parameter is varied at the time, while the others are kept at their SM values. If two TGC parameters 
are varied at the time while the third is kept at its SM value we set the two-dimensional limits. 

In the following sections we review the most relevant TGC results from the CDF and D0 experiments obtained 
from the Run II dataset.

\subsection{Experimental Results from CDF}
\begin{figure}[htpb]
  \begin{centering}
    \includegraphics[width=8.0cm]{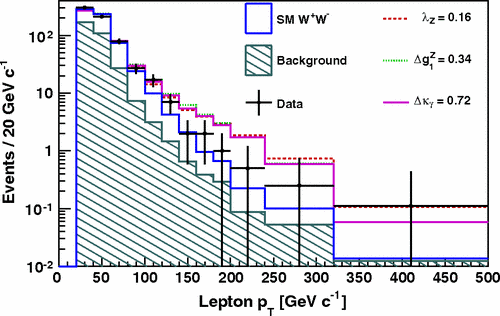}
    \caption{Leading-lepton $p_{T}^{l}$ distribution for $WW\rightarrow \ell\nu \ell\nu$ CDF data selected from 
      3.6~fb$^{-1}$ of integrated luminosity compared to the SM expectation and to expectations modified 
      by anomalous TGCs near the observed limits at $\Lambda=2$~TeV (from Ref.~\citen{cdf-ww-xs1}).}
    \label{fig:cdf-tgc1} 
  \end{centering} 
\end{figure}

At CDF, the $WW\rightarrow \ell\nu \ell\nu$ final states were analyzed in 3.6~fb$^{-1}$ of integrated luminosity and the 
reconstructed leading lepton $p_{T}$ spectrum shown in Figure~\ref{fig:cdf-tgc1} has been used for 
a comparison to Monte Carlo models to asses the sensitivity to different anomalous values of $\Delta\kappa_\gamma$, 
$\lambda$ and $\Delta g_1^Z$~\cite{cdf-ww-xs1}. The one-dimensional 95\% C.L. limits on anomalous TGCs, where one TGC parameter 
is varied at a time while the others are kept at their SM values, in the LEP parameterization 
from $\ell\nu \ell\nu$ final states were found to be of 
$-0.57<\Delta\kappa_{\gamma}<0.65$, $-0.14<\lambda <0.15$ and $-0.22<\Delta{g_{1}^{Z}}<0.30$ for $\Lambda=2$~TeV. 

Anomalous effects were also studied in $\ell\nu{\ell\ell}$ events from $WZ$ production~\cite{cdf-wz-xs1}. The $Z$ boson $p_{T}$ 
distribution $p_{T}^{Z}$, is used to set the 95\% C.L. limits of $-0.39<\Delta\kappa_{Z}<0.90$, $-0.08<\lambda <0.10$ and 
$-0.08<\Delta{g_{1}^{Z}}<0.20$ for $\Lambda = 2$~TeV.

\begin{figure}[htpb]
  \begin{centering}
  \includegraphics[width=8.0cm]{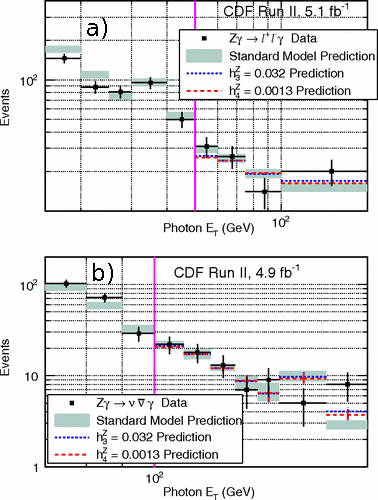}
  \caption{Comparison of the measured $E_{T}^{\gamma}$ distribution from $Z\gamma$ CDF data with the predicted 
    distributions from both the SM and beyond the SM scenarios for (a) $Z\gamma\rightarrow \ell\ell\gamma$ and 
    (b) $Z\gamma\rightarrow \nu\nu\gamma$ samples selected from 5.0~fb$^{-1}$ of integrated luminosity. 
    $\Lambda=1.5$~TeV (from Ref.~\citen{cdf-wz-xs1}).}
  \label{fig:cdf-tgc2} 
  \end{centering} 
\end{figure}

Significant sensitivity to neutral TGCs $h_{3}$ and $h_{4}$ is achieved by using both $Z\gamma\rightarrow\nu\nu\gamma$ 
and $Z\gamma\rightarrow \ell\ell\gamma$ final 
states and optimizing the $E_{T}^{\gamma}$ cut-off values~\cite{cdf-zgamma}. These were determined to be $E_{T}^{\gamma}>50$~GeV and 
$E_{T}^{\gamma}>100$~GeV for the $\ell\ell\gamma$ and $\nu\nu\gamma$ final states, respectively. Selected $E_{T}^{\gamma}$ 
distributions for both final states are shown in Figure~\ref{fig:cdf-tgc2} where data is compared to 
the SM and predictions in the presence of anomalous TGCs. The one-dimensional 95$\%$ C.L. limits on 
${h_{3,4}^{\gamma,Z}}$ at $\Lambda = 1.5$~TeV are $-0.017<h_{3}^{Z,\gamma}<0.016$, $-0.0006<h_{4}^{Z}<0.0005$ and 
$|h_{4}^{\gamma}|<0.0006$ and were the best limits on $h_{3}$ and $h_{4}$ at the time obtained from combination of two 
final states.

The only combined study of anomalous charged TGCs at the CDF experiments in Run II was performed on a data sample of 0.35~fb$^{-1}$
of integrated luminosity~\cite{cdf-tgc-combo}. The combined final states are $WW+WZ\rightarrow \ell\nu{jj}$ and 
$W\gamma\rightarrow \ell\nu\gamma$~\cite{cdf-wgamma-xs} with $p_{T}^{W/Z\rightarrow jj}$ and $p_{T}^{\gamma}$ distributions 
used to set the one-dimensional 95\% C.L. limits of $-0.46<\Delta\kappa <0.39$, $-0.18<\lambda <0.17$ at $\Lambda = 1.5$~TeV. 
Constraints between the couplings given by Eq.~(\ref{eq:lepparam}),~(\ref{eq:eqparam}) and~(\ref{eq:lepeq}) were not applied. 

\subsection{Experimental Results from D0}
Due to a large number of TGC studies performed at the D0 experiment we are going to review only the latest measurements 
and mention those previously performed. Since the very beginning of the Tevatron Run II data taking period the final states 
with a photon produced in association with $W$ or $Z$ boson were studied in great detail. The $W\gamma$ production in 
$\ell\nu\gamma$ final states was regularly tested, first in 0.16~fb$^{-1}$ of integrated luminosity~\cite{d0-wgamma1} 
and later with 0.7~fb$^{-1}$~\cite{d0-wgamma2} and 4.2~fb$^{-1}$~\cite{d0-wgamma3}. These analyses applied somewhat 
different selection criteria but eventually most of the sensitivity to the anomalous TGCs was statistically driven. The 
$\ell\nu\gamma$ final state is of particular importance because it tests only the $\gamma{WW}$ vertex and thus can be studied 
independently of the $ZWW$ vertex, unlike $WW$ interactions. The photon $E_{T}$ spectra for candidate events 
were used to probe data for the presence of anomalous TGCs, $\Delta\kappa_\gamma$ and $\lambda$. 
The one-dimensional 95\% C.L. limits on anomalous TGCs set in data corresponding to 4.2~fb$^{-1}$ of integrated luminosity are 
-0.4~$<\Delta\kappa_{\gamma}<$~0.4 and -0.08~$<\lambda <$~0.07 for $\Lambda=$~2~TeV~\cite{d0-wgamma3}.

Anomalous effects in $Z\gamma$ events have been also regularly tested at the D0 experiment. The $\ell\ell\gamma$ and 
$\nu\nu\gamma$ final states were analyzed and treated individually, then combined to set the limits on ${h_{3,4}^{\gamma,Z}}$ 
TGCs. Only the 0.30~fb$^{-1}$ analysis probed ${h_{1,2}^{\gamma,Z}}$ TGCs utilizing $\ell\ell\gamma$ events~\cite{d0-zgamma1}. 
Later on, limits on these TGCs were not derived as there was not significant sensitivity to those couplings. The first 
anomalous TGC analysis to use $\nu\nu\gamma$ final states used 3.6~fb$^{-1}$ of integrated luminosity~\cite{d0-zgamma3}. 
\begin{figure}[htpb]
  \begin{centering}
  \includegraphics[width=8.0cm]{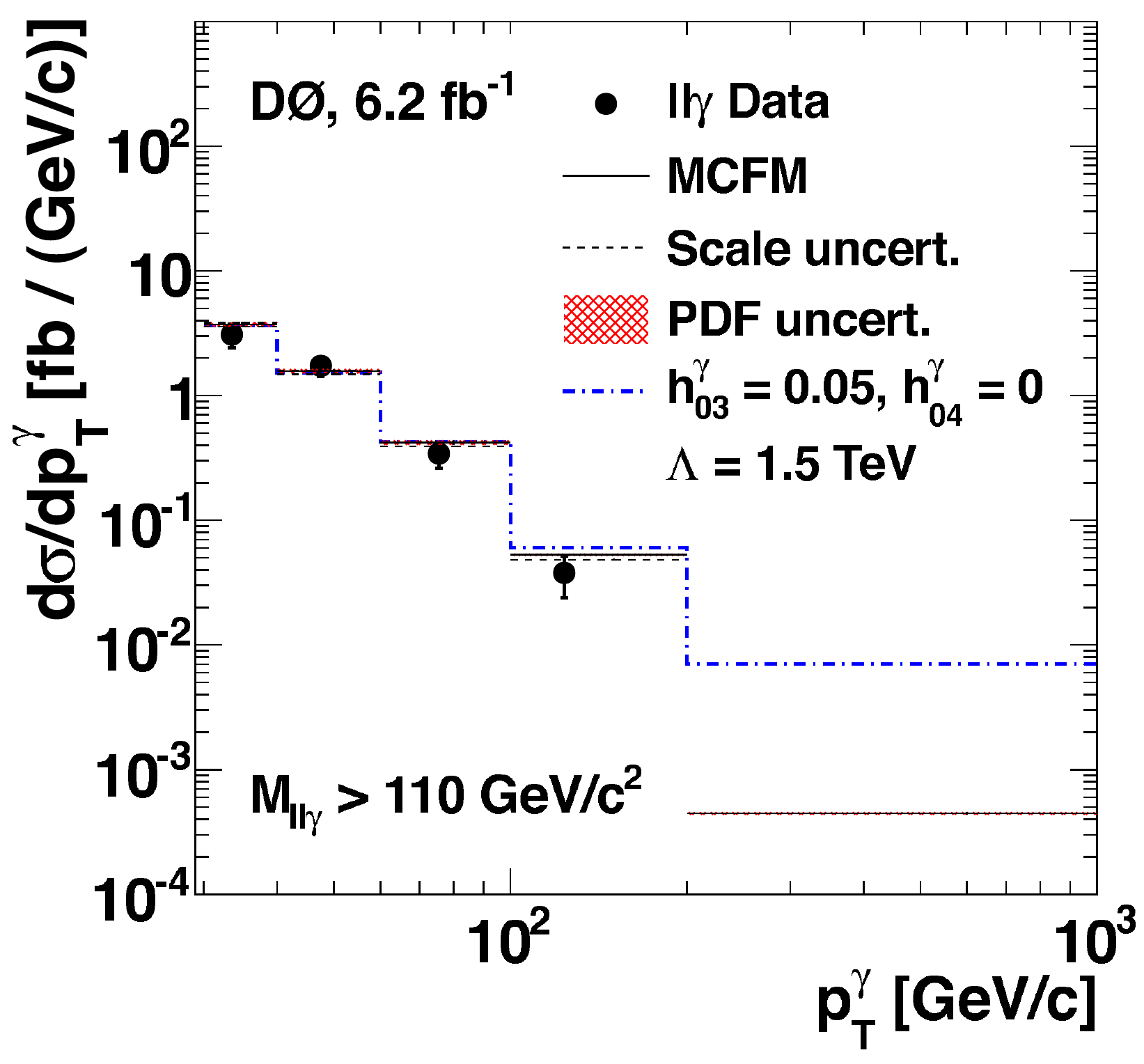}
  \caption{The SM prediction and anomalous $Z\gamma$ coupling production at $\Lambda=1.5$~TeV compared with the unfolded 
    $d\sigma/dp_{T}^{\gamma}$ for $\ell\ell\gamma$ in 6.2~fb$^{-1}$ of integrated luminosity (from Ref.~\citen{d0-zgamma4}).}
  \label{fig:d0-neutgc} 
  \end{centering} 
\end{figure}
The tightest D0 limits on ${h_{3,4}^{\gamma,Z}}$ TGCs describing the $\gamma{Z}\gamma$ and $ZZ\gamma$ vertices are derived 
from combining previous measurements with the results from 6.2~fb$^{-1}$~\cite{d0-zgamma4} of integrated luminosity in 
$\ell\ell\gamma$ final states. The unfolded differential cross section for $Z\gamma\rightarrow{\ell\ell}\gamma$ production as a 
function of a photon $p_{T}^{\gamma}$, $d(\sigma\times{BR})/dp_{T}$, shown in Figure~\ref{fig:d0-neutgc}, has been 
used to probe neutral TGCs. After combining these with previous limits from 1.0~fb$^{-1}$ Run II data~\cite{d0-zgamma2} for 
$\ell\ell\gamma$ and 3.6~fb$^{-1}$ of Run II data~\cite{d0-zgamma3} for $\nu\nu\gamma$, the one-dimensional 95\%\ C.L. limits were set at 
$|h_{30}^{\gamma}|<$~0.027, $|h_{30}^{Z}|<$~0.026, $|h_{40}^{\gamma}|<$~0.0014 and $|h_{40}^{Z}|<$~0.0013 for $\Lambda=$~1.5~TeV. 

\par The only time when the D0 experiment probed $ZZZ$ and $\gamma ZZ$ vertices in $ZZ$ production and set limits on 
${f_{40,50}^{\gamma,Z}}$ TGCs only one $\ell\ell\ell\ell$ data event, consistent with the SM prediction, was selected from 1.0~fb$^{-1}$ of 
integrated luminosity~\cite{d0-zz}. Due to insufficient statistics the number of events expected for each choice of 
anomalous couplings was used to form a likelihood for that point. The one-dimensional 95\%\ C.L. limits for $CP$ violating 
couplings are $|f_{40}^{\gamma}|<0.26$ and $|f_{40}^{Z}|<0.28$, and $-0.30<f_{50}^{\gamma}<0.28$ and $-0.31<f_{50}^{Z}<0.29$ 
for $CP$ conserving couplings, all with $\Lambda = 1.2$~TeV.

\par The $WWZ$ vertex can be probed for anomalous TGC contributions independently of the $WW\gamma$ vertex in 
$WZ\rightarrow \ell\nu{\ell\ell}$ production. Although the production cross section for these fully leptonic final states is relatively 
small they have very little background contamination, making them specially sensitive to $\Delta\kappa_{Z}$ and $g_{1}^{Z}$ TGCs. 
In the first iteration of this analysis only 3 events were selected from a 0.30~fb$^{-1}$ dataset~\cite{d0-wz1-lnull}. Due to 
the low statistics the production cross section has been used to set the limits on anomalous TGCs. The lack of shape information 
from the $Z$ boson $p_{T}$ distribution reduced the sensitivity to anomalous effects. This issue was overcome as the dataset grew. 
The one-dimensional 95\%\ C.L. limits on the coupling parameters were obtained without any coupling relation and with 
the HISZ parameterization at $\Lambda = 2.0$~TeV exploiting the $p_{T}^{Z}$ lineshape~\cite{d0-wz2-lnull}.
\begin{figure}[htpb]
  \begin{centering}
    \includegraphics[width=9.0cm]{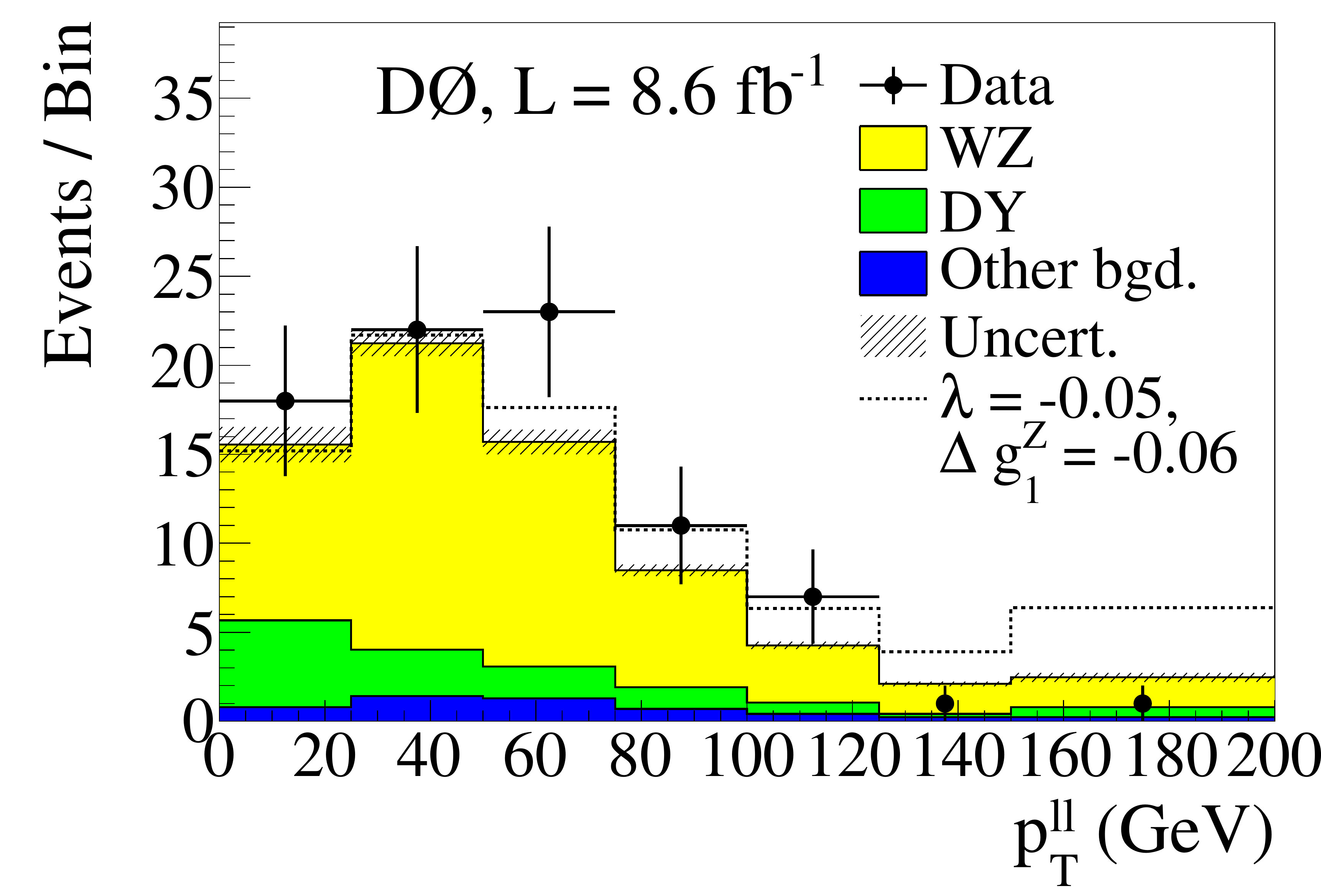}
    \caption{The $Z$ boson $p_{T}$ spectra for $WZ\rightarrow \ell\nu{\ell\ell}$ production from D0 data, SM predictions and 
      anomalous TGC models at $\Lambda=2$~TeV as selected from 8.6~fb$^{-1}$ of integrated luminosity (from Refs.~\citen{d0-zz2-xs,d0-combo-tgc}).}
    \label{fig:d0-wztgc} 
  \end{centering} 
\end{figure}
The former scenario gives limits of -0.376~$<\Delta\kappa_{Z}<$~0.686, -0.075~$<\lambda <$~0.093 and -0.053~$<\Delta{g_{1}^{Z}}<$~0.156 
while the latter gives -0.027~$<\Delta\kappa_{Z}<$~0.080 and -0.075~$<\lambda <$~0.093. The ``LEP parameterization'' relation 
was imposed when setting limits in 8.6~fb$^{-1}$ of integrated luminosity~\cite{d0-combo-tgc}. The corresponding $Z$ boson 
$p_{T}$ shown in Figure~\ref{fig:d0-wztgc} yields limits of -0.077~$<\lambda <$~0.089 and -0.055~$<\Delta{g_{1}^{Z}}<$~0.117.

\par
\begin{figure}[htpb]
  \begin{centering}
  \includegraphics[width=6.0cm]{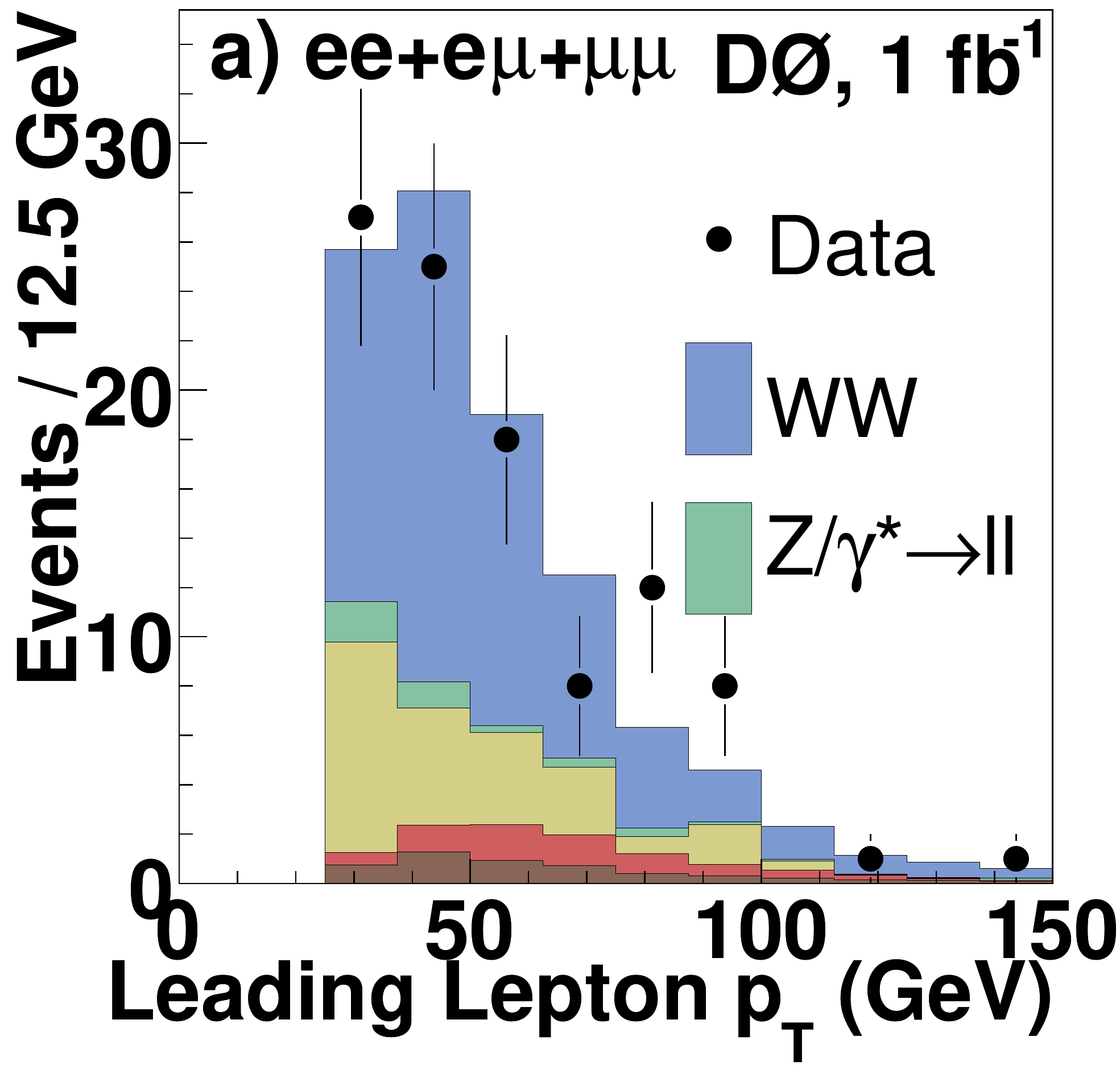}
  \includegraphics[width=6.0cm]{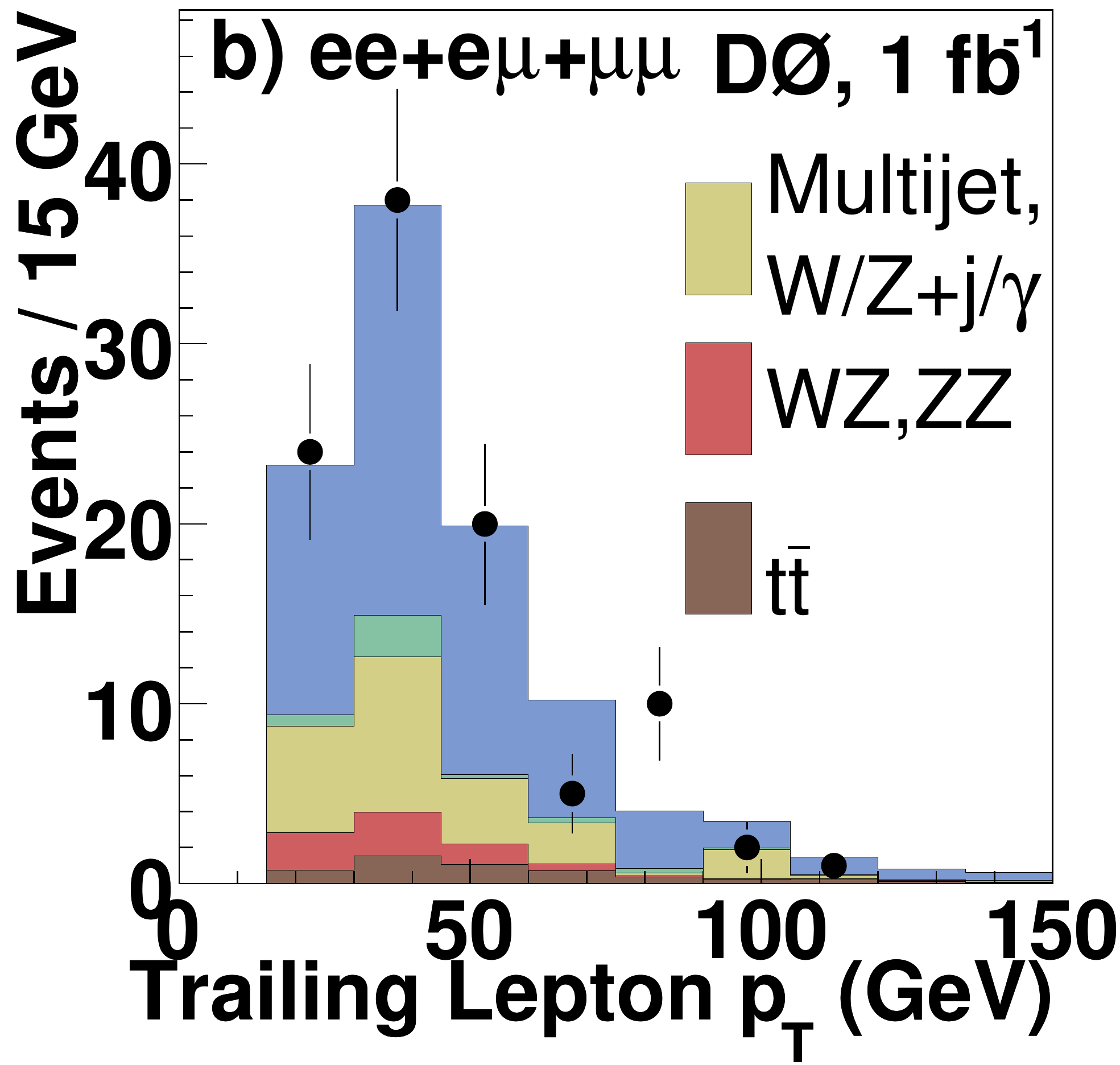}
  \caption{Distributions of (a) the leading and (b) trailing lepton $p_{T}$ after final $WW\rightarrow \ell\nu \ell\nu$ selection 
    at D0 (from Ref.~\citen{d0-ww-xs1}).}
  \label{fig:d0-ww-lnulnu-tgc} 
  \end{centering} 
\end{figure}

 The interference between the $ZWW$ and $\gamma{WW}$ vertices in $WW$ production allows relating the TGCs via the ``LEP 
parameterization'' as given by Eq.~(\ref{eq:lepparam}). The first analysis in $\ell\nu \ell\nu$ final states~\cite{d0-lnulnu-first} 
performed with a 0.25~fb$^{-1}$ dataset was superseded when 100 $\ell\nu \ell\nu$ events were selected from 1.1~fb$^{-1}$ and when, 
instead of the leading $p_{T}$ lepton distribution, both the leading and trailing $p_{T}$ lepton distributions, shown in 
Figure~\ref{fig:d0-ww-lnulnu-tgc}, were used to set the limits~\cite{d0-ww-xs1}. The one-dimensional 95\% C.L. limits for 
$\Lambda=2$~TeV are $-0.54<\Delta\kappa_{\gamma}<0.83$, $-0.14<\lambda <0.18$ and $-0.14<\Delta{g_{1}^{Z}}<0.30$ under the 
``LEP parameterization'' constraints, and $-0.12<\Delta\kappa_{\gamma}=\Delta\kappa_{Z}<0.35$ and 
$-0.14<\lambda <0.18$ under the assumption that $\gamma{WW}$ and $ZWW$ couplings are equal.

\begin{figure}[htpb]
  \begin{centering}
  \includegraphics[width=10.0cm]{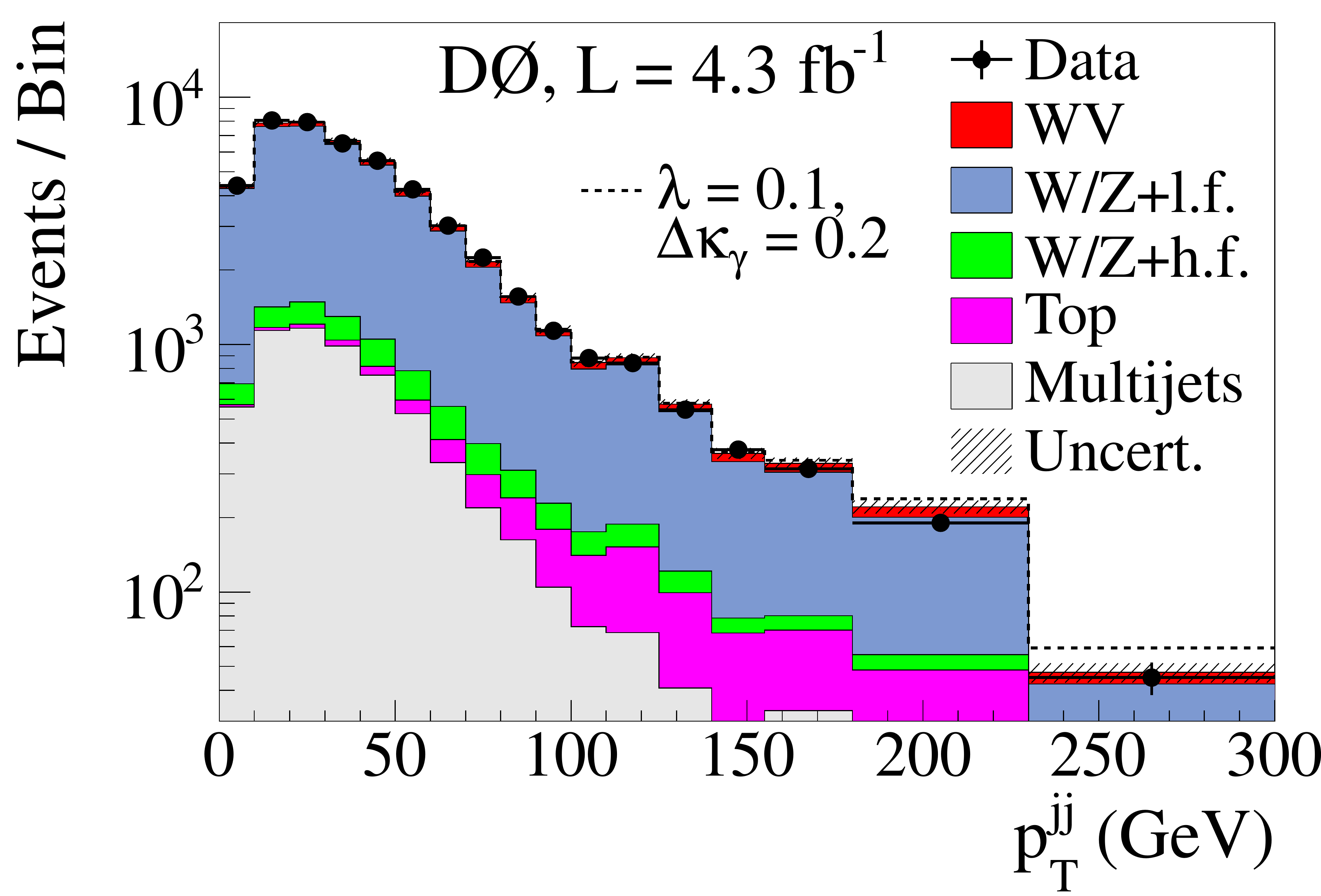}
  \caption{The $p_{T}^{W/Z\rightarrow{jj}}$ distribution from $WW+WZ\rightarrow \ell\nu{jj}$ production for D0 data and SM Monte 
    Carlo predictions. Also shown are expected distributions for an anomalous TGC models at $\Lambda=2$~TeV.}
  \label{fig:d0-combo-tgc1} 
  \end{centering} 
\end{figure}

\begin{figure}[htpb]
  \begin{centering}
    \includegraphics[width=6.2cm]{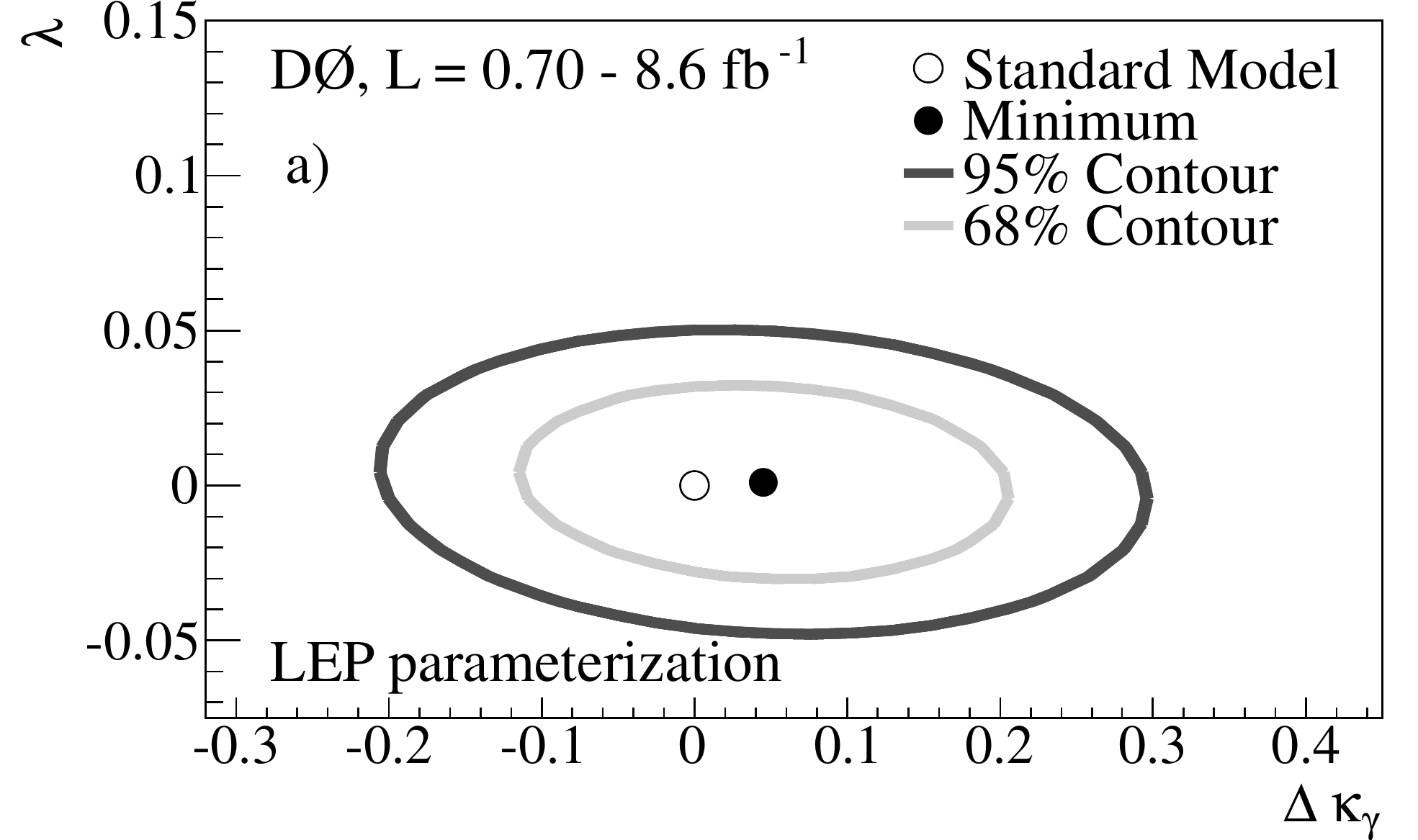}
    \includegraphics[width=6.2cm]{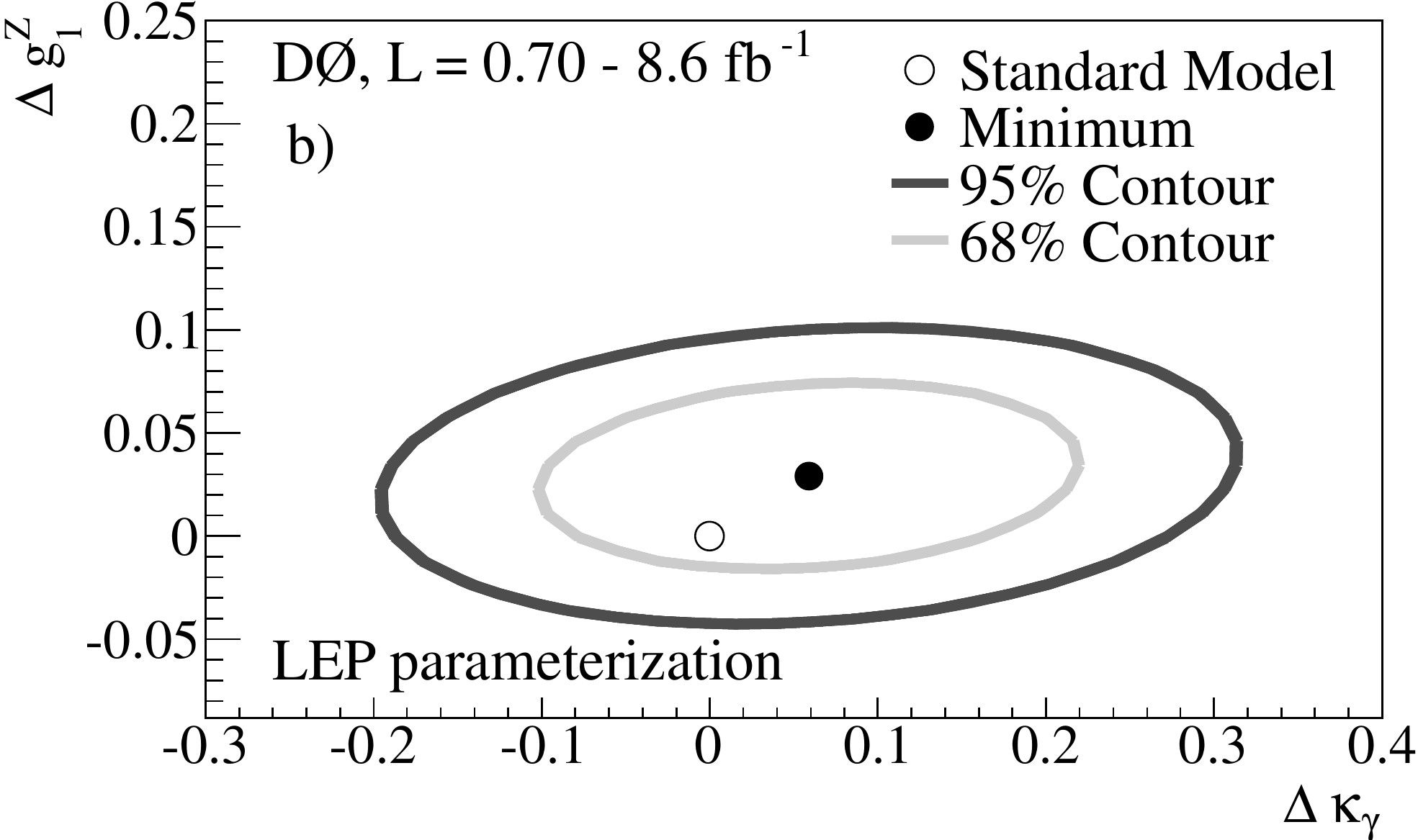}
    \includegraphics[width=6.2cm]{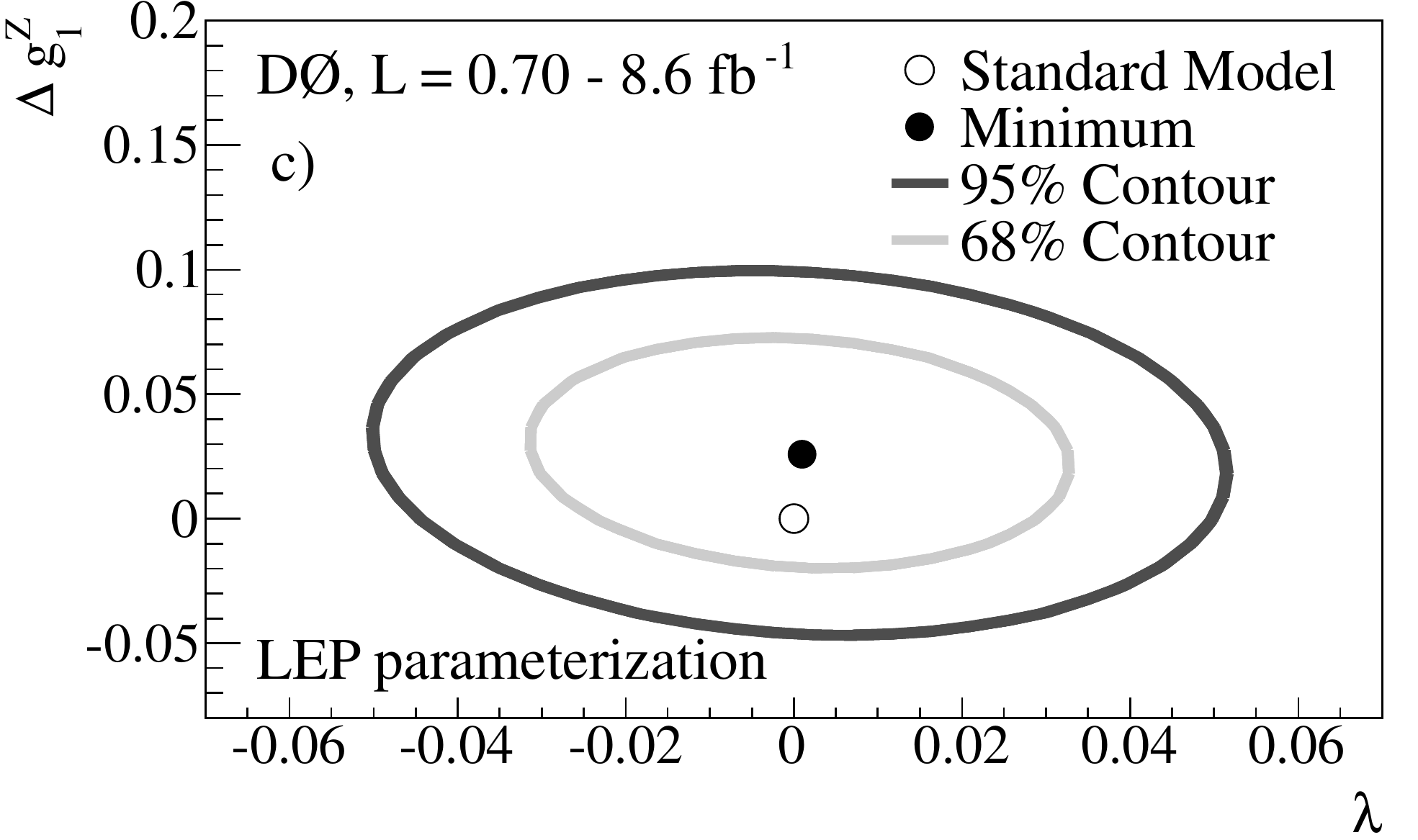}
    \caption{The 68\% and 95\% C.L. two-parameter limits on the $ZWW$ and $\gamma{WW}$ coupling parameters assuming the LEP 
      parameterization with $\Lambda=2$~TeV. Black circles indicate the most probable values of an anomalous TGCs from 
      the two-parameter fit (from Ref.~\citen{d0-combo-tgc}).}
    \label{fig:d0-combo-tgc} 
  \end{centering} 
\end{figure}

The final state most sensitive to anomalous TGC effects, $WW/WZ\rightarrow \ell\nu{jj}$, has been also studied by the D0 
experiment~\cite{d0-combo-tgc,d0-lnujj-tgc}. The dijet $p_{T}$ spectrum for data selected from 4.3~fb$^{-1}$ of integrated 
luminosity and Monte Carlo predictions are shown in Figure~\ref{fig:d0-combo-tgc1}. 
This analysis extends the previous cross section analysis~\cite{d0-wwwz-xs1} with an additional selection of events with the 
dijet mass between $55$ and $110$~GeV to increase the sensitivity to anomalous effects. Although this final state is heavily 
contaminated by $W+$jets and other processes the full reconstruction of the $W$ boson, high energy tail and high $WW+WZ\rightarrow \ell\nu{jj}$ 
statistics significantly boost the sensitivity to $\Delta\kappa_\gamma$, $\lambda$ and $\Delta g_1^Z$ couplings relative to 
fully leptonic final states. The one-dimensional 95\% C.L. limits for $\Lambda=2$~TeV are $-0.27<\Delta\kappa_{\gamma}<0.37$, 
$-0.075<\lambda <0.080$ and $-0.071<\Delta{g_{1}^{Z}}<0.137$ assuming the ``LEP parameterization''. Combining with previous 
analyses in $\ell\nu\gamma$, $\ell\nu \ell\nu$ and $\ell\nu{\ell\ell}$ final states resulted in the most stringent limits on $\Delta\kappa_\gamma$, $\lambda$ and 
$\Delta g_1^Z$ couplings at a hadron collider to that date. These one-dimensional 95\% C.L. limits are presented in Table~\ref{tgcCombo} 
and two-dimensional limits in Figure~\ref{fig:d0-combo-tgc} for three different planes, $\Delta\kappa_\gamma - \lambda$ (a), 
$\Delta\kappa_\gamma - \Delta{g_{1}^{Z}}$ (b) and $\Delta{g_{1}^{Z}} - \lambda$ (c).
Table~\ref{tgcCombo} also presents 68\%\ and 95\%~C.L. limits on the $W$ boson magnetic dipole moment, $\mu_W$, and electric 
quadrupole moment, $q_W$, extracted from combined TGC limits assuming the LEP parameterization with $g^Z_1=1$. The quantities 
$\mu_W$ and $q_W$ are related to the coupling parameters by:
\begin{equation}
  \mu_W = \frac{e}{2M_W} (1 + \kappa_\gamma + \lambda_\gamma),~ 
  q_W = - \frac{e}{M^2_W} (\kappa_\gamma - \lambda_\gamma). 
  \label{eq:moments}
\end{equation}

The corresponding 68\%~C.L. intervals are $\mu_W=2.012^{+0.035}_{-0.034} \, (e/2M_W)$ and $q_W=-0.995^{+0.042}_{-0.043} \, (e/M^2_W)$ 
which are the most stringent 68\%~C.L. constraints on these parameters to date. 
\begin{table}[htpb]
\tbl{One-dimensional $\chi^2$ minimum and 68\% and 95\%~C.L. allowed intervals on anomalous values of $\gamma WW/ZWW$ anomalous 
  TGCs from the combined fit of $WW+WZ\rightarrow \ell\nu{jj}$, $WZ\rightarrow \ell\nu{\ell\ell}$, $W\gamma\rightarrow \ell\nu\gamma$, 
  and $WW\rightarrow \ell\nu \ell\nu$ final states.}
{\begin{tabular}{@{}lrcc@{}} \toprule
\multicolumn{4}{c}{Results for LEP parameterization} \\
Parameter & Minimum & 68\%~C.L. & 95\%~C.L. \\ \colrule
$\Delta\kappa_\gamma$ \hphantom{00} & \hphantom{0}$0.048$ & \hphantom{0}$[-0.057, 0.154]$ & $[-0.158, 0.255]$ \\
$\Delta g_1^Z$ \hphantom{00} & \hphantom{0}$0.022$ & \hphantom{0}$[-0.008, 0.054]$  & $[-0.034, 0.084]$ \\
$\lambda$\hphantom{00} & \hphantom{0}$0.007$  & \hphantom{0}$[-0.015, 0.028]$  & $[-0.036, 0.044]$ \\ \\
$\mu_W$~$(e/2M_W)$\hphantom{00} & \hphantom{0}$2.012$  & \hphantom{0}$[1.978, 2.047]$   & $[1.944, 2.080]$ \\
$q_W$~$(e/M^2_W)$\hphantom{00} & \hphantom{0}$-0.995$ & \hphantom{0}$[-1.038, -0.953]$ & $[-1.079, -0.916]$ \\ \botrule
\end{tabular} \label{tgcCombo}}
\end{table}
%


\subsubsection{Study of Quartic Gauge Boson Couplings at D0}
If there is a third gauge boson produced in the final state the gauge boson self-interactions are described by 
quartic gauge couplings (QGC)~\cite{qgc}. At the Tevatron the production cross sections for three-boson final states
are very small because of the relatively low center-of-mass energy. The only QGC study at the Tevatron 
has been focused on $WW\gamma\gamma$ couplings, $a_{0}^{W}$ and $a_{C}^{W}$, in $WW\gamma$ production with both $W$ 
bosons decaying leptonically~\cite{d0-qgc}. This analysis of 9.7~fb$^{-1}$ of integrated luminosity uses a 
Boosted Decision Tree discriminant to search for anomalous QGCs. As no evidence for new physics was found the 95\% C.L. upper limits 
on $a_{0}^{W}$ and $a_{C}^{W}$ were set to be $|a_{0}^{W}/\Lambda|<0.0025$~GeV$^{-2}$ and 
$|a_{C}^{W}/\Lambda|<0.0092$~GeV$^{-2}$ for $\Lambda=0.5$~TeV.


\section{Summary and Conclusions}	

The Tevatron collider dataset has been a very rich source of measurements pertaining to electroweak physics and testing higher-order QCD 
calculations. Differential distributions of electroweak gauge bosons have constrained PDFs. Precision measurements of $M_W$, $\Gamma_W$ 
and the $Z$ boson forward-backward asymmetry $A_{FB}$, have tested the electroweak theory at loop-level at a similar level of precision 
as the LEP and SLD measurements. Studies of diboson production have confirmed the $SU(2)_L \times U(1)_Y$ gauge structure in a manner 
complementary to, and with a precision similar to, that achieved at LEP II. Tables~\ref{tgcTable1} and~\ref{tgcTable2} summarize the best 
one-dimensional 95\% C.L. limits on anomalous charged and neutral TGCs set using individual final states at the Tevatron experiments. 
Some of these measurements will be legacy measurements from the Tevatron.

\begin{table}[htpb]
  \tbl{The best one-dimensional 95\%~C.L. limits on charged TGCs from individual channels analized at the Tevatron experiments in Run II. 
    In all cases $\Lambda=2$~TeV.}
      {\begin{tabular}{@{}llccccc@{}} \toprule
Anomalous & Scenario & Luminosity & $W\gamma\rightarrow \ell\nu\gamma$ & $WW\rightarrow \ell\nu \ell\nu$ & $WW+WZ\rightarrow \ell\nu{jj}$ & $WZ\rightarrow l\nu{\ell\ell}$ \\
TGC & & [fb$^{-1}$] & & & & \\ \colrule

$\Delta\kappa_\gamma$ \hphantom{00} & $-$ &  4.2~fb$^{-1}$ (D0) & \hphantom{0}$[-0.4,0.4]$ & & &   \\ 
$\lambda$ \hphantom{00}             & $-$ &  4.2~fb$^{-1}$ (D0) & \hphantom{0}$[-0.08,0.07]$ & & &  \\
$g_{1}^{Z}$ \hphantom{00}           & $-$ & & $-$ & & &  \\ \\

$\Delta\kappa_\gamma$ \hphantom{00} & LEP &  3.6~fb$^{-1}$ (CDF) & & $[-0.57,0.65]$ & &   \\
$\lambda$ \hphantom{00}             & LEP &  3.6~fb$^{-1}$ (CDF) & & $[-0.14,0.15]$ & &  \\
$g_{1}^{Z}$ \hphantom{00}           & LEP &  1.0~fb$^{-1}$ (D0) & & $[-0.14,0.30]$ & &  \\ \\

$\Delta\kappa_\gamma$ \hphantom{00} & EQUAL &  1.0~fb$^{-1}$ (D0) & & $[-0.12,0.35]$ & &  \\
$\lambda$ \hphantom{00}             & EQUAL &  1.0~fb$^{-1}$ (D0) & & $[-0.14,0.18]$ & &  \\ 
$g_{1}^{Z}$ \hphantom{00}           & EQUAL & & & $-$ & &  \\ \\

$\Delta\kappa_\gamma$ \hphantom{00} & LEP &  4.3~fb$^{-1}$ (D0) & & & $[-0.27,0.37]$ &  \\
$\lambda$ \hphantom{00}             & LEP &  4.3~fb$^{-1}$ (D0) & & & $[-0.075,0.080]$ &  \\
$g_{1}^{Z}$ \hphantom{00}           & LEP &  4.3~fb$^{-1}$ (D0) & & & $[-0.071,0.137]$ &  \\ \\

$\Delta\kappa_\gamma$ \hphantom{00} & EQUAL &  4.3~fb$^{-1}$ (D0) & & & $[-0.078,0.153]$ &  \\
$\lambda$ \hphantom{00}             & EQUAL &  4.3~fb$^{-1}$ (D0) & & & $[-0.074,0.079]$ &  \\
$g_{1}^{Z}$ \hphantom{00}           & EQUAL & & & & $-$ &  \\ \\

$\Delta\kappa_\gamma$ \hphantom{00} & LEP & & & & & $-$  \\
$\lambda$ \hphantom{00}             & LEP &  8.6~fb$^{-1}$ (D0) & & & & $[-0.077,0.089]$ \\
$\lambda$ \hphantom{00}             & EQUAL &  8.6~fb$^{-1}$ (D0) & & & & $[-0.077,0.090]$ \\
$g_{1}^{Z}$ \hphantom{00}           & LEP &  8.6~fb$^{-1}$ (D0) & & & & $[-0.055,0.117]$ \\ \\

$\Delta\kappa_{Z}$ \hphantom{00}    & LEP &  4.1~fb$^{-1}$ (D0) & & & & $[-0.376,0.686]$ \\
$\lambda$ \hphantom{00}             & LEP &  4.1~fb$^{-1}$ (D0) & & & & $[-0.075,0.093]$ \\
$g_{1}^{Z}$ \hphantom{00}           & LEP &  4.1~fb$^{-1}$ (D0) & & & & $[-0.053,0.156]$ \\ \\

$\Delta\kappa_{Z}$ \hphantom{00}    & HISZ &  4.1~fb$^{-1}$ (D0) & & & & $[-0.027,0.080]$ \\
$\lambda$ \hphantom{00}             & HISZ &  4.1~fb$^{-1}$ (D0) & & & & $[-0.075,0.093]$ \\
$g_{1}^{Z}$ \hphantom{00}           & HISZ &  & & & & $-$ \\ \botrule

\end{tabular} 
\label{tgcTable1}}
\end{table}

\clearpage

\begin{table}[htpb]
\tbl{The best one-dimensional 95\%~C.L. limits on neutral TGCs from individual channels analized at the Tevatron experiments in Run II.}
{\begin{tabular}{@{}lccccc@{}} \toprule
Anomalous & $\Lambda$ & Luminosity & $Z\gamma\rightarrow \ell\ell\gamma$ & $Z\gamma\rightarrow \nu\nu\gamma$ & $ZZ\rightarrow{\ell\ell\ell\ell}$ \\
TGC & [TeV] & [fb$^{-1}$] & & & \\ \colrule
$h_{30}^{\gamma}$ \hphantom{00}      & 1.5 & 6.2~fb$^{-1}$ (D0) & \hphantom{0}$[-0.044,0.044]$ & &  \\
$h_{30}^{Z}$ \hphantom{00}           & 1.5 & 6.2~fb$^{-1}$ (D0) & \hphantom{0}$[-0.041,0.041]$ & &  \\
$h_{40}^{\gamma}$ \hphantom{00}      & 1.5 & 6.2~fb$^{-1}$ (D0) & \hphantom{0}$[-0.0023,0.0023]$ & &  \\
$h_{40}^{Z}$ \hphantom{00}           & 1.5 & 6.2~fb$^{-1}$ (D0) & \hphantom{0}$[-0.0023,0.0023]$ & &  \\ \\

$h_{30}^{\gamma}$ \hphantom{00}      & 1.5 & 3.6~fb$^{-1}$ (D0) & & \hphantom{0}$[-0.036, 0.036]$ &  \\
$h_{30}^{Z}$ \hphantom{00}           & 1.5 & 3.6~fb$^{-1}$ (D0) & & \hphantom{0}$[-0.035, 0.035]$ &  \\
$h_{40}^{\gamma}$ \hphantom{00}      & 1.5 & 3.6~fb$^{-1}$ (D0) & & \hphantom{0}$[-0.0019, 0.0019]$ &  \\ 
$h_{40}^{Z}$ \hphantom{00}           & 1.5 & 3.6~fb$^{-1}$ (D0) & & \hphantom{0}$[-0.0019, 0.0019]$ &  \\ 
\colrule
$f_{40}^{\gamma}$ \hphantom{00}      & 1.2 & 1.0~fb$^{-1}$ (D0) &  &  & $[-0.26,0.26]$ \\
$f_{40}^{Z}$ \hphantom{00}           & 1.2 & 1.0~fb$^{-1}$ (D0) &  &  & $[-0.28,0.28]$ \\
$f_{50}^{\gamma}$ \hphantom{00}      & 1.2 & 1.0~fb$^{-1}$ (D0) &  &  & $[-0.30,0.28]$ \\
$f_{50}^{Z}$ \hphantom{00}           & 1.2 & 1.0~fb$^{-1}$ (D0) &  &  & $[-0.31,0.29]$ \\ \botrule
\end{tabular} 
\label{tgcTable2}}
\end{table}
%
\section*{Acknowledgements}

The authors would like to thank the following: Paul Grannis, Willis Sakumoto, Marco Verzocchi and Hang Yin for useful comments and discussions.

We thank the Fermilab staff and technical staffs of
the participating institutions for their vital contributions.
We acknowledge support from the DOE and NSF
(USA), ARC (Australia), CNPq, FAPERJ, FAPESP
and FUNDUNESP (Brazil), NSERC (Canada), NSC,
CAS and CNSF (China), Colciencias (Colombia), MSMT
and GACR (Czech Republic), the Academy of Finland,
CEA and CNRS/IN2P3 (France), BMBF and DFG (Germany),
DAE and DST (India), SFI (Ireland), INFN
(Italy), MEXT (Japan), the KoreanWorld Class University
Program and NRF (Korea), CONACyT (Mexico),
FOM (Netherlands), MON, NRC KI and RFBR (Russia),
the Slovak R\&D Agency, the Ministerio de Ciencia
e InnovaciÂ´on, and Programa Consolider--Ingenio 2010
(Spain), The Swedish Research Council (Sweden), SNSF
(Switzerland), STFC and the Royal Society (United
Kingdom), the A.P. Sloan Foundation (USA), and the
EU community Marie Curie Fellowship contract 302103.


%


\begin{thebibliography}{000}

\bibitem{GWS} S.~Glashow, Nucl.\,Phys.\, {\bf 22}, 579 (1961); 
A.~Salam and J.~C.~Ward, Phys.\,Lett.\, {\bf 13}, 168 (1964); 
S.~Weinberg, Phys.\,Rev.\,Lett.\, {\bf 19}, 1264 (1967).
\bibitem{lhchiggs} G.~Aad {\it et al.} (The ATLAS Collaboration), Phys.\,Lett.\,B {\bf 716}, 1 (2012); 
S.~Chatrchyan {\it et al.} (The CMS Collaboration), Phys.\, Lett.\,B {\bf 716}, 30 (2012).
\bibitem{ewsb} P.~W.~Anderson, Phys.\,Rev.\, {\bf 130}, 439 (1960);
F.~Englert and R.~Brout, Phys.\,Rev.\,Lett.\, {\bf 13}, 321 (1964); 
P.~W.~Higgs, Phys.\,Rev. \,Lett.\, {\bf 13}, 508 (1964);  
G.~S.~Guralnik, C.~R.~Hagen, and T.~W.~B.~Kibble, Phys.\,Rev.\,Lett.\, {\bf 13}, 585 (1964).

\bibitem{CDFZmass1989} CDF Collaboration (F.~Abe {\it et al.}), Phys.\ Rev.\ Lett.\  {\bf 63}, 720 (1989).
\bibitem{NIMS} CDF Collaboration (F.~Abe {\it et al.}), Nucl.\,Instrum.\,Meth.\, A {\bf 271}, 387 (1988) and references contained therein;
CDF Collaboration (F.~Abe {\it et al.}), Nucl.\,Instrum.\,Meth.\, A {\bf 271}, 387 (1988);
CDF Collaboration (F.~Abe {\it et al.}), Phys.\,Rev.\, D {\bf 50}, 2966 (1994);
D.~Amidei {\it et al.}, Nucl.\,Instrum.\,Meth.\, A {\bf 350}, 73 (1994); 
P.~Azzi {\it et al.}, Nucl.\,Instrum.\,Meth.\, A {\bf 360}, 137 (1995);
F.~Snider {\it et al.}, Nucl.\,Instrum.\,Meth.\, A {\bf 268}, 75 (1988). 
F.~Bedeschi {\it et al.}, Nucl.\,Instrum.\,Meth.\, A {\bf 268}, 50 (1988);
L.~Balka {\it et al.}, Nucl.\,Instrum.\,Meth.\, A {\bf 267}, 272 (1988); 
S.~R.~Hahn {\it et al.}, Nucl.\,Instrum.\,Meth.\, A {\bf 267}, 351 (1988); 
K.~Yasuoka {\it et al.}, Nucl.\,Instrum.\,Meth.\, A {\bf 267}, 315 (1988); 
R.~G.~Wagner {\it et al.}, Nucl.\,Instrum.\,Meth.\, A {\bf 267}, 330 (1988); 
T.~Devlin {\it et al.}, Nucl.\,Instrum.\,Meth.\, A {\bf 267}, 24 (1988);
S.~Bertolucci {\it et al.}, Nucl.\,Instrum.\,Meth.\, A {\bf 267}, 301 (1988);
Y.~Fukui {\it et al.}, Nucl.\,Instrum.\,Meth.\, A {\bf 267}, 280 (1988); 
W.~C.~Carithers {\it et al.}, ``Proceedings of the Gas Sampling Calorimetry Workshop II,'' Batavia, Illinois, 54 (1985); 
S.~Cihangir {\it et al.}, Nucl.\,Instrum.\,Meth.\, A {\bf 267}, 249 (1988); 
G.~Brandenburg {\it et al.}, Nucl.\,Instrum.\,Meth.\, A {\bf 267}, 257 (1988);
G.~Ascoli {\it et al.}, Nucl.\,Instrum.\,Meth.\, A {\bf 268}, 33 (1988); 
G.~Ascoli {\it et al.}, Nucl.\,Instrum.\,Meth.\, A {\bf 268}, 41 (1988);
D.~Amidei {\it et al.}, Nucl.\,Instrum.\,Meth.\, A {\bf 269}, 51 (1988);
T.~Carroll {\it et al.}, Nucl.\,Instrum.\,Meth.\, A {\bf 263}, 199 (1988);
G.~W.~Foster {\it et al.}, Nucl.\,Instrum.\,Meth.\, A {\bf 269}, 93 (1988);
K.~Byrum {\it et al.}, Nucl.\,Instrum.\,Meth.\, A {\bf 364}, 144 (1995).

\bibitem{D0NIMS} D0 Collaboration (S.~Abachi {\it et al.}), Nucl.\,Instrum.\,Meth.\,  A {\bf 338}, 185 (1994). 

\bibitem{RunIINIMS}
T.~LeCompte and H.~T.~Diehl, Ann.\,Rev.\,Nucl.\,Part.\,Sci. 50, 71 (2000);
H.~Minemura {\it et al.}, Nucl.\,Instrum.\,Meth.\,Phys.\,Res., Sect.\, A {\bf 238}, 18 (1985);
K.~Yasuoka, S.~Mikamo, T.~Kamon, and A.~Yamashita, Nucl.\,Instrum.\,Meth.\,Phys.\,Res., Sect.\, A {\bf 267}, 315 (1987);
L.~Balka {\it et al.}, Nucl.\,Instrum.\,Meth.\,Phys.\,Res., Sect.\, A {\bf 267}, 272 (1988);
S.~Bertolucci {\it et al.}, Nucl.\,Instrum.\,Meth.\,Phys.\,Res., Sect.\, A {\bf 267}, 301 (1988);
G.~Ascoli {\it et al.}, Nucl.\,Instrum.\,Meth.\,Phys.\,Res., Sect.\, A {\bf 268}, 33 (1988);
G.~Apollinari {\it et al.}, Nucl.\,Instrum.\,Meth.\,Phys.\,Res., Sect.\, A {\bf 412}, 515 (1998);
J.~Elias {\it et al.}, Nucl.\,Instrum.\,Meth.\,Phys.\,Res., Sect.\, A {\bf 441}, 366 (2000);
A.~Sill {\it et al.}, Nucl.\,Instrum.\,Meth.\,Phys.\,Res., Sect.\, A {\bf 447}, 1 (2000);
M.~Albrow {\it et al.}, Nucl.\,Instrum.\,Meth.\,Phys.\,Res., Sect.\, A {\bf 480}, 524 (2002);
E.~J.~Thomson {\it et al.}, IEEE Trans.\,Nucl.\,Sc. {\bf 49}, 1063 (2002);
D.~Acosta {\it et al.}, Nucl.\,Instrum.\,Meth.\,Phys.\,Res., Sect.\, A {\bf 518}, 605 (2004);
T.~Affolder {\it et al.}, Nucl.\,Instrum.\,Meth.\,Phys.\,Res., Sect.\, A {\bf 526}, 249 (2004);.
CDF Collaboration (D.~Acosta {\it et al.}), Phys.\,Rev.\, D {\bf 71}, 032001 (2005);
CDF Collaboration (A.~Abulencia {\it et al.}), J.\,Phys.\, G {\bf 34} 2457 (2007); 
G.~D.~Alexeev {\it et al.}, 
Nucl.\,Instrum.\,Meth.\,Phys.\,Res., Sect.\, A {\bf 473}, 269 (2001);
T.~Zhao {\it et al.}, 
IEEE Trans\,Nucl.\,Sci.\, 49, 1092 (2002);
M.~Abolins {\it et al.}, 
IEEE Trans\,Nucl.\,Sci.\, 51, 340 (2004);
V.~M.~Abazov {\it et al.}, 
Nucl.\,Instrum.\,Meth.\,Phys.\,Res., Sect.\, A {\bf 552}, 372 (2005);
M.~Abolins {\it et al.}, 
IEEE Trans\,Nucl.\,Sci.\,52, 3233 (2005);
M.~Abolins {\it et al.}, 
Nucl.\,Instrum.\,Meth.\,Phys.\,Res., Sect.\, A  {\bf 584}, 75 (2008);.
R.~Angstadt {\it et al.}, 
Nucl.\,Instrum.\,Meth.\,Phys.\,Res., Sect.\, A {\bf 622}, 298 (2010);
S.~N.~Ahmed {\it et al.}, 
Nucl.\,Instrum.\,Meth.\,Phys.\,Res., Sect.\, A {\bf 634}, 8 (2011);
B.~Abbott {\sl et al.}, Nucl.\,Instrum.\,Methods\, Phys.\, Res.\, A {\bf 565}, 463 (2006); 
M.~Abolins {\sl et al.}, Nucl\,Instrum.\,Methods\, A {\bf 584}, 75 (2007); 
R.~Angstadt {\sl et al.}, Nucl\, Instrum.\,Methods\,Phys.\, Res.\, A {\bf 622}, 298 (2010).

\bibitem{LEPEWWG} [ALEPH and CDF and D0 and DELPHI and L3 and OPAL and SLD and LEP Electroweak Working Group and Tevatron Electroweak Working Group and SLD Electroweak and Heavy Flavour Groups Collaborations], ``Precision Electroweak Measurements and Constraints on the Standard Model,'' arXiv:1012.2367 [hep-ex].

\bibitem{D0ElectronID} D0 Collaboration (V.~M.~Abazov {\it et al.}), 
Nucl.\,Instrum.\,Meth.\, A {\bf 750}, 78 (2014). 

\bibitem{DrellYan} S.~D.~Drell and T.~-M.~Yan, Phys.\, Rev.\, Lett.\, {\bf 25}, 316 (1970) [Erratum-ibid.\, {\bf 25}, 902 (1970)].

\bibitem{Hamberg1990} R.~Hamberg, W.~L.~van Neerven and T.~Matsuura, 
Nucl.\, Phys.\, B {\bf 359}, 343 (1991) [Erratum-ibid.\, B {\bf 644}, 403 (2002)].

\bibitem{CDFCross1992} CDF Collaboration (F.~Abe {\it et al.}), 
Phys.\, Rev.\, Lett.\, {\bf 69}, 28 (1992).

\bibitem{CDFCross1994} CDF Collaboration (F.~Abe {\it et al.}), 
Phys.\, Rev.\, Lett.\, {\bf 73}, 220 (1994).

\bibitem{CDFCross1995} CDF Collaboration (F.~Abe {\it et al.}), 
Phys.\, Rev.\, D {\bf 52}, 2624 (1995).

\bibitem{CDFCross1996} CDF Collaboration (F.~Abe {\it et al.}), 
 Phys.\, Rev.\, Lett.\, {\bf 76}, 3070 (1996).

\bibitem{CDFCross1998} CDF Collaboration (F.~Abe {\it et al.}),
Phys.\, Rev.\, D {\bf 59}, 052002 (1999).

\bibitem{CDFCross1999} CDF Collaboration (T.~Affolder {\it et al.}), 
Phys.\ Rev.\, Lett.\, {\bf 84}, 845 (2000). 

\bibitem{D0Cross1995} D0 Collaboration (S.~Abachi {\it et al.}), 
Phys.\, Rev.\, Lett.\, {\bf 75}, 1456 (1995). 

\bibitem{D0WZCross1999} D0 Collaboration (B.~Abbott {\it et al.}), 
Phys.\, Rev.\, D {\bf 60}, 052003 (1999). 

\bibitem{D0WZWidth1999} D0 Collaboration (B.~Abbott {\it et al.}), 
Phys.\, Rev.\, D {\bf 61}, 072001 (2000). 

\bibitem{CDFCross2004} CDF Collaboration (D.~Acosta {\it et al.}), 
Phys.\, Rev.\, Lett.\, {\bf 94}, 091803 (2005). 

\bibitem{CDFCross2005} CDF Collaboration (A.~Abulencia {\it et al.}), 
J.\, Phys.\, G {\bf 34}, 2457 (2007). 

\bibitem{MSTW2008}A.~D.~Martin, W.~J.~Stirling, R.~S.~Thorne and G.~Watt, 
Eur.\, Phys.\, J.\, C {\bf 63}, 189 (2009). 

\bibitem{CDFCross1989} CDF Collaboration (F.~Abe {\it et al.}), 
Phys.\, Rev.\, Lett.\, {\bf 64}, 152 (1990).

\bibitem{CDFCross1990} CDF Collaboration (F.~Abe {\it et al.}), 
Phys.\, Rev.\, D {\bf 44}, 29 (1991).

\bibitem{CDFZPT1991} CDF Collaboration (F.~Abe {\it et al.}), 
Phys.\, Rev.\, Lett.\, {\bf 67}, 2937 (1991).

\bibitem{D0WPT1998} D0 Collaboration (B.~Abbott {\it et al.}), 
Phys.\, Rev.\, Lett.\, {\bf 80}, 5498 (1998). 

\bibitem{D0ZPT1999} D0 Collaboration (B.~Abbott {\it et al.}), 
Phys.\, Rev.\, D {\bf 61}, 032004 (2000). 

\bibitem{D0ZPT2000} D0 Collaboration (B.~Abbott {\it et al.}), 
Phys.\, Rev.\, Lett.\, {\bf 84}, 2792 (2000). 

\bibitem{D0WPT2001} D0 Collaboration (B.~Abbott {\it et al.}), 
Phys.\, Lett.\, B {\bf 513}, 292 (2001). 

\bibitem{D0WZPT2001} D0 Collaboration (V.~M.~Abazov {\it et al.}), 
Phys.\, Lett.\, B {\bf 517}, 299 (2001). 

\bibitem{D0ZPT2008} D0 Collaboration (V.~M.~Abazov {\it et al.}), %
Phys.\, Rev.\, Lett.\, {\bf 100}, 102002 (2008). 

\bibitem{aT_papers}
M. Vesterinen, T. R. Wyatt, Nucl.\,Instrum.\, Methods\, Phys.\,Res. A
{\bf 602}, 432 (2009). A. Banfi, M. Dasgupta, R. Delgado, J. High Energy Phys. {\bf 12}, 022 (2009).
  A.~Banfi, S.~Redford, M.~Vesterinen, P.~Waller and T.~R.~Wyatt,
  Eur.\ Phys.\ J.\ C {\bf 71}  1600 (2011).


\bibitem{D0ZPT2010} D0 Collaboration (V.~M.~Abazov {\it et al.}), 
Phys.\, Rev.\, Lett.\, {\bf 106}, 122001 (2011). 



\bibitem{CDFPT2012} CDF Collaboration (T.~Aaltonen {\it et al.}), 
Phys.\, Rev.\, D {\bf 86}, 052010 (2012).

\bibitem{D0ZY2007} D0 Collaboration (V.~M.~Abazov {\it et al.}), 
Phys.\, Rev.\, D {\bf 76}, 012003 (2007).

\bibitem{CDFZY2010} CDF Collaboration (T.~A.~Aaltonen {\it et al.}), 
Phys.\, Lett.\, B {\bf 692}, 232 (2010). 

\bibitem{ADMP} C.~Anastasiou, L.~J.~Dixon, K.~Melnikov and F.~Petriello, 
Phys.\, Rev.\, D {\bf 69}, 094008 (2004).

\bibitem{Catani2007} S.~Catani and M.~Grazzini, 
Phys.\, Rev.\, Lett.\, {\bf 98}, 222002 (2007). 

\bibitem{MRST2004} A.~D.~Martin, R.~G.~Roberts, W.~J.~Stirling and R.~S.~Thorne, 
Phys.\, Lett.\, B {\bf 604}, 61 (2004).

\bibitem{CTEQ61}
 J.~Pumplin, D.~R.~Stump, J.~Huston, H.~L.~Lai, P.~Nadolsky, W.~K.~Tung, JHEP 0207 (2002) 012;  J.~Pumplin, D.~R.~Stump, J.~Huston, H.~L.~Lai, W.~K.~Tung, S.~Kuhlmann, J.~F.~Owens, JHEP 0310:046 (2003).
\bibitem{CTEQ66}
  P.~Nadolsky, H.~L.~Lai, Q.~H.~Cao, J.~Huston, J.~Pumplin, D.~Stump, W.~K. Tung, and C.~P.~Yuan, Phys.\, Rev.\, {\bf D78}, 013004 (2008).

\bibitem{Alekhin02}
\bibitem{Alekhin:2002fv}
  S.~Alekhin,
  Phys.\, Rev.\, D {\bf 68}, 014002 (2003). 


\bibitem{ArnoldReno} P.~B.~Arnold and M.~H.~Reno,
Nucl.\, Phys.\, B {\bf 319}, 37 (1989) [Erratum-ibid.\, B {\bf 330}, 284 (1990)].

\bibitem{ResBos} F.~Landry, R.~Brock, P.~M.~Nadolsky, and C.-P.~Yuan, Phys.\,Rev.\, D {\bf 67}, 073016 (2003); C.~Balazs and C.-P.~Yuan, Phys.\,Rev.\, D {\bf 56}, 5558 (1997); G.~A.~Ladinsky and C.-P.~Yuan, Phys.\,Rev.\, D {\bf 50}, 4239 (1994).

\bibitem{smalls}
S.~Berge, P.~M.~Nadolsky, F.~Olness and C.~-P.~Yuan, 
Phys.\, Rev.\, D {\bf 72}, 033015 (2005). 


\bibitem{CDFZPT2012} CDF Collaboration (T.~Aaltonen {\it et al.}), 
Phys.\, Rev.\, D {\bf 86}, 052010 (2012). 

\bibitem{FEWZ2}
  R.~Gavin, Y.~Li, F.~Petriello and S.~Quackenbush,
  Comput.\ Phys.\ Commun.\  {\bf 182}, 2388 (2011).
  
    
  
\bibitem{ijmpaQCD} ``Review of physics results from the Tevatron: QCD Physics,'' C.~Mesropian, D.~Bandurin, to be submitted to IJMPA [arXiv:XXXX.XXXX].

\bibitem{berger}
E.~L.~Berger, F.~Halzen, C.~S.~Kim and S.~Willenbrock,
Phys.\, Rev.\, D {\bf 40}, 83 (1989) [Erratum-ibid\, D {\bf 40}, 3789 (1989)].

\bibitem{CDFWasym1994} CDF Collaboration (F.~Abe {\it et al.}), 
Phys.\, Rev.\, Lett.\, {\bf 74}, 850 (1995). 

\bibitem{CDFWasym1998} CDF Collaboration (F.~Abe {\it et al.}), 
Phys.\, Rev.\, Lett.\, {\bf 81}, 5754 (1998).

\bibitem{CDFWasym2005} CDF Collaboration (D.~Acosta {\it et al.}), 
Phys.\, Rev.\, D {\bf 71}, 051104 (2005).

\bibitem{D0MUASYM} D0 Collaboration (V.~M.~Abazov {\it et al.}), 
Phys.\, Rev.\, D {\bf 77}, 011106 (2008). 

\bibitem{D0Wasym2008} D0 Collaboration (V.~M.~Abazov {\it et al.}), 
Phys.\, Rev.\, Lett.\, {\bf 101}, 211801 (2008). 

\bibitem{NNPDF2.3}
  R.~D.~Ball, V.~Bertone, S.~Carrazza, C.~S.~Deans, L.~Del Debbio, S.~Forte, A.~Guffanti and N.~P.~Hartland {\it et al.},
  Nucl.\ Phys.\ B {\bf 867}, 244 (2013).

\bibitem{D0WAsymMuon2013} D0 Collaboration (V.~M.~Abazov {\it et al.}), 
Phys.\, Rev.\, D {\bf 88}, 091102 (2013). 

\bibitem{Wasym} A.~Bodek, Y.~Chung, B.~-Y.~Han, K.~S.~McFarland and E.~Halkiadakis,
Phys.\, Rev.\, D {\bf 77}, 111301 (2008). 

\bibitem{CDFWasym2009} CDF Collaboration (T.~Aaltonen {\it et al.}), 
Phys.\, Rev.\, Lett.\, {\bf 102}, 181801 (2009). 

\bibitem{D0Wasym2014} D0 Collaboration (V.~M.~Abazov {\it et al.}),
Phys.\, Rev.\, Lett.\, {\bf 112}, 151803 (2014). 

\bibitem{CDFAFB1991} CDF Collaboration (F.~Abe {\it et al.}), 
Phys.\, Rev.\, Lett.\, {\bf 67}, 1502 (1991).

\bibitem{CDFAFB1996} CDF Collaboration (F.~Abe {\it et al.}), 
Phys.\, Rev.\, Lett.\, {\bf 77}, 2616 (1996).

\bibitem{CDFAFB2013} CDF Collaboration (T.~Aaltonen {\it et al.}), 
Phys.\, Rev.\, D {\bf 88}, 7 (2013),  072002 [Erratum-ibid.\, D {\bf 88}, 7 (2013) 079905]].

\bibitem{CDFAFB2014} CDF Collaboration (T.~Aaltonen {\it et al.}), 
Phys.\, Rev.\, D {\bf 89}, 072005 (2014). 

\bibitem{D0AFB2011} D0 Collaboration (V.~M.~Abazov {\it et al.}), 
Phys.\, Rev.\, D {\bf 84}, 012007 (2011). 

\bibitem{CSframe} J.~C.~Collins and D.~E.~Soper, Phys.\,Rev.\, D {\bf 16}, 2219 (1977).

 \bibitem{PYTHIA}
   T. Sj\"ostrand  \etal, Comp. Phys. Comm. {\bf 135}, 238 (2001). T.~Sjostrand, S.~Mrenna and P.~Z.~Skands,
  JHEP {\bf 0605}, 026 (2006).
  



\bibitem{Mirkes1992} E.~Mirkes, 
Nucl.\, Phys.\, B {\bf 387}, 3 (1992).


\bibitem{npconstraints} M.~Ciuchini, E.~Franco, S.~Mishima, and L.~Silvestrini, J.\,High\,Energy\,Phys.\, 08, 106 (2013).

\bibitem{sirlin} A.~Sirlin, Phys.\,Rev.\, D {\bf 22}, 971 (1980).  

\bibitem{STUreference} M.~E.~Peskin and T.~Takeuchi, Phys.\, Rev.\, Lett. {\bf 65}, 964 (1990); M.~Golden and L.~Randall, Nucl.\, Phys. B {\bf 361}, 3 (1990); 
 B.~Holdom and J.~Terning, Phys.\, Lett.\, B {\bf 247}, 88 (1990);
 M.~E.~Peskin and T.~Takeuchi, Phys.\, Rev.\, D {\bf 46}, 381 (1992); 
 G.~Altarelli and R.~Barbieri, Phys.\, Lett.\, B {\bf 253}, 161 (1991); G.~Altarelli, R.~Barbieri and S.~Jadach, Nucl.\, Phys.\, B {\bf 369}, 3 (1992) [Erratum-ibid. B {\bf 376}, 444 (1992)].

\bibitem{Baak:2014ora} M.~Baak, J.~Cuth, J.~Haller, A.~Hoecker, R.~Kogler, K.~Moenig, M.~Schott and J.~Stelzer,
arXiv:1407.3792 [hep-ph], submitted for publication in Eur.\,Phys.\,J.\,C. 

\bibitem{npGfitter} M~Baak, M.~Goebel, J.~Haller, A.~Hoecker, D.~Kennedy, K.~M\"{o}nig, M.~Schott, and J.~Stelzer (The Gfitter Group), Eur.\,Phys.\,J.\,C {\bf 72}, 2003 (2012).

\bibitem{WZdiscovery} G.~Arnison {\it et al.} (UA1 Collaboration), Phys.\,Lett.\,B {\bf 122}, 103 (1983);
UA2 Collaboration (M.~Banner {\it et al.}), Phys.\,Lett.\, B {\bf 122}, 476 (1983);
UA1 Collaboration (G.~Arnison {\it et al.}), Phys.\,Lett.\, B {\bf 126}, 398 (1983);
UA2 Collaboration (P.~Bagnaia {\it et al.}), Phys.\,Lett.\, B {\bf 129}, 130 (1983).

\bibitem{CDF} CDF Collaboration (T.~Affolder {\it et al.}), Phys.\,Rev.\, D {\bf 64}, 052001 (2001).

\bibitem{DZERO} D0 Collaboration (V.~M.~Abazov {\it et al.}), Phys.\,Rev.\, D {\bf 66}, 012001 (2002);
D0 Collaboration (B.~Abbott {\it et al.}), Phys.\,Rev.\, D {\bf 62}, 092006 (2000);
D0 Collaboration (B.~Abbott {\it et al.}), Phys.\,Rev.\, D {\bf 58}, 092003 (1998).

\bibitem{run1combo} D0 Collaboration (V.~M.~Abazov {\it et al.}) (CDF Collaboration, The D0 Collaboration, Tevatron Electroweak Working Group), Phys.\,Rev.\, D {\bf 70}, 092008 (2004). 

\bibitem{lepewwg} ALEPH, CDF, D0, DELPHI, L3, OPAL, and SLD Collaborations, the LEP Electroweak Working Group, the Tevatron Electroweak Working Group, and the SLD Electroweak and Heavy Flavour Groups, CERN Report No. CERN-PH-EP-2010-095, and Fermilab Report No. FERMILAB-TM-2480-PPD, 2010.

\bibitem{ALEPH} ALEPH Collaboration (S.~Schael {\it et al.}), Eur.\,Phys.\,Jour.\, C {\bf 47}, 309 (2006). 

\bibitem{DELPHI} DELPHI Collaboration (J.~Abdallah {\it et al.}), Eur.\,Phys.\,Jour.\, C {\bf 55}, 1 (2008).

\bibitem{L3} L3 Collaboration (P.~Achard {\it et al.}), Eur.\,Phys.\,Jour.\, C {\bf 45}, 569 (2006).

\bibitem{OPAL} OPAL Collaboration (G.~Abbiendi {\it et al.}), Eur.\,Phys.\,Jour.\, C {\bf 45}, 307 (2006).

\bibitem{CDF2} CDF Collaboration (T.~Aaltonen {\it et al.}), Phys.\,Rev.\,Lett.\, {\bf 99}, 151801 (2007); 
CDF Collaboration (T.~Aaltonen {\it et al.}), Phys.\,Rev.\, D {\bf 77}, 112001 (2008).

\bibitem{cdf2fbprl} CDF Collaboration (T.~Aaltonen {\it et al.}), Phys.\,Rev.\,Lett.\, {\bf 108}, 151803 (2012).

\bibitem{DZERO2} D0 Collaboration (V.~M.~Abazov {\it et al.}), Phys.\,Rev.\,Lett.\, {\bf 103}, 141801 (2009).

\bibitem{dzero5fbprd} D0 Collaboration (V.~M.~Abazov {\it et al.}), Phys.\,Rev.\, D {\bf 89} 012005 (2014). 

\bibitem{run2combo} CDF and D0 Collaborations (T.~Aaltonen {\it et al.}), Phys.\,Rev.\, D {\bf 88}, 052018 (2013).

\bibitem{kotwalStark} A.~V.~Kotwal and J.~Stark, Ann.\,Rev.\,Nucl.\,Part.\,Sci.\, {\bf 58}, 147 (2008).

\bibitem{pdg} Particle Data Group (K.~Nakamura {\it et al.}), J.\,Phys.\, G {\bf 37}, 075021 (2010).

\bibitem{GEANT} R.~Brun and F.~Carminati, CERN Program Library Long Writeup, W5013, 1993 (unpublished), version 3.15.

\bibitem{cdf2fbprd} CDF Collaboration (T.~Aaltonen {\it et al.}), Phys.\,Rev.\, D {\bf 89} 072003 (2014). 

\bibitem{COT} T. Affolder {\it et al.}, Nucl.\,Instrum.\,Methods\,Phys.\,Res., Sect.\, A {\bf 526}, 249 (2004).

\bibitem{solenoid} H.~Minemura {\it et al.}, Nucl.\,Instrum.\,Methods\,Phys.\,Res., Sect.\, A {\bf 238}, 18 (1985).

\bibitem{CEM} L.~Balka {\it et al.}, Nucl.\,Instrum.\,Methods\,Phys.\,Res., Sect.\,A {\bf 267}, 272 (1988).

\bibitem{cemresponse} K.~Yasuoka, S.~Mikamo, T.~Kamon, and A.~Yamashita, Nucl.\,Instrum.\,Meth.\,Phys.\, Res., Sect.\, A {\bf 267}, 315 (1987).

\bibitem{cosmic} A.~V.~Kotwal, H.~K.~Gerberich and C.~Hays, Nucl.\,Instrum.\,Methods\, Phys.\, Res., Sect.\, A {\bf 506}, 110 (2003).

\bibitem{cosmicAlignment} A.~V.~Kotwal and C.~Hays, Nucl.\,Instrum.\,Methods\,Phys.\,Res., Sect.\, A {\bf 762}, 85 (2014).

\bibitem{migdal} A.~B.~Migdal, Phys.\,Rev {\bf 103}, 1811 (1956); L.~D.~Landau and I.~J.~Pomeranchuk, Dokl.\,Akad.\,Nauk.\, SSSR {\bf 92}, 535 (1953); {\bf 92}, 735 (1953).

\bibitem{bichsel} H.~Bichsel, Rev.\,Mod.\,Phys. {\bf 60}, 663 (1988).

\bibitem{ms} G.~R.~Lynch and O.~I.~Dahl, Nucl.\,Instrum.\,Methods\, Phys.\, Res., Sect.\, B {\bf 58}, 6 (1991).

\bibitem{mstail} MuScat Collaboration (D.~Attwood {\it et al.}), Nucl.\,Instrum.\,Methods\,Phys.\, Res., Sect.\, B {\bf 251}, 41 (2006).

\bibitem{calgeantnim} A.~V.~Kotwal and C.~Hays, Nucl.\,Intrum.\,Methods\, Phys.\, Res., Sect.\, A {\bf 729}, 25 (2013).

\bibitem{CTEQ} P.~Nadolsky, H.~Lai, Q.~Cao, J.~Huston, J.~Pumplin, D.~Stump, W.~Tung, and C.-P.~Yuan, Phys.\,Rev.\, D {\bf 78}, 013004 (2008);
J.~Pumplin, D.~Stump, J.~Huston, H.~Lai, P.~Nadolsky, and W.~Tung, J.\,High Energy Phys. 0207, 012 (2002).


\bibitem{blny}  F.~Landry, R.~Brock, P.~Nadolsky, and C.-P.~Yuan, Phys.\,Rev.\, D {\bf 67}, 073016 (2003).

\bibitem{photos} P.~Golonka and Z.~Was, Eur.\,Phys.\,J.\, C {\bf 45}, 97 (2006).


\bibitem{horace} C.~M.~Carloni~Calame, G.~Montagna, O.~Nicrosini, and A.~Vicini, J.\,High Energy Phys.\, 10 (2007) 109.

\bibitem{cdfWwidth} CDF Collaboration (T.~Aaltonen {\it et al.}), Phys.\,Rev.\,Lett.\, {\bf 100}, 071801 (2008). 
%
\bibitem{d0Wwidth} D0 Collaboration (V.~M.~Abazov {\it et al.}), Phys.\,Rev.\,Lett.\, {\bf 103}, 231802 (2009).

\bibitem{cdf-runI-tgc1} CDF Collaboration (F.~Abe {\it et al}), Phys.\,Rev.\,Lett.\, {\bf{74}}, 1941 (1995).
\bibitem{cdf-runI-tgc2} CDF Collaboration (F.~Abe {\it et al}), Phys.\,Rev.\,Lett.\, {\bf{74}}, 1936 (1995).
\bibitem{cdf-runI-tgc3} CDF Collaboration (F.~Abe {\it et al}), Phys.\,Rev.\,Lett.\, {\bf{75}}, 1017 (1995).
\bibitem{d0-runI-tgc1} D0 Collaboration (V.~M.~Abachi {\it et al}), Phys.\,Rev.\,Lett.\, {\bf{75}}, 1028 (1995).
\bibitem{d0-runI-tgc2} D0 Collaboration (V.~M.~Abachi {\it et al}), Phys.\,Rev.\,Lett.\, {\bf{75}}, 1034 (1995).
\bibitem{d0-runI-tgc3} D0 Collaboration (V.~M.~Abachi {\it et al}), Phys.\,Rev.\,Lett.\, {\bf{77}}, 3303 (1996).
\bibitem{d0-runI-tgc4} D0 Collaboration (V.~M.~Abachi {\it et al}), Phys.\,Rev.\,Lett.\, {\bf{78}}, 3634 (1997).
\bibitem{d0-runI-tgc5} D0 Collaboration (V.~M.~Abachi {\it et al}), Phys.\,Rev.\,Lett.\, {\bf{78}}, 3640 (1997).
\bibitem{d0-runI-tgc6} D0 Collaboration (V.~M.~Abachi {\it et al}), Phys.\,Rev.\, D {\bf{56}}, 6742 (1997).
\bibitem{d0-runI-tgc7} D0 Collaboration (V.~M.~Abbott {\it et al}), Phys.\,Rev.\,Lett.\, {\bf{79}}, 1441 (1997).
\bibitem{d0-runI-tgc8} D0 Collaboration (V.~M.~Abbott {\it et al}), Phys.\,Rev.\, D {\bf{57}}, 3817(R) (1998).
\bibitem{d0-runI-tgc9} D0 Collaboration (V.~M.~Abbott {\it et al}), Phys.\,Rev.\, D {\bf{58}}, 031102 (1998).
\bibitem{d0-runI-tgc10} D0 Collaboration (V.~M.~Abbott {\it et al}), Phys.\,Rev.\, D {\bf{58}}, 051101 (1998).
\bibitem{d0-runI-tgc11} D0 Collaboration (V.~M.~Abbott {\it et al}), Phys.\,Rev.\, D {\bf{60}}, 072002 (1999).
\bibitem{d0-runI-tgc12} D0 Collaboration (V.~M.~Abbott {\it et al}), Phys.\,Rev.\, D {\bf{62}}, 052005 (2000).
\bibitem{cdf-runI-xs1} CDF Collaboration (F.~Abe {\it et al}), Phys.\,Rev.\,Lett.\, {\bf{78}}, 4537 (1997). 
\bibitem{d0-runI-xs1} D0 Collaboration (V.~M.~Abachi {\it et al}), Phys.\,Rev.\,Lett.\, {\bf{75}}, 1023 (1995).
\bibitem{mcfm} J.~Campbell, R.~Ellis, and C.~Williams, J.\, High\, Energy\, Phys.\, {\bf 07}, 018 (2011).
\bibitem{cdf-wgamma-xs} CDF Collaboration (T.~Acosta {\it et al}), Phys.\,Rev.\,Lett.\, {\bf{94}}, 041803 (2005).
\bibitem{d0-wgamma1} D0 Collaboration (V.~M.~Abazov {\it et al}), Phys.\,Rev.\, D {\bf{71}}, 091108(R) (2005).
\bibitem{d0-wgamma3} D0 Collaboration (V.~M.~Abazov {\it et al}), Phys.\,Rev.\,Lett.\, {\bf{107}}, 241803 (2011).
\bibitem{cdf-zgamma1} CDF Collaboration (T.~Aaltonen {\it et al}), Phys.\,Rev.\, D {\bf{82}}, 031103 (2010).
\bibitem{d0-zgamma1} D0 Collaboration (V.~M.~Abazov {\it et al}), Phys.\,Rev.\,Lett.\, {\bf{95}}, 051802 (2005).
\bibitem{d0-zgamma2} D0 Collaboration (V.~M.~Abazov {\it et al}), Phys.\,Lett.\, {\bf{B 653}}, 378 (2007).
\bibitem{d0-zgamma3} D0 Collaboration (V.~M.~Abazov {\it et al}), Phys.\,Rev.\,Lett.\, {\bf{102}}, 201802 (2009).
\bibitem{d0-zgamma4} D0 Collaboration (V.~M.~Abazov {\it et al}), Phys.\,Rev.\, D {\bf{85}}, 052001 (2012).

\bibitem{d0-zz1-xs} D0 Collaboration (V.~M.~Abazov {\it et al}), Phys.\,Rev.\,Lett.\, {\bf{101}} 171803 (2008).
\bibitem{cdf-zz1-xs} CDF Collaboration (T.~Aaltonen {\it et al}), Phys.\,Rev.\,Lett.\, {\bf{100}}, 201801 (2008).
\bibitem{d0-zz-rayan} D0 Collaboration (V.~M.~Abazov {\it et al}), Phys.\,Rev.\, D {\bf{84}}, 011103 (2011). 
\bibitem{zzphi} Q.~Cao, C.~B.~Jackson, W.~Keung, I.~Low, and J.~Shu, Phys.\,Rev.\, D {\bf{81}}, 015010 (2010).
\bibitem{d0-zz3-xs} D0 Collaboration (V.~M.~Abazov {\it et al}), Phys.\,Rev\, D {\bf{88}}, 032008 (2013).

\bibitem{cdf-zz2-xs} CDF Collaboration (T.~Aaltonen {\it et al}), Phys.\,Rev.\,Lett. {\bf{108}} 101801 (2012).
\bibitem{d0-zz2-xs} D0 Collaboration (V.~M.~Abazov {\it et al}), Phys.\,Rev\, D {\bf{85}}, 112005 (2012).
\bibitem{d0-zz-emanuel} D0 Collaboration (V.~M.~Abazov {\it et al}), Phys.\,Rev\, D {\bf{78}}, 072002 (2008).
\bibitem{cdf-zz3-xs} CDF Collaboration (T.~Aaltonen {\it et al}), Phys.\,Rev.\, D {\bf{89}}, 112001 (2014).

\bibitem{cdf-ww1-xs} CDF Collaboration (T.~Aaltonen {\it et al}), Phys.\,Rev.\,Lett.\, {\bf{94}}, 211801 (2005).
\bibitem{cdf-wz-xs} CDF Collaboration (T.~Aaltonen {\it et al}), Phys.\,Rev.\,Lett.\, {\bf{98}}, 161801 (2007).
\bibitem{d0-ww1-xs} D0 Collaboration (V.~M.~Abazov {\it et al}), Phys.\,Rev.\,Lett.\, {\bf{94}}, 151801 (2005).
\bibitem{d0-ww-xs1} D0 Collaboration (V.~M.~Abazov {\it et al}), Phys.\,Rev.\,Lett.\, {\bf{103}}, 191801 (2009).
\bibitem{d0-wz1-xs} D0 Collaboration (V.~M.~Abazov {\it et al}), Phys.\,Rev.\,Lett.\, {\bf{95}}, 141802 (2005).
\bibitem{d0-wz2-xs} D0 Collaboration (V.~M.~Abazov {\it et al}), Phys.\,Rev.\, D {\bf{76}}, 111104 (2007).
\bibitem{d0-wz2-lnull} D0 Collaboration (V.~M.~Abazov {\it et al}), Phys.\,Lett.\, B {\bf{695}}, 67 (2011).


\bibitem{cdf-ww-xs1} CDF Collaboration (T.~Aaltonen {\it et al}), Phys.\,Rev.\,Lett.\, {\bf{104}}, 201801 (2010).
\bibitem{cdf-wz-xs1} CDF Collaboration (T.~Aaltonen {\it et al}), Phys.\,Rev.\, D {\bf{86}}, 031104(R) (2012).
\bibitem{cdf-wwwz-xs1} CDF Collaboration (T.~Aaltonen {\it et al}), Phys.\,Rev.\,Lett.\, {\bf{103}}, 091803 (2009).
\bibitem{d0-wwwz-xs1} D0 Collaboration (V.~M.~Abazov {\it et al}), Phys.\,Rev.\,Lett.\, {\bf{102}}, 161801 (2009).
\bibitem{cdf-wwwz-xs2} CDF Collaboration (T.~Aaltonen {\it et al}), Phys.\,Rev.\,Lett.\, {\bf{104}}, 101801 (2010).
\bibitem{cdf-wwwz-xs3} CDF Collaboration (T.~Aaltonen {\it et al}),  Phys.\,Rev\, D {\bf{82}}, 112001 (2010).
\bibitem{d0-wwwz-xs2} D0 Collaboration (V.~M.~Abazov {\it et al}), Phys.\,Rev.\,Lett.\, {\bf{108}}, 181803 (2012).
\bibitem{cdf-wwwz-xs4} CDF Collaboration (T.~Aaltonen {\it et al}),  Phys.\,Rev\, D {\bf{85}}, 012002 (2012).
\bibitem{d0-vz-xs1} D0 Collaboration (V.~M.~Abazov {\it et al}), Phys.\,Rev.\,Lett.\, {\bf{109}}, 121802 (2012).
\bibitem{d0-vz-xs2} D0 Collaboration (V.~M.~Abazov {\it et al}), Phys.\,Rev\, D {\bf{88}}, 052011 (2013).
\bibitem{cdf-vz-xs2} CDF Collaboration (T.~Aaltonen {\it et al}), Phys.\,Rev\, D {\bf{88}}, 092002 (2013).
\bibitem{tev-combo1} CDF and D0 Collaborations (T.~Aaltonen {\it et al}), Phys.\,Rev.\,Lett.\, {\bf{109}}, 071804 (2012). 
\bibitem{tev-combo2} CDF and D0 Collaborations (T.~Aaltonen {\it et al}), Phys.\,Rev.\, D {\bf{88}}, 052014 (2013).

\bibitem{lagrangian} K.~Hagiwara, R.~D.~Peccei, and D.~Zeppenfeld, Nucl.\, Phys.\, {\bf B282}, 253 (1987).
\bibitem{HWZ} K.~Hagiwara, J.~Woodside, and D.~Zeppenfeld, Phys.\, Rev.\, D {\bf 41}, 2113 (1990).
\bibitem{strong} S.~Weinberg, Phys.\,Rev.\,D {\bf 13}, 974 (1976); L.~Susskind, Phys.\,Rev.\, D {\bf 20}, 2619 (1979); 
H.~P.~Nilles, Phys.\, Rep.\, {\bf 110}, 1 (1984); H.~E.~Haber and G.~L.~Kane, Phys.\, Rep.\, {\bf 117}, 75 (1985); 
A.~G.~Cohen, D.~B.~Kaplan, and A.~E.~Nelson, Phys.\, Lett.\, B {\bf{388}}, 588 (1996); 
C.~Csaki, C.~Grojean, L.~Pilo, and J.~Terning, Phys.\, Rev.\, Lett.\, {\bf 92}, 101802 (2004); 
R.~Foadi, S.~Gopalakrishna, and C.~Schmidt, JHEP {\bf 0403}, 042 (2004).
\bibitem{su2u1} C.~Grosse-Knetter, I.~Kuss, and D.~Schildknecht, Z.\, Phys.\, C {\bf{60}}, 375 (1993); 
M.~Bilenky, J.~L.~Kneur, F.~M.~Renard, and D.~Schildknecht, Nucl.\, Phys.\, {\bf{B409}}, 22 (1993); Nucl.\,Phys.\, {\bf{B419}}, 240 (1994).
\bibitem{hisz} K.~Hagiwara {\sl et al.}, Phys.\,Rev.\, D {\bf{48}} (1993).
\bibitem{zgammaBaur} U.~Baur and E.~Berger, Phys.\,Rev.\, D {\bf{47}}, 4889 (1993); U.~Baur, T.~Han, J.~Ohnemus,  Phys.\,Rev.\, D {\bf{57}}, 2823 (1998).
\bibitem{zzBaur} U.~Baur, D.~Rainwater, Phys.\,Rev.\, D {\bf{62}}, 113011 (2000); U.~Baur, D.~Rainwater, Int.\,J.\,Mod.\,Phys.\, A16S1A, 315 (2001).
\bibitem{hwz} K.~Hagiwara, J.~Woodside, and D.~Zeppenfeld, Phys.\, Rev.\, D {\bf 41}, 2113 (1990).
\bibitem{baurs} U.~Baur, T.~Han, and J.~Ohnemus, Phys.\, Rev\, D {\bf 48}, 5140 (1993).
\bibitem{cdf-zgamma} CDF Collaboration (T.~Aaltonen {\it et al}), Phys.\,Rev.\,Lett.\, {\bf{107}}, 051802 (2011).
\bibitem{cdf-tgc-combo} CDF Collaboration (T.~Aaltonen {\it et al}), Phys.\,Rev\, D {\bf{76}}, 111103 (2007).

\bibitem{d0-wgamma2} D0 Collaboration (V.~M.~Abazov {\it et al}), Phys.\,Rev.\,Lett.\, {\bf{100}}, 241805 (2008).
\bibitem{d0-zz} D0 Collaboration (V.~M.~Abazov {\it et al}), Phys.\,Rev.\,Lett.\, {\bf{100}}, 131801 (2008).
\bibitem{d0-wz1-lnull} D0 Collaboration (V.~M.~Abazov {\it et al}), Phys.\,Rev.\,Lett.\, {\bf{95}}, 141802 (2005).
\bibitem{d0-combo-tgc} D0 Collaboration (V.~M.~Abazov {\it et al}), Phys.\,Lett.\, B {\bf{718}}, 451 (2012).
\bibitem{d0-lnulnu-first} D0 Collaboration (V.~M.~Abazov {\it et al}), Phys.\,Rev.\, D {\bf{74}}, 057101 (2006).
\bibitem{d0-lnujj-tgc} D0 Collaboration (V.~M.~Abazov {\it et al}), Phys.\,Rev.\, D {\bf{80}}, 053012 (2009).
\bibitem{qgc} G.~Belanger and F.~Boudjema, Phys.\,Lett.\. B {\bf{288}}, 210 (1992).
\bibitem{d0-qgc} D0 Collaboration (V.~M.~Abazov {\it et al}), Phys.\,Rev.\, D {\bf{88}}, 012005 (2013).

\end{thebibliography}
\end{document}